\documentclass[useAMS,usenatbib]{mn2e}

\usepackage{verbatim,graphicx,epsfig,dcolumn,xcolor,amsmath}
\newcolumntype{.}{D{.}{.}{4}}
\newcolumntype{,}{D{.}{.}{2}}
\newcolumntype{;}{D{.}{.}{1}}
\newcommand{\nodata}{$\cdot\cdot\cdot$}
\newcommand{\lesssim}{{\lower-1.2pt\vbox{\hbox{\rlap{$<$}\lower5pt\vbox{\hbox{$\sim$}}}}}}
\newcommand{\gtrsim}{{\lower-1.2pt\vbox{\hbox{\rlap{$>$}\lower5pt\vbox{\hbox{$\sim$}}}}}}
\def\clap#1{\hbox to0pt{\hss#1\hss}}
\newcommand{\appropto}{\mathrel{\vcenter{
  \offinterlineskip\halign{\hfil$##$\cr
    \propto\cr\noalign{\kern2pt}\sim\cr\noalign{\kern-2pt}}}}}

\title[Gaia DR1]{Fundamental parameters and infrared excesses of Tycho--\emph{Gaia} stars}
\author[I. McDonald et al.]{I.~McDonald$^{1}$\thanks{E-mail: mcdonald@jb.man.ac.uk}, A.~A.~Zijlstra$^{1,2}$, R.~A.~Watson$^{1}$\\
$^{1}$Jodrell Bank Centre for Astrophysics, Alan Turing Building, Manchester, M13 9PL, UK\\
$^{2}$Department of Physics, The University of Hong Kong, Pokfulam Road, Hong Kong}

\begin{document}

\date{Accepted 9999 December 32. Received 9999 December 32; in original form 9999 December 32}

\pagerange{\pageref{firstpage}--\pageref{lastpage}} \pubyear{9999}

\maketitle

\label{firstpage}

\begin{abstract}
Effective temperatures and luminosities are calculated for 1\,475\,921 Tycho-2 and 107\,145 \emph{Hipparcos} stars, based on distances from \emph{Gaia} Data Release 1. Parameters are derived by comparing multi-wavelength archival photometry to {\sc bt-settl} model atmospheres. The 1$\sigma$ uncertainties for the Tycho-2 and \emph{Hipparcos} stars are $\pm$137 K and $\pm$125 K in temperature and $\pm$35 per cent and $\pm$19 per cent in luminosity. The luminosity uncertainty is dominated by that of the \emph{Gaia} parallax. Evidence for infrared excess between 4.6 and 25 $\mu$m is found for 4256 stars, of which 1883 are strong candidates. These include asymptotic giant branch (AGB) stars, Cepheids, Herbig Ae/Be stars, young stellar objects, and other sources. We briefly demonstrate the capabilities of this dataset by exploring local interstellar extinction, the onset of dust production in AGB stars, the age and metallicity gradients of the solar neighbourhood and structure within the Gould Belt. We close by discussing the potential impact of future \emph{Gaia} data releases.
\end{abstract}



\section{Introduction}
\label{IntroSect}

Modern precision astrometry has recovered distances to large samples of nearby stars, the pinnacles of which are the catalogues returned by the \emph{Hipparcos} \citep{Perryman89} and \emph{Gaia} satellites \citep{Perryman01,GaiaMission,GaiaDR1}. These catalogues provide the basic measurements of colour, brightness and parallactic distance. They do not contain fundamental parameters, such as temperature or luminosity. Hence, `value added' catalogues are often computed, e.g.\ \citep{AF12} and \citep{MZB12} for the \emph{Hipparcos} dataset. The latter of these papers provides a catalogue of stellar fundamental parameters, which is replicated here using the \emph{Gaia} satellite's Data Release 1\footnote{\tt{http://gea.esac.esa.int/archive/}}.

\emph{Gaia} DR1 is based on the first six months of \emph{Gaia} operations. It lists parallaxes for 2\,057\,050 stars contained in the \emph{Hipparcos} Tycho-2 catalogue \citep{HFM00,MLH15}. We use spectral energy distribution (SED) fitting of pre-existing photometry to place those stars on the true Hertzsprung--Russell (H--R) diagram. We also identify the stars among them with infrared excess: i.e.\ excess flux in the mid-infrared ($\sim$3--30 $\mu$m) when compared to the spectral energy distribution from a stellar atmosphere model.

SED fitting to determine stellar parameters has its advantages and limitations. Compared to simple, single-colour bolometric corrections, it can be more robust against bad photometric data. It can also be more accurate, due to the larger number of data points included, and it can be effective over a wider range of stellar effective temperatures.  Secondary effects, such as binary companions or reprocessing of stellar light, can sometimes be identified where simple bolometric corrections would not be able to do so. Both bolometric corrections and SED fitting are equally limited by prior assumptions of stellar metallicity, surface gravity and interstellar extinction, which determine the properties of the stellar atmosphere models that the stars are compared against. Stellar temperatures and luminosities from SED fitting are most accurate if both the short- (Wien) and long-wavelength (Rayleigh--Jeans) tails of the SED are covered with good-quality photometry.

Spectroscopic temperature determinations generally have greater accuracy than those obtained through SED fitting. They can also measure metallicity and surface gravity, and are not affected by extinction. However, SED fitting is observationally and computationally much cheaper, allowing it both to be used on fainter stars, and to more effectively survey a larger number of stars. SED fitting provides a more accurate luminosity than can be derived via spectroscopic measurements. This allows SED fits to be used to be used to select targets for more expensive follow-up campaigns.

Both photometric colours and spectroscopy often fail to identify infrared excess. Infrared excess is typically caused by warm dust in the circumstellar environment. It is therefore a good tracer of objects at both ends of stellar evolution: young and pre-main-sequence stars that have yet to clear their circumstellar environments of their proto-planetary discs, and evolved stars that are undergoing the terminal process of stellar mass loss (e.g.\ \citealt{CS16}). Other mass-losing or mass-gaining stars can also be identified, such as interacting binary stars containing an accretion disc, Wolf--Rayet stars, and Herbig A[e]/B[e] stars. Unlike simple photometric colours, infrared excess can also trace unresolved, non-interacting binary companions and physically separate line-of-sight binary stars, if the contrast ratio is sufficiently close to unity and the colours are sufficiently different.

In this paper, we cross-reference catalogues of multi-wavelength literature photometry to construct SEDs for stars in the Tycho-2 and \emph{Hipparcos} catalogues \citep{HFM00,vanLeeuwen07}, supplemented by the Tycho--\emph{Gaia} astrometric solution from \emph{Gaia} DR1. These are compared against stellar atmosphere models to derive effective temperatures for each star. When combined with the parallax information from \emph{Gaia} DR1, this allows us to derive the luminosity of each star (Section \ref{SEDSect}) and to place it on the H--R diagram. The H--R diagram is presented, and the uncertainties in individual measurements discussed (Section \ref{HRSect}). A catalogue of stars which likely exhibit excess infrared flux is presented, and their categorisation and location in the H--R diagram is discussed (Section \ref{DiscXSSect}). Here, we also explore dust production by evolved stars. Further applications and details are presented in the online appendices which accompany this paper.


\section{The SED fitting process}
\label{SEDSect}

\subsection{Methodology}
\label{SEDMethodSect}

\subsubsection{Cross-referencing photometric source data}
\label{SEDDataSect}

This section describes the methodology used to create the SEDs and fit the data. The practical application is detailed in Section \ref{SEDAnalSect}.

A cross-reference catalogue was intended to form part of the \emph{Gaia} Data Release 1 but was not provided with the data release itself. For this paper, photometric data was collected using the CDS `X-Match' cross-matching service\footnote{http://cdsxmatch.u-strasbg.fr/xmatch}, which provides fast, effective cross-matching across a variety of photometric catalogues.

While fast and efficient, the VizieR cross-matching service contains some limitations. For example, in the following analysis, SDSS Data Release 7 was used in preference to Data Release 9: although DR9 is more complete, the VizieR implementation also matches child objects\footnote{Sources which SDSS notes as resolved or overlapping are assigned a parent object, then deblended and decomposed into child objects. This process can also occur with saturated stars and artifacts associated with them. Further details are given on the SDSS webpages: http://www.sdss.org/dr12/algorithms/deblend/} instead of their parents, resulting in improper photometric matches. Flagging data from DR7 was not passed to the cross-matching service\footnote{We thank the staff at Centre de Donn\'ees astronomiques de Strasbourg for later including these on our suggestion.}, meaning (e.g.) saturated stars cannot automatically be removed.

A further limitation is that source proper motion is not accounted for during the cross-matching process. Already, many nearby stars are not in the \emph{Gaia} DR1 sample due to their proper motion cutoff of 750 mas yr$^{-1}$. Unfortunately, this lack of accounting for proper motion appears to remove considerably more. The effect depends both on the 3$\sigma$ tolerance and the temporal spacing between catalogues. For recent ($\sim$2012) catalogues like AllWISE, comparison to the $\sim$1991 Tycho photometry with a limit of 1.2$^{\prime\prime}$ risks removing any object with proper motion greater than 57 mas yr$^{-1}$, or 5 per cent of the combined Tycho--\emph{Hipparcos} sample.

From this compiled list, we removed stars where the photometric parallax is too uncertain to obtain a meaningful luminosity. We dictated this to be when the uncertainty in the parallax ($\delta \varpi$)\footnote{In the remainder of this work, we use $\delta$ to denote the uncertainty on an individual object, and $\sigma$ to denote the standard deviation, uncertainty, or any other noted derivative of variance in a statistical ensemble.} led to a factor of two uncertainty in the stellar luminosity, i.e.\ when $\delta \varpi / \varpi > 0.414$. This reduced the number of Tycho--\emph{Gaia} sources to from 2\,057\,050 to 1\,535\,006. We explicitly note that the parallax cut-off we have made means that this is not a volume-selected or volume-limited sample. It should not be considered complete for any given set of stars, and retains the biases and limitations present in the \emph{Gaia} and \emph{Tycho} catalogues, and the other photometric catalogues used later.

The bespoke, iterative methods by which we removed bad data from the compiled SEDs are detailed later, in Section \ref{SEDBadSect} and the online Appendix.

We stress that this sample of stars is subject to the Lutz--Kelker bias \citep{LK73}. The fractional parallax uncertainty we have used is still relatively lax, and we encourage users to adopt stricter criteria for volume-limited samples. The minimum suggested criterion we can recommend is the $\delta \varpi / \varpi < 0.2$ limit we use in parts of our analysis below \citep[cf.][]{BailerJones15}. Further discussion on Lutz--Kelker-related effects can be found in Section \ref{HRErrorLutzKelkerSect}.

\subsubsection{SED-fitting methology}
\label{SEDMethodSect}

Once the source data is collated to provide an SED for each star, the fitting procedure can determine the best-fit spectral model and derive the stellar temperature and luminosity. The {\sc getsed} SED-fitting pipeline used here was first described in \citet{MvLD+09} and updated in \citet{MZB12}. The pipeline has been altered slightly for this paper to improve efficiency and reduce artefacts in the final H--R diagrams caused by discrepant data. The following provides an account of the fitting procedure, including these alterations.

The pipeline begins with an SED from observed photometry in the form of $\lambda,\,F_\nu$. Required meta-data are the (\emph{Gaia}) distance, the interstellar extinction to the star, and the stellar metallicity. Unless stated otherwise, in the following discussion we use an assumption of solar metallicity and zero extinction.

{\it Step 1:} The best-fitting blackbody is calculated to provide a first estimate of stellar parameters. Each filter is reduced to a single, representative wavelength. The flux of a blackbody at these wavelengths is calculated for a grid of temperatures with 400 K spacing over the range 2600--7400 K. The blackbody is normalised to the wavelength-integrated (bolometric) flux of the observed SED, and a $\chi^2$ minimum is computed. This and later $\chi^2$ minima are determined in magnitudes, rather than fluxes, to avoid giving undue weight to points around the SED peak. If the best-fitting temperature is 7400 K, the temperature range is extended up to 20\,000 K, then 60\,000 K. A sub-grid is defined at $\pm$200 K from the best-fitting temperature, and a $\chi^2$ minimum computed, then iterated down to 100 K and 50 K, thus fitting a blackbody temperature between 2250 and 60350 K with 50 K resolution.

The apparent bolometric flux of the blackbody fit is used in combination with the input distance to determine the luminosity of the fitted blackbody. This identifies whether the star is a main-sequence star or a giant. A mass is estimated using the procedure described in \citet{MZB12}, and this mass is used to obtain a surface gravity, $\log(g)$. The temperature change caused by an imperfect mass and log($g$) estimate is small compared to the total error budget (Section \ref{HRErrorSect}), provided the mass is within a factor of $\sim$10 of the true value. For main-sequence and red giant branch (RGB) stars, we expect our masses to be correct to well within a factor of two, and for asymptotic giant branch (AGB) stars within a factor of four to ten (depending on their luminosity).

{\it Step 2:} Unlike previous implementations, we now repeat this process with a grid of model atmospheres. For this paper, we use the {\sc bt-settl} models of \citet{AGL+03}. We use these in preference to the more widely used {\sc marcs} models \citep{DMA+04,GEE+08} because of their greater completeness. While there are substantial and astrophysically important differences between these models, tests performed in \citet{MZB12} showed that the choice of model atmosphere has negligible impact on the final temperature derived for a variety of types of star.

Each model in the grid is reddened, using the procedure described in \citet[][see also Section \ref{HRErrorEBVSect}]{MvLD+09}, and convolved with a list of filter transmission functions. The flux that would be observed in each filter, and the relative reddening in that filter ($A_{\lambda}/A_V$), are tabulated.

Models are selected from the grid, bracketting the star's assumed metallicity and log($g$). This creates a selection of four models at each temperature point. A two-dimensional, linear interpolation is made to obtain a single photometric flux for each band at each gridded temperature point. The luminosity of each model is then normalised to the luminosity of the SED, and a $\chi^2$ minimum performed to determine the best-fitting temperature. A new value for $\log(g)$ is determined.

{\it Step 3:} We interpolate within the now-one-dimensional temperature model grid, modify $\log(g)$, and iterate to a solution. This last two-stage interpolation is the most computationally expensive part of the analysis: unlike before, this interpolation is performed for each point on each filter transmission function, therefore better accounting for wavelength-dependent effects such as molecular band strength changes and interstellar reddening. The two stages of this interpolation are as follows.

(a) We begin our initial temperature interpolation by computing two models, above and below the best-fit temperature. The deviation above and below is taken as the largest power of two which is numerically less than the temperature grid spacing of the stellar atmosphere models: e.g.\ if the grid spacing is 100 K, the models are computed at the gridded best-fitting temperature $\pm$ 64 K; if the grid spacing is 250 K, a deviation of $\pm$ 128 K is applied. If one of these interpolated models is a better $\chi^2$ fit than the original, its temperature is adopted as the new best fit, otherwise the old best-fitting temperature remains. Models are computed at the new best-fitting temperature $\pm$ half the previous value, and the process iterated. In our example, that is namely $\pm$ 32 K, then $\pm$ 16 K, $\pm$ 8 K, $\pm$ 4 K, $\pm$ 2 K, and $\pm$ 1 K, allowing the new best-fit temperature to deviate from the original by up to 127 K.

(b) A new $\log(g)$ is now determined, and the temperature iteration begun again. To optimise the system, the process begins at the smallest power of two above the deviation from the original value. For example, a star may be initially fit at 5800 K, and interpolated to 5776 K, the difference being 24 K. The interpolation would then start by interpolating new models at 5776 $\pm$ 32 K, rather than $\pm$ 64 K as previously.

These two steps (a \& b) are iterated until a solution is found. In a small fraction of cases, the solution can oscillate between two solutions, or run towards zero or infinity. To prevent this, the starting deviation of each interpolation is tapered. It is allowed to run at the initial value for three times, then is limited by half at each step. In our example, this limits the interpolation to a maximum deviation to $\pm$ 64, 64, 64, 32, 16, 8, 4, 2 and 1 K on subsequent iterations. This allows our example model to deviate by no more than 255 K from its initial best-fit value (for a grid spacing of 100 K). Investigation showed that this was sufficient to account for any difference in temperature caused by a revised log($g$).

{\it Step 4:} Once a best-fit temperature, luminosity and log($g$) have been determined, the final interpolated model atmosphere is integrated in frequency and a final luminosity produced. The normalised $\chi^2$ minimum is calculated. For each of the $n$ observed filters, the ratio of the observed to modelled flux ($R_{\rm n} = F_{\rm o}/F_{\rm m}$) is computed. A goodness-of-fit metric ($Q$) is calculated, based on the number of points ($n$):
\begin{equation}
Q = \sum_n \frac{\left( R^\ast_{\rm n} - 1 \right)}{n},
\end{equation}
where $R^\ast_{\rm n} = R_{\rm n}$ if $R_{\rm n} > 1$ or $R^{-1}_{\rm n}$ otherwise. This metric gives $Q = 0$ for a perfectly fit dataset and (e.g.) reaches $Q = 1$ for a dataset where the average deviation from the model fit is a factor of two.

\subsection{Data analysis}
\label{SEDAnalSect}

The data were divided into two subsets, the first corresponding to stars in the original Tycho-2 astrometric and proper-motion catalogue, the second to stars in the mission's primary \emph{Hipparcos} catalogue, which also includes parallax data of its own. This separation was motivated by the comparative optical brightness of the \emph{Hipparcos} stars, and the greater accuracy in their \emph{Gaia} DR1 parallax.

\subsubsection{The Tycho-2 dataset}
\label{SEDTychoSect}

We used the original Tycho-2 catalogue as the astrometric reference, as it is temporally closer to the epoch of the surveys we cross-reference against. A number of catalogues were cross-correlated against the Tycho-2 catalogue, allowing matches within an initial tolerance of 5$^{\prime\prime}$.

For certain catalogues, a 5$^{\prime\prime}$ tolerance allows one or more spurious sources to be wrongly matched to the Tycho-2 source. To circumvent this, each matched catalogue was sorted by the distance of the match from the Tycho-2 position, and the 1$\sigma$ deviation in distance was determined, corresponding to the matching radius at which 68.3 per cent of the sources cross-matched at 5$^{\prime\prime}$ tolerance were included. For each catalogue, cross-matches were retained if they fell within 3$\sigma$ of the Tycho-2 source. The cross-matched source catalogues and their adopted 3$\sigma$ tolerances (in brackets\footnote{Tolerances for IPHAS and \emph{IRAS} are set manually, rather than using the 3$\sigma$ cutoff.}) are given below:
\begin{itemize}
\item The American Association of Variable Star Observers (AAVSO) Photometric All-Sky Survey (APASS) Data Release 9 (1.65$^{\prime\prime}$; released as VizieR catalogue II/336/apass9: \citealt{HTT+16}, paper in prep.)\footnote{http://www.aavso.org/apass};
\item The Sloan Digital Sky Survey (SDSS) Data Release 7 (1.94$^{\prime\prime}$; \citealt{AAA+09});
\item The Issac Newton Telescope (INT) Photometric H{$\alpha$} Survey of the Northern Galactic Plane (IPHAS) Data Release 2 (0.70$^{\prime\prime}$; \citealt{BFD+14});
\item The United Kingdom Infra-Red Telescope (UKIRT) Infrared Deep Sky Survey (UKIDSS) Large Area Survey (LAS) Data Release 9 (4.62$^{\prime\prime}$);
\item The Deep Near Infrared Survey of the Southern Sky (DENIS) Third Data Release (1.15$^{\prime\prime}$; released as VizieR catalogue B/denis);
\item The Two-Micron All Sky Survey (2MASS) all-sky catalogue (0.71$^{\prime\prime}$; \citealt{CSvD+03});
\item The \emph{Akari} / Infrared Camera (IRC) all-sky survey (2.34$^{\prime\prime}$; \citealt{IOK+10});
\item The \emph{Wide-Field Infrared Survey Explorer} `\emph{AllWISE}' all-sky catalogue (abbreviated WISE; 1.20$^{\prime\prime}$; \citealt{CWC+13}); and
\item The \emph{Infrared Astronomical Satellite} (\emph{IRAS}) all-sky survey (5$^{\prime\prime}$; \citealt{NHvD+84}).
\end{itemize}

\subsection{The \emph{Hipparcos} dataset}
\label{SEDHipparcosSect}

This procedure was broadly repeated for the \emph{Hipparcos} data. Here, parallaxes were taken from the Tycho--\emph{Gaia} DR1 catalogue if they had been updated, or the `new' \emph{Hipparcos} reduction of \citet{vanLeeuwen07} if they had not. In the combined catalogue, 88\,417 objects had revised parallaxes, while 18\,915 parallaxes come from the original dataset. This include objects with high proper motions and very red colours, which are known to be missing from the \emph{Gaia} dataset (Section \ref{SEDDataSect}). Objects were removed if they had negative parallaxes, or if they had parallax uncertainties greater than $\delta \varpi / \varpi > 0.414$, totalling 6\,399 objects. 

The \emph{Hipparcos} stars are typically much brighter than the Tycho-2 stars, resulting in severe saturation problems which rendered several catalogues unusable. A significant number of brighter stars have insufficient photometry to make a good fit: often only Tycho $B_{\rm T}$ and $V_{\rm T}$, and the \emph{Hipparcos} $H_{\rm p}$ data, which together do not cover a sufficiently large range of wavelengths to constrain the SED. For this reason, we have incorporated a number of additional optical and infrared catalogues of bright stars. This increased dataset makes us more robust against bad data (as it is easier to flag), at the expense of maintaining a homogeneous catalogue between the \emph{Hipparcos} and Tycho-2 stars. The extra catalogues are namely:
\begin{itemize}
\item Mermilliod's ``Photoelectric Photometric Catalogue of Homogeneous Means in the UBV System'' (see \citet{Warren91}).
\item \citet{MM78}, containing $UBVRIJHKLMN$-band photometry.
\item The \emph{Cosmic Background Explorer} (\emph{COBE}) \emph{Diffuse Infrared Background Experiment} (\emph{DIRBE}) Point Source Catalogue \citep{SPB04}.
\item The \emph{Midcourse Space Experiment} (\emph{MSX}) Astrometric Catalogue \citep{EP96}.
\end{itemize}

Astrometric matching tolerences for the four catalogues were set respectively to 0.7$^{\prime\prime}$, 0.47$^{\prime\prime}$, 0.66$^{\prime\prime}$ and 5$^{\prime\prime}$. Data were fitted with the SED fitter as above. A detailed discussion of the methods used to remove bad data are listed in the online Appendix. We stress again that proper motions have not been taken account of in our simple matching exercise: the limited astrometric matching radius means that photometric data will not always be matched for stars with proper motions which are significant on the $\sim$15-year timescales between the \emph{Hipparcos} observations and the relevant catalogue observations. In many cases, a faint, unrelated source may be matched instead. Care has been taken to remove these from the catalogue where they stand out.

\subsubsection{Interstellar extinction}
\label{SEDExtinctionSect}

The line-of-sight interstellar extinction was estimated using maps from the \emph{Planck} Legacy Archive. \emph{Planck} provides visible extinction maps based on the \citet{DL07} dust model in {\sc healpix} format in Galactic coordinates. To facilitate cross-referencing, the Galactic longitude and latitude for each star in the \emph{Hipparcos} and Tycho-2 catalogues were derived via the VizieR portal, and the {\sc python healpy ang2pix} routine was used to locate {\sc healpix} pixels corresponding to catalogue positions, providing the extinction for each object.

Without assuming a prior model for Galactic extinction, there is no ready means to tell whether the extinction lies behind or in front of the object of interest. We must therefore compute two estimates, one with zero and one with full line-of-sight extinction, to bracket the possible range of model fits. Further information on the use of these interstellar extinction data is given in Section \ref{HRErrorEBVSect}.

\subsubsection{Removing bad data}
\label{SEDBadSect}

The data quality of the fitted photometry can be tested using both the goodness-of-fit of individual data points, and the overall goodness-of-fit of a star's SED. These can be used as a basis for removing bad data from the sample. Due to the extensive nature of these tests, and the complex way in which bad data is deleted from the dataset, we have moved the detailed discussion of this topic to the online Appendices. Sources with three or more remaining photometric points were retained for the catalogue: this reduced the number of fitted stars to 1\,475\,921.


\section{The final catalogue and Hertzsprung--Russell diagram}
\label{HRSect}

\begin{figure*}
\centerline{\includegraphics[height=0.92\textwidth,angle=-90]{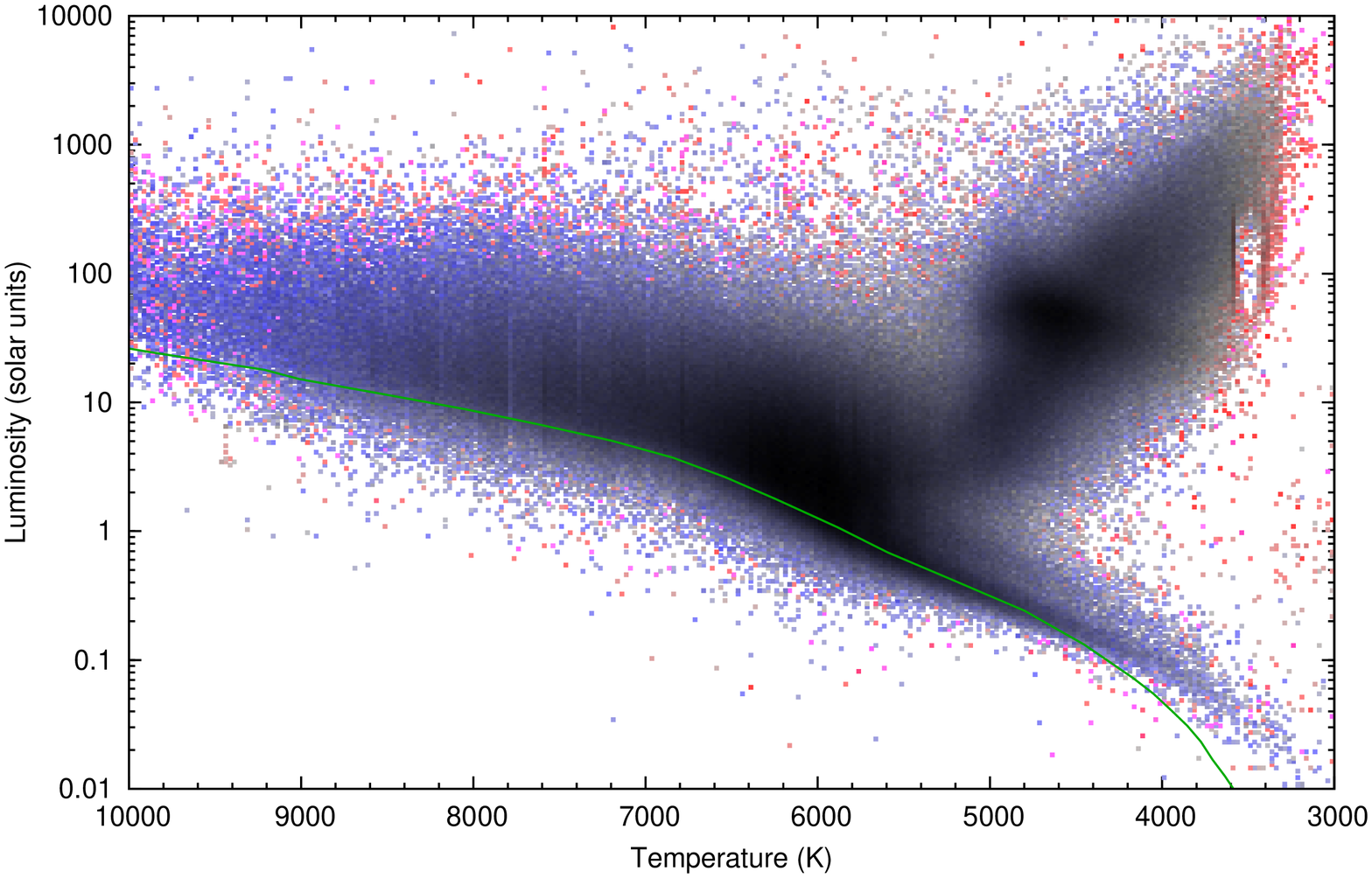}}
\centerline{\includegraphics[height=0.92\textwidth,angle=-90]{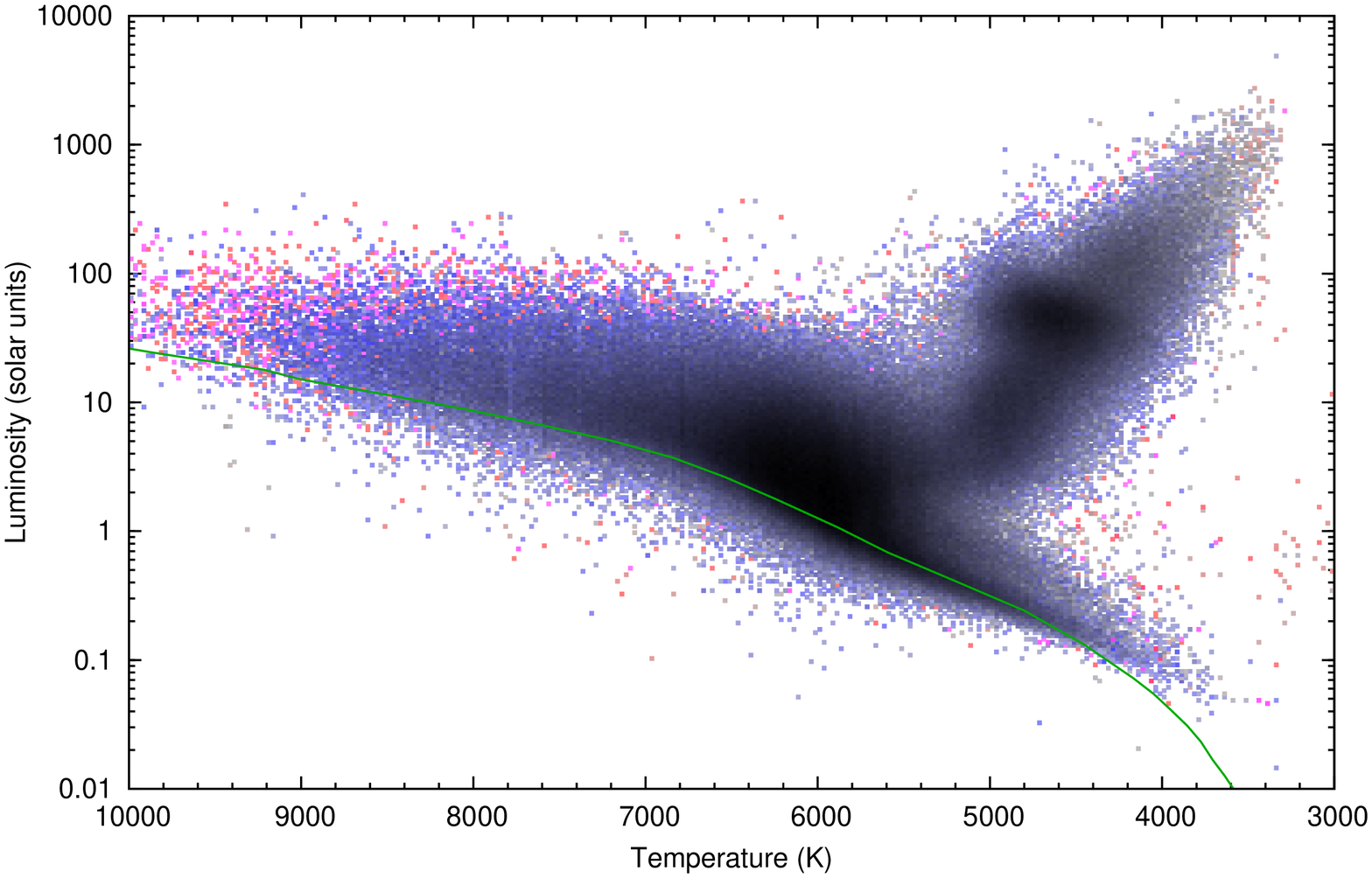}}
\caption{The Hertzsprung--Russell diagram of nearby stars. Darker points represent a greater density of stars. The average value of log($Q$) for each bin is indicated by colour: blue colours denote the best fits, grey colours denote intermediate fits, and red colours denote the worst fits. Systematic deviations from unity can be caused by poor-quality input photometry, or poor fitting by the model atmospheres. The zero-age main sequence is shown in green \citep{MGB+08}. The bottom panel shows a restricted set (40 per cent) of objects, with $<$25 per cent parallax uncertainty, line-of-sight $A_{\rm V} < 3$ mag, and goodness-of-fit $Q < 0.5$.}
\label{HRDFig}
\end{figure*}

\begin{center}
\begin{table*}
\caption{Fundamental parameters and infrared excess for Tycho-2 stars. A portion of the online table is shown here, where table columns are numbered for clarity. The columns are described in full in the text, but can briefly be described as: (1) Tycho-2 reference number; (2) Tycho-2 right ascension; (3) Tycho-2 declination; (4,5) Tycho-2 Galactic latitude and longitude; (6,7) distance and fractional uncertainty; (8,9) extinction and absolute uncertainty; (10,11) effective temperature and absolute uncertainty; (12,13) luminosity and fractional uncertainty; (14) implied stellar radius; (15) assumed surface gravity; (16,17) fitted temperature and luminosity when full line-of-sight reddening is applied; (18,19) fitted temperature and luminosity under the Lutz--Kelker correction of \citet{ABJ16}; (20) fit quality; (21--24) number of datapoints in (respectively) the full SED, and optical, near-IR and mid-IR regions; (25--28) average fit deviation in the total SED, and optical/near-IR/mid-IR regions, respectively; (29) mid-IR excess; (30) mid-IR excess with most-excessive datapoint removed; (31) (uncalibrated) significance of the excess; (32) deviation of most-excessive datapoint; (33) luminosity of the infrared excess; (34) fraction of reprocessed infrared light; (35) peak wavelength of infrared excess; (36--55) deviation of individual datapoints; (56--75) fluxes of datapoints used in final fit. Complete versions are to be made available through the Centre de Donn\'ees astronomiques de Strasbourg (CDS).}
\label{TychoTable}
\begin{tabular}{c@{\ \ }c@{\ }c@{\ \ }c@{\ }c@{\ \ }c@{\ }c@{\ \ }c@{\ }c}
    \hline \hline
\multicolumn{1}{c}{(1)} & \multicolumn{1}{c}{(2)} & \multicolumn{1}{c}{(3)} & \multicolumn{1}{c}{(4)} & \multicolumn{1}{c}{(5)} & \multicolumn{1}{c}{(6)} & \multicolumn{1}{c}{(7)} & \multicolumn{1}{c}{(8)} & \multicolumn{1}{c}{(9)} \\
\multicolumn{1}{c}{TYC} & \multicolumn{1}{c}{RA} & \multicolumn{1}{c}{Dec.} & \multicolumn{1}{c}{G.\ Lat.} & \multicolumn{1}{c}{G.\ Long.} & \multicolumn{1}{c}{$d$} & \multicolumn{1}{c}{$\delta \varpi / \varpi$} & \multicolumn{1}{c}{$A_V$} & \multicolumn{1}{c}{$\delta A_V$} \\
\multicolumn{1}{c}{\ } & \multicolumn{1}{c}{(J2000)} & \multicolumn{1}{c}{(J2000)} & \multicolumn{1}{c}{(deg)} & \multicolumn{1}{c}{(deg)} & \multicolumn{1}{c}{(pc)} & \multicolumn{1}{c}{\ } & \multicolumn{1}{c}{(mag)} & \multicolumn{1}{c}{(mag)} \\
    \hline
1000-1016-1 & 264.019440 & 11.275677 & 34.759265 & 21.778061 & 575.585 & 0.137 & 0.897 & 0.037 \\
1000-1018-1 & 262.982107 & 11.568592 & 34.585083 & 22.823855 & 347.823 & 0.094 & 0.816 & 0.016 \\
1000-1043-1 & 264.093473 & 12.636898 & 36.126451 & 22.280018 & 465.817 & 0.120 & 1.365 & 0.066 \\
\nodata&\nodata&\nodata&\nodata&\nodata&\nodata&\nodata&\nodata&\nodata\\
    \hline
\\
\end{tabular}
\begin{tabular}{@{}c@{}cc@{}ccccc@{\ \ }cc@{\ \ }cc@{\ \ }c@{\ \ }c@{\ \ }c@{\quad}c@{\ \ }c@{\ \ }c@{\ \ }c@{}}
    \hline \hline
\multicolumn{1}{c}{(10)} & \multicolumn{1}{c}{(11)} & \multicolumn{1}{c}{(12)} & \multicolumn{1}{c}{(13)} & \multicolumn{1}{c}{(14)} & \multicolumn{1}{c}{(15)} & \multicolumn{1}{c}{(16)} & \multicolumn{1}{c}{(17)} & \multicolumn{1}{c}{(18)} & \multicolumn{1}{c}{(19)} & \multicolumn{1}{c}{(20)} & \multicolumn{4}{c}{(21 $\cdots$ 24)} & \multicolumn{4}{c}{(25 $\cdots$ 28)}\\
\multicolumn{1}{c}{$T_{\rm eff}$} &\multicolumn{1}{c}{$\delta T_{\rm eff}$} & \multicolumn{1}{c}{$L$} & \multicolumn{1}{c}{$\delta L/L$} & \multicolumn{1}{c}{$r$} & \multicolumn{1}{c}{log($g$)} & \multicolumn{1}{c}{$T_{\rm Av}$} &\multicolumn{1}{c}{$L_{\rm Av}$} & \multicolumn{1}{c}{$T_{\rm ABJ}$} &\multicolumn{1}{c}{$L_{\rm ABJ}$} & \multicolumn{1}{c}{\clap{$Q$}} & \multicolumn{4}{c}{$N$} & \multicolumn{4}{c}{\clap{$\Re$}} \\
\multicolumn{1}{c}{(K)} & \multicolumn{1}{c}{(K)} & \multicolumn{1}{c}{(L$_\odot$)} & \multicolumn{1}{c}{\ } & \multicolumn{1}{c}{(R$_\odot$)} & \multicolumn{1}{c}{(dex)} & \multicolumn{1}{c}{(K)} & \multicolumn{1}{c}{(L$_\odot$)} & \multicolumn{1}{c}{(K)} & \multicolumn{1}{c}{(L$_\odot$)} & \multicolumn{1}{c}{\ } & \multicolumn{4}{c}{$\overbrace{\qquad\qquad}$} & \multicolumn{4}{c}{$\overbrace{\qquad\qquad\qquad\qquad\qquad}$} \\
    \hline
7182 & 212 & 12.524 & 0.147 &  2.289 & 3.881 & 7818 & 16.360 &     0 &  0.000 & 0.039 & 13 & 6 & 5 & 2 & 1.009 & 1.014 & 0.989 & 1.040 \\
6020 & 138 &  1.945 & 0.104 &  1.284 & 4.273 & 6399 &  2.384 &  6020 &  1.945 & 0.070 & 13 & 6 & 5 & 2 & 1.024 & 1.012 & 1.016 & 1.083 \\
4769 & 125 & 57.760 & 0.131 & 11.148 & 2.341 & 5138 & 73.964 &  4769 & 57.760 & 0.052 & 14 & 5 & 5 & 4 & 1.019 & 1.002 & 1.011 & 1.050\\
\nodata&\nodata&\nodata&\nodata&\nodata&\nodata&\nodata&\nodata&\nodata&\clap{\nodata}&\clap{\nodata}&\clap{\nodata}&\clap{\nodata}&\nodata&\nodata&\nodata&\nodata\\
    \hline
\\
\end{tabular}
\begin{tabular}{c@{}c@{}c@{}c@{}c@{}c@{}c@{}c@{}c@{}c@{}cc@{\ }c@{}c}
    \hline \hline
\multicolumn{1}{c}{(29)} & \multicolumn{1}{c}{(30)} & \multicolumn{1}{c}{(31)} & \multicolumn{1}{c}{(32)} & \multicolumn{1}{c}{(33)} & \multicolumn{1}{c}{(34)} & \multicolumn{1}{c}{(35)} & \multicolumn{1}{c}{(36)} & \multicolumn{1}{c}{\clap{\nodata}} & \multicolumn{1}{c}{(55)} & \multicolumn{1}{c}{(56)} & \multicolumn{1}{c}{\clap{\nodata}} & \multicolumn{1}{c}{(75)} \\
\multicolumn{1}{c}{$X_{\rm MIR}$} & \multicolumn{1}{c}{$X^\prime_{\rm MIR}$} & \multicolumn{1}{c}{$S_{\rm MIR}$} & \multicolumn{1}{c}{$R_{\rm max}$} & \multicolumn{1}{c}{$L_{\rm XS}$} & \multicolumn{1}{c}{$f_{\rm XS}$} & \multicolumn{1}{c}{$\lambda_{\rm XS,peak}$} & \multicolumn{3}{c}{\clap{$(F_{\rm o}/F_{\rm m})_{\{BT\,...\,[25]\}}$}} & \multicolumn{3}{c}{$F_{{\rm o},\{BT\,...\,[25]\}}$} \\
\multicolumn{1}{c}{\ } & \multicolumn{1}{c}{\ } & \multicolumn{1}{c}{\ } & \multicolumn{1}{c}{\ } & \multicolumn{1}{c}{(L$_\odot$)}  & \multicolumn{1}{c}{\ }  & \multicolumn{1}{c}{($\mu$m)} & \multicolumn{1}{c}{(Jy)} & \multicolumn{1}{c}{\ } & \multicolumn{1}{c}{(Jy)} & \multicolumn{1}{c}{(Jy)} & \multicolumn{1}{c}{\ } & \multicolumn{1}{c}{(Jy)} \\
    \hline
1.037 & 1.052 & 1.471 & 1.090 & 0.0002 & 0.000015 & 3.4 & 1.010 & \clap{\nodata} & 0.000 & 139.132 & \clap{\nodata} & 0.000 \\
1.068 & 1.066 & 1.669 & 1.269 & 0.0001 & 0.000077 & 2.3 & 1.269 & \clap{\nodata} & 0.000 &  50.842 & \clap{\nodata} & 0.000 \\
1.043 & 1.039 & 1.916 & 1.135 & 0.0064 & 0.000111 & 2.2 & 1.043 & \clap{\nodata} & 0.000 & 306.356 & \clap{\nodata} & 0.000 \\
\nodata&\nodata&\nodata&\nodata&\nodata&\nodata&\nodata&\nodata&\clap{\nodata}&\nodata&\nodata&\clap{\nodata}&\nodata\\
    \hline
\end{tabular}
\end{table*}
\end{center}

\begin{center}
\begin{table*}
\caption{Fundamental parameters and infrared excess for \emph{Hipparcos} stars. A portion of the online table is shown here, where table columns are numbered for clarity. The columns are described in full in the text, but can briefly be described as: (1) \emph{Hipparcos} reference number; (2) \emph{Hipparcos} right ascension; (3) \emph{Hipparcos} declination; (4,5) \emph{Hipparcos} Galactic latitude and longitude; (6--35) as Table \ref{TychoTable}; (36) source of parallax (\emph{Hipparcos}/\emph{Gaia}); (37--62) deviation of individual datapoints; (63--90) fluxes of datapoints used in final fit. Complete tables are to be found at CDS.}
\label{HipparcosTable}
\begin{tabular}{c@{\ \ }c@{\ }c@{\ \ }c@{\ }c@{\ \ }c@{\ }c@{\ \ }c@{\ }c}
    \hline \hline
\multicolumn{1}{c}{(1)} & \multicolumn{1}{c}{(2)} & \multicolumn{1}{c}{(3)} & \multicolumn{1}{c}{(4)} & \multicolumn{1}{c}{(5)} & \multicolumn{1}{c}{(6)} & \multicolumn{1}{c}{(7)} & \multicolumn{1}{c}{(8)} & \multicolumn{1}{c}{(9)} \\
\multicolumn{1}{c}{HIP} & \multicolumn{1}{c}{RA} & \multicolumn{1}{c}{Dec.} & \multicolumn{1}{c}{G.\ Lat.} & \multicolumn{1}{c}{G.\ Long.} & \multicolumn{1}{c}{$d$} & \multicolumn{1}{c}{$\delta \varpi / \varpi$} & \multicolumn{1}{c}{$A_V$} & \multicolumn{1}{c}{$\delta A_V$} \\
\multicolumn{1}{c}{\ } & \multicolumn{1}{c}{(J2000)} & \multicolumn{1}{c}{(J2000)} & \multicolumn{1}{c}{(deg)} & \multicolumn{1}{c}{(deg)} & \multicolumn{1}{c}{(pc)} & \multicolumn{1}{c}{\ } & \multicolumn{1}{c}{(mag)} & \multicolumn{1}{c}{(mag)} \\
    \hline
3 & 0.005024 & 38.859279 & 112.090026 & --22.927558 & 350.804 & 0.344 & 0.929 & 0.083 \\
4 & 0.008629 & --51.893546 & 320.793090 & --63.415309 & 135.654 & 0.039 & 0.124 & 0.039 \\
5 & 0.009973 & --40.591202 & 337.897763 & --72.861671 & 381.080 & 0.092 & 0.057 & 0.019 \\
\nodata&\nodata&\nodata&\nodata&\nodata&\nodata&\nodata&\nodata&\nodata\\
    \hline
\\
\end{tabular}
\begin{tabular}{@{}c@{}cc@{}ccccc@{\ \ }cc@{\ \ }cc@{\ \ }c@{\ \ }c@{\ \ }c@{\quad}c@{\ \ }c@{\ \ }c@{\ \ }c@{}}
    \hline \hline
\multicolumn{1}{c}{(10)} & \multicolumn{1}{c}{(11)} & \multicolumn{1}{c}{(12)} & \multicolumn{1}{c}{(13)} & \multicolumn{1}{c}{(14)} & \multicolumn{1}{c}{(15)} & \multicolumn{1}{c}{(16)} & \multicolumn{1}{c}{(17)} & \multicolumn{1}{c}{(18)} & \multicolumn{1}{c}{(19)} & \multicolumn{1}{c}{(20)} & \multicolumn{4}{c}{(21 $\cdots$ 24)} & \multicolumn{4}{c}{(25 $\cdots$ 28)}\\
\multicolumn{1}{c}{$T_{\rm eff}$} &\multicolumn{1}{c}{$\delta T_{\rm eff}$} & \multicolumn{1}{c}{$L$} & \multicolumn{1}{c}{$\delta L/L$} & \multicolumn{1}{c}{$r$} & \multicolumn{1}{c}{log($g$)} & \multicolumn{1}{c}{$T_{\rm Av}$} &\multicolumn{1}{c}{$L_{\rm Av}$} & \multicolumn{1}{c}{$T_{\rm ABJ}$} &\multicolumn{1}{c}{$L_{\rm ABJ}$} & \multicolumn{1}{c}{\clap{$Q$}} & \multicolumn{4}{c}{$N$} & \multicolumn{4}{c}{\clap{$\Re$}} \\
\multicolumn{1}{c}{(K)} & \multicolumn{1}{c}{(K)} & \multicolumn{1}{c}{(L$_\odot$)} & \multicolumn{1}{c}{\ } & \multicolumn{1}{c}{(R$_\odot$)} & \multicolumn{1}{c}{(dex)} & \multicolumn{1}{c}{(K)} & \multicolumn{1}{c}{(L$_\odot$)} & \multicolumn{1}{c}{(K)} & \multicolumn{1}{c}{(L$_\odot$)} & \multicolumn{1}{c}{\ } & \multicolumn{4}{c}{$\overbrace{\qquad\qquad}$} & \multicolumn{4}{c}{$\overbrace{\qquad\qquad\qquad\qquad\qquad}$} \\
    \hline
7096 & 2561 & 194.076 & 0.732 &  9.230 & 2.642 & 7261 & 210.793 & 7093 & 210.805 & 0.618 & 10 & 3 & 4 & 3 & 1.281 & 1.229 & 0.670 & 2.147\\
6777 &  168 &   8.373 & 0.059 &  2.102 & 3.930 & 6834 & 8.523 & 6777 & 8.425 & 0.058 & 14 & 6 & 5 & 3 & 1.015 & 1.021 & 0.992 & 1.042 \\
4885 &  125 &  56.536 & 0.106 & 10.512 & 2.364 & 4897 & 56.882 & 4885 & 55.987 & 0.039 & 13 & 5 & 4 & 4 & 1.021 & 1.015 & 0.999 & 1.050 \\
\nodata&\nodata&\nodata&\nodata&\nodata&\nodata&\nodata&\nodata&\nodata&\clap{\nodata}&\clap{\nodata}&\clap{\nodata}&\clap{\nodata}&\nodata&\nodata&\nodata&\nodata\\
    \hline
\\
\end{tabular}
\begin{tabular}{@{}c@{}c@{}c@{}ccc@{}cc@{}c@{}c@{}c@{}c@{}c@{}c@{}}
    \hline \hline
\multicolumn{1}{c}{(29)} & \multicolumn{1}{c}{(30)} & \multicolumn{1}{c}{(31)} & \multicolumn{1}{c}{(32)} & \multicolumn{1}{c}{(33)} & \multicolumn{1}{c}{(34)} & \multicolumn{1}{c}{(35)} & \multicolumn{1}{c}{(36)} & \multicolumn{1}{c}{(37)} & \multicolumn{1}{c}{\clap{\nodata}} & \multicolumn{1}{c}{(62)} & \multicolumn{1}{c}{(63)} & \multicolumn{1}{c}{\clap{\nodata}} & \multicolumn{1}{c}{(90)} \\
\multicolumn{1}{c}{\clap{$X_{\rm MIR}$}} & \multicolumn{1}{c}{\clap{$X^\prime_{\rm MIR}$}} & \multicolumn{1}{c}{$S_{\rm MIR}$} & \multicolumn{1}{c}{$R_{\rm max}$} & \multicolumn{1}{c}{$L_{\rm XS}$} & \multicolumn{1}{c}{$f_{\rm XS}$} & \multicolumn{1}{c}{$\lambda_{\rm XS,peak}$} & \multicolumn{1}{c}{\clap{$G/H$}} & \multicolumn{3}{c}{\clap{$(F_{\rm o}/F_{\rm m})_{\{...\}}$}} & \multicolumn{3}{c}{$F_{{\rm o},\{...\}}$} \\
\multicolumn{1}{c}{\ } & \multicolumn{1}{c}{\ } & \multicolumn{1}{c}{\ } & \multicolumn{1}{c}{\ } & \multicolumn{1}{c}{(L$_\odot$)}  & \multicolumn{1}{c}{\ }  & \multicolumn{1}{c}{($\mu$m)} & \multicolumn{1}{c}{\ } & \multicolumn{1}{c}{(Jy)} & \multicolumn{1}{c}{\ } & \multicolumn{1}{c}{(Jy)} & \multicolumn{1}{c}{(Jy)} & \multicolumn{1}{c}{\ } & \multicolumn{1}{c}{(Jy)} \\
    \hline
2.361 & 3.206 & 3.217 & 3.523 & 0.0171 & 0.000088 & 17.1 & G & 0.000 & \clap{\nodata} & 0.000 & 0.000 & \clap{\nodata} & 0.000 \\
1.034 & 1.050 & 1.584 & 1.128 & 0.0001 & 0.000013 &  8.2 & G & 0.000 & \clap{\nodata} & 0.000 & 0.000 & \clap{\nodata} & 0.000 \\
1.042 & 1.051 & 2.667 & 1.103 & 0.0027 & 0.000048 &  8.6 & G & 0.000 & \clap{\nodata} & 0.000 & 0.000 & \clap{\nodata} & 0.000 \\
\nodata&\nodata&\nodata&\nodata&\nodata&\nodata&\nodata&\nodata&\clap{\nodata}&\nodata&\nodata&\clap{\nodata}&\nodata\\
    \hline
\end{tabular}
\end{table*}
\end{center}

\subsection{The catalogue and diagram}
\label{HRCatSect}

Figure \ref{HRDFig} shows the main Hertzsprung--Russell diagram of the combined Tycho--\emph{Gaia} and \emph{Hipparcos--Gaia} datasets, under the assumption of zero interstellar extinction. The top panel contains the entire dataset, while the bottom panel shows a restricted subset of well-fit objects. This data is tabulated in Tables \ref{TychoTable} and \ref{HipparcosTable}, for the Tycho-2 and \emph{Hipparcos} stars, respectively.

The upper panel of Figure \ref{HRDFig} shows several artefacts. The main sequence is broad, reflecting the higher extinction and greater parallax uncertainties in some of the data. Vertical bands of red symbols (poorly fit stars) in the most luminous regions of the diagram come mainly from \emph{Hipparcos} stars which are not well modelled by a single stellar atmosphere model. The vertical stripe between 3400 and 3500 K on the upper giant branch seems largely occupied by stars which have a combination of high reddening and uncertain distances: these are mostly normal giant branch stars that have been pushed onto this artificial sequence by interstellar reddening.

The giant branch also has a significant overdensity about halfway along its length: this is a real feature, representing the merged features of the RGB bump and red clump\footnote{The RGB bump is a concentration of stars on the hydrogen-burning RGB, caused by the transition of the hydrogen-burning shell into material that has previously been convectively mixed. The red clump is the high mass equivalent of the horizontal branch, and represents the core-helium-burning phase of giant-branch evolution \citep[e.g.][]{KL14}.}.

The lower panel of Figure \ref{HRDFig} shows a subset of same data, but with poor quality data removed (objects on highly extincted lines of sight, with large parallax uncertainties, or where the SEDs are not well fit by a single stellar model). In this lower panel, the main sequence stands out clearly, being best populated for solar-like stars, but with distributions tailing off towards very hot temperatures (rare stars which cannot be well modelled without good UV data and extinction corrections) and towards very low temperatures (faint stars missing due to photometric incompleteness).

Both panels include a zero-age main sequence (ZAMS) model, derived from the Padova stellar evolution models of \citet{MGB+08}. The lower main sequence, between $\sim$4600 and $\sim$5400 K, fits the ZAMS model very well. At temperatures $>$5400 K, scatter above the ZAMS line indicates the presence of more-evolved main-sequence stars, which are approaching the main-sequence turn-off. This can be used to extract age information about the solar neighbourhood. The bottom end of the main sequence is not well fit by a zero-age main sequence model, but this deviation is substantially reduced in the lower panel. This suggests it results from a combination of photometric inaccuracy or incompleteness near the sensitivity limit of photometric databases (including Tycho-2 itself), biased scatter upward in the diagram due to uncertain parallaxes (possibly a manifestation of the Lutz--Kelker bias; \citealt{LK73}), and (in a limited number of cases) heavy reddening of lower main-sequence stars.

Many cool stars on the upper giant branch are not included in the lower panel of Figure \ref{HRDFig}. Several factors contribute to this. (1) Despite their luminosity, these are often red, optically faint stars, which consequently have significant uncertainties in their Tycho-2 positions, hence also in their \emph{Gaia} parallaxes. (2) Being luminous stars, these stars are visible at large distances from the Earth, and congregate in the Galactic Plane, so are more often subject to strong interstellar extinction than nearby stars. (3) Variability of stars in this part of the H--R diagram leads to variability induced motion (see \citealt{vanLeeuwen07}), which increases the uncertainty in their parallax. Variability also worsens (increases) the SED quality estimator, $Q$. (4) A substantial fraction of these stars have circumstellar dust, which reprocesses their light from the optical into the infrared, resulting them in being poorly fit by a simple stellar SED.

\subsection{Limitations and uncertainties}
\label{HRErrorSect}

For well-fit stars, the three primary sources of uncertainty in this analysis are: (1) random and systematic uncertainties in the source data; (2) Lutz--Kelker effects when converting parallax to distance; (3) systematic `cooling' of the SEDs caused by interstellar reddening; and (4) the effect on the stellar temperature of the unknown metallicity of each star.

\subsubsection{Random versus systematic uncertainties}
\label{HRErrorRandSect}

Formal uncertainties for SED fitting of this nature are difficult to determine. The published photometric uncertainties for many of the public surveys can grossly underestimate the true uncertainties involved, both within individual catalogues, across catalogues, and across different epochs. For example, the 2MASS photometric uncertainties can be as low as a few millimagnitudes, and represent the internal error in the catalogue, yet the photometric zero points are uncertain by $\sim$2 per cent\footnote{{\tt http://www.ipac.caltech.edu/2mass/releases/allsky/faq.html}}. Different surveys take these uncertainties into account in different ways, and to different degrees. Across catalogues, source blending and astrophysical sky background can become important, particularly in crowded regions and in the infrared. Across different epochs, stellar variability or proper motion can become significant.

This means that quantifying uncertainties on photometry and assigning appropriate weights is non-trivial. For this reason, no weighting was applied to the photometry during the fitting process. This can cause problems, particularly when observations are near the limit of photometric completeness. However, in such cases, fits can generally be improved simply by removing these photometric datapoints from the catalogue, as described in the online Appendices.

\begin{figure}
\centerline{\includegraphics[height=0.47\textwidth,angle=-90]{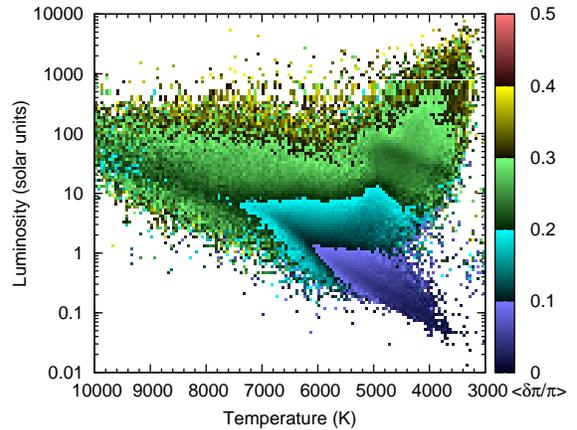}}
\caption{The fractional parallax uncertainty of stars in different regions of the H--R diagram. A binned average is displayed for each pixel.}
\label{HRDPlxErrFig}
\end{figure}

For the luminosity measurement, in the vast majority of cases, the largest uncertainty is from the photometric parallax of the star (Figure \ref{HRDPlxErrFig}).

\subsubsection{Lutz--Kelker effects}
\label{HRErrorLutzKelkerSect}

The derived luminosity of a star is subject to the uncertainty in its distance and hence its parallax as $L \propto d^2 \propto \varpi^{-2}$. The probability distribution function (PDF) in parallax is normally expected to be Gaussian \citep[e.g.][]{LK73,BailerJones15}. However, when inverting parallax to distance, the PDF becomes non-Gaussian and asymmetric. For stars with small fractional uncertainties, this is a relatively minor effect, but at large uncertainties it manifests itself in a variety of phenomena that can be broadly termed Lutz--Kelker effects, after \citet{LK73}.

The full range of Lutz--Kelker effects are complex, and there is no definitively appropriate way to correct for them. The magnitude by which Lutz--Kelker effects affect quantities derived from this dataset varies according to the sub-sample chosen, particularly in respect to any limiting fractional parallax uncertainty ($\delta \varpi$/$\varpi$).

To account for the Lutz--Kelker effect, we present two sets of temperatures and luminosities. In the first, we present temperatures and luminosities derived from a simple inversion of parallax to obtain distance ($T_{\rm naive}$, $L_{\rm naive}$). For comparison, we also present temperatures and luminosities derived from distances quoted by \citet{ABJ16}, who model the Lutz--Kelker effects on the \emph{Gaia} DR1 sample using a population model of the Milky Way ($T_{\rm ABJ}$, $L_{\rm ABJ}$). We strongly advise the reader to explore which of these is most appropriate for their individual application, and to use the difference between the ``na\"ive'' and ``ABJ'' parameters as a qualitative estimate of how much the Lutz--Kelker bias could affect their data.

A detailed comparison of these two sets of data is presented in online Appendix \ref{AppendixLK}. In summary, roughly 35 per cent of our stars are estimated to suffer some level of Lutz--Kelker bias in their na\"ive distances. The corrected luminosities for the remainder are almost all only modestly (a few per cent) different from the na\"ive assumptions. Barring a small number of stars, the corrections are all negligible compared to the luminosity uncertainties applied from other sources. The resulting distance changes also affect the assumed stellar gravity and (in many cases) stellar mass, resulting in a marginally different temperature distribution that is generally within the temperature uncertainties of the source in question and, for the vast majority of stars, within 200 K of the na\"ive estimate. While a detailed comparison of the two approaches is beyond the scope of this work, the corrected distances from \citet{ABJ16} result in either no clear improvement or a slightly \emph{worse} fit to specific features on the H--R diagram, therefore we retain the na\"ive estimates for use in the remainder of this paper.

\subsubsection{Interstellar reddening}
\label{HRErrorEBVSect}

The interstellar reddening towards each star is unknown. The \emph{Planck} data we use provide the line-of-sight reddening, which will be partly in front of, and partly behind the star. To estimate the uncertainty this creates, we have de-reddened the input photometry, assuming that the full \emph{Planck} line-of-sight reddening is in front of the star, and re-run the SED-fitting code. For stars with large reddening, we also compute fits for $A_V$ = 1, 2 and 3 mag. The photometry is dereddened using the Milky Way $R_V = 3.1$ extinction curve of \citet{Draine03}. Dereddening is performed for each point in the model SED, before it is convolved with the filter transmission functions, ensuring accurate dereddening for sources with high extinction.

\begin{figure}
\centerline{\includegraphics[height=0.47\textwidth,angle=-90]{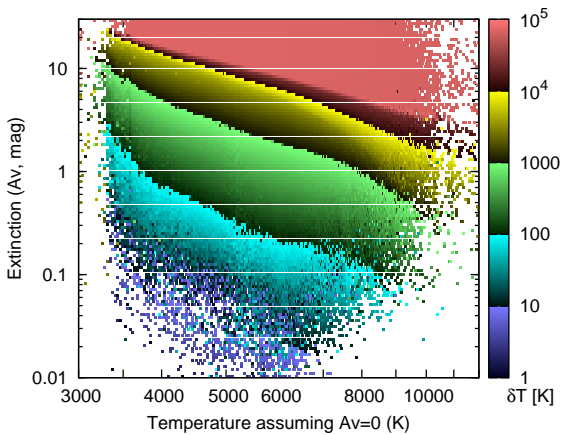}}
\caption{The correction ($\delta T$) applied to a star of given photometrically derived temperature behind a given column density of extinction. Stars with $\delta T \gtrsim T_{\rm eff}$ cannot reliably be fit, even if the extinction is known.}
\label{AVCorrFig}
\end{figure}

Figure \ref{AVCorrFig} shows the increase in temperature that must be applied to a star which is subject to a given amount of interstellar reddening. Taking the whole dataset, the average star is 6000 K and lies in a line of sight with an extinction $A_V = 1.0$ mag. If we assume half of this extinction to lie between us and the star, the \emph{average} under-estimation of the temperature for these stars is $\sim$240 K.

For most stars, this value should be conservatively large. At higher extinctions, there is a progressively greater chance that the star will be made too faint to be found in the Tycho-2 catalogue. The significant majority of stars in the Tycho-2 catalogue are below the completeness limit\footnote{The 90 per cent completeness limit in $V_{\rm T}$ is $\sim$11.5 mag, and 86 per cent of stars are fainter than this. The 10 per cent completeness limit is reached about a magnitude below this, and few stars are found at $V_{\rm T} > 12.5$ mag.}. Due to the steep increase in number of stars per magnitude ($N {\rm d} V_{\rm T} \appropto V_{\rm T}^9$), the vast majority of stars suffering significant extinction ($A_V \gtrsim 1$ mag) will be reddened out of the Tycho-2 catalogue. This corrolary should hold strongest for stars which are optically faint, hence stars of later spectral types (which need less correction), and more distant stars (which are likely to suffer from more reddening anyway). Therefore, the average star in our final catalogue should have a reddening correction which is $>$240 K. However, care should be taken for luminous stars and hot stars, where larger corrections could be required.

Further discussion on interstellar extinction and its spatial correlation can be found in Appendix \ref{AppendixEBV} (online version only).

\subsubsection{Metallicity}
\label{HRErrorFeHSect}

\begin{figure}
\centerline{\includegraphics[height=0.47\textwidth,angle=-90]{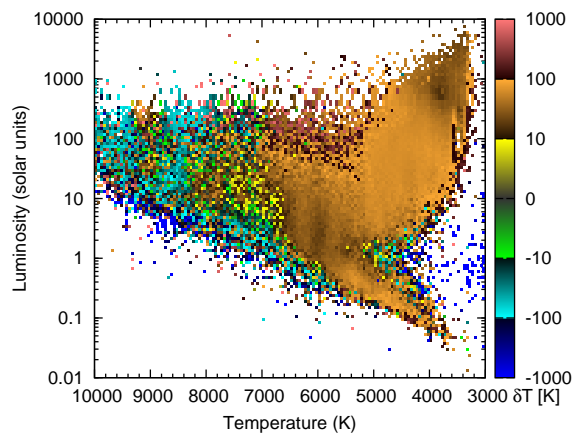}}
\caption{Difference in temperature between models assuming [Fe/H] = 0 and [Fe/H] = --0.5 dex, coded in the sense that positive numbers (red colours) denote cooler stellar temperatures in the metal-poor models, while negative numbers (blue colours) denote warmer stellar temperatures in the metal-poor models.}
\label{FeHCorrFig}
\end{figure}

Figure \ref{FeHCorrFig} shows the correction to our fitted stellar temperatures that must be applied to stars of [Fe/H] = --0.5 dex. Note that the {\sc bt-settl} elemental abundance ratios also change during this step, from [$\alpha$/Fe] = 0 to [$\alpha$/Fe] = +0.2 dex. The majority of stars below $\sim$6500 K require a temperature adjustment of between --10 and --100 K if the metallicity is decreased to [Fe/H] = --0.5 dex. The majority of stars warmer than $\sim$6500 K require a temperature change of +10 to +100 K. Stars lying outside the main regions of the H--R diagram tend to be stars which are poorly fit. Here, temperature changes of 1000 K are not uncommon, as a better fit can often result from relatively minor changes to the poorly constrained SED.

Different studies using differing methods yield different metallicity distributions for stars in the Local Neighbourhood \citep[e.g.][]{TC05,RTT+07,BFO14,HTY+14}. The large majority of stars fall in the range --0.3 $\lesssim$ [Fe/H] $\lesssim$ +0.2 dex, although significant tails make substantial contributions to --0.9 $\lesssim$ [Fe/H] $\lesssim$ +0.6 dex. While age plays a factor in this spread, it is also location dependent, with metal-poor stars being further from the Galactic Plane. It is expected that the typical star in this sample requires a metallicity correction to its temperature of $<$100 K, and much less than this in most cases.

\subsubsection{Comparison to literature data}
\label{HRErrorLitSect}

\begin{figure}
\centerline{\includegraphics[height=0.47\textwidth,angle=-90]{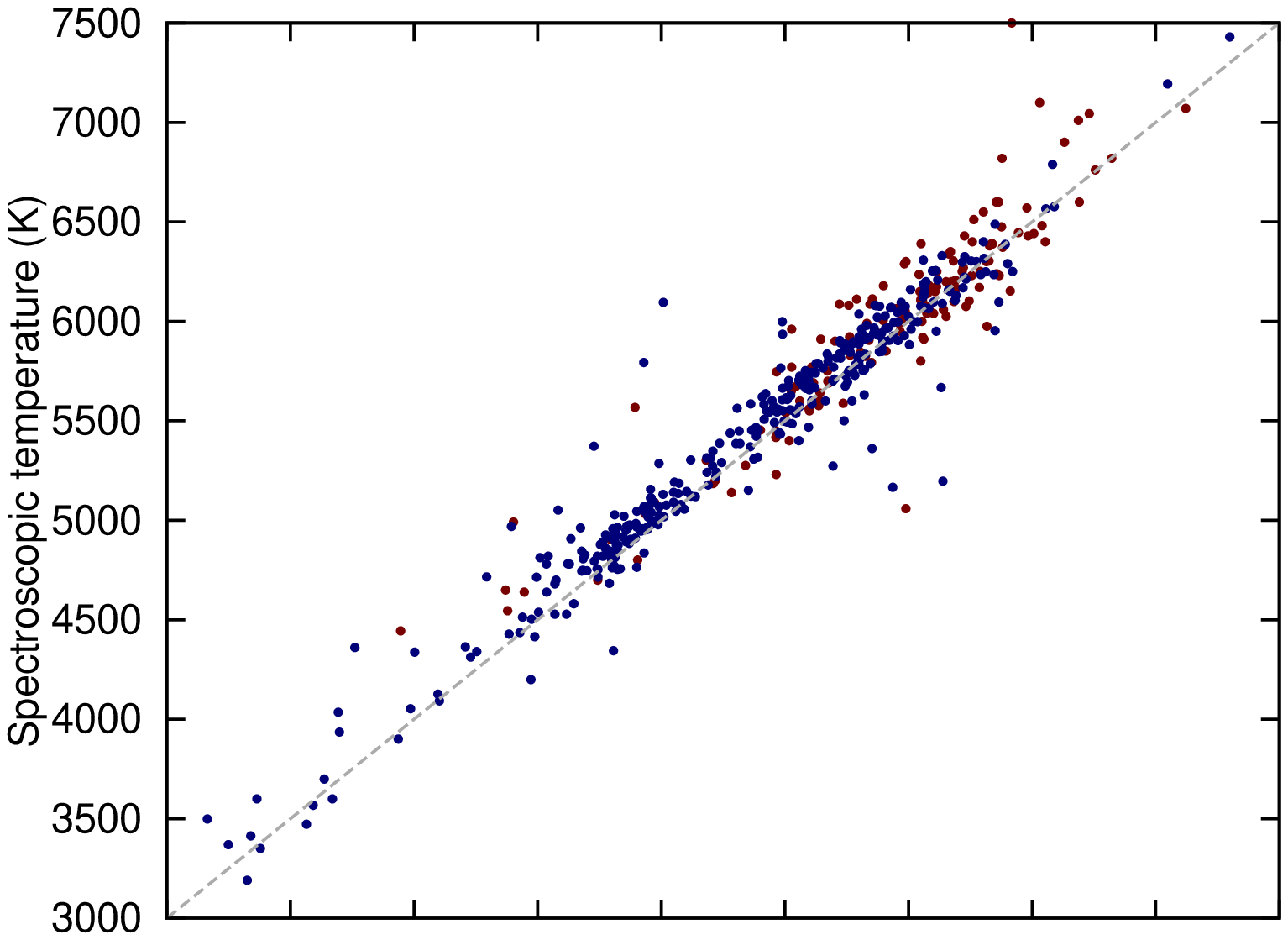}}
\centerline{\includegraphics[height=0.47\textwidth,angle=-90]{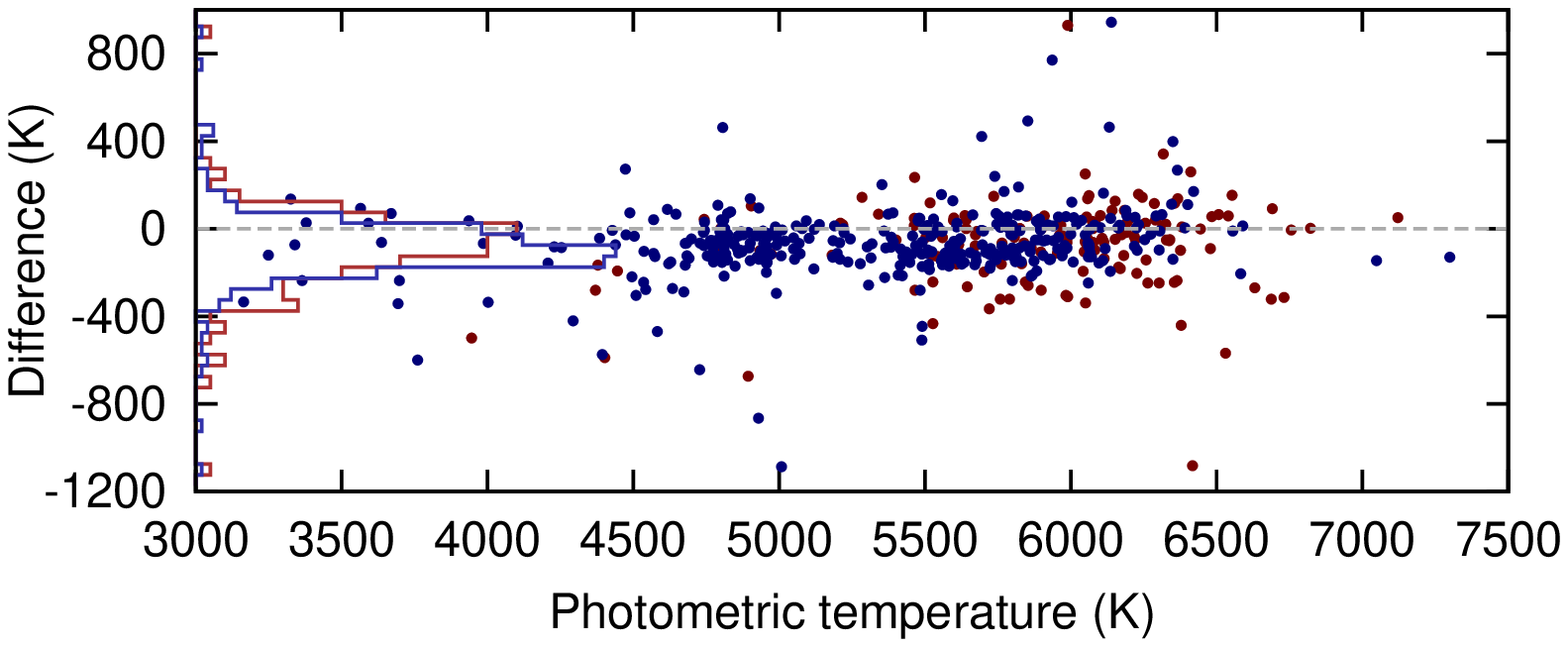}}
\caption{Comparison between literature temperatures of exoplanet hosts (mostly derived from spectroscopy) compared with temperatures derived from photometry in this work. The top panel plots both temperatures, while the bottom panel shows the difference between them, with a histogram shown on the left. Blue points represent \emph{Hipparcos} stars, while red points represent Tycho-2 stars.}
\label{ExoFig}
\end{figure}

In order to better estimate the combined uncertainties inherent in our temperatures, we compare to published literature measurements. One of the most accurate sets of stellar temperatures comes from the exoplanet community: radial velocity confirmations of exoplanets require high signal-to-noise spectra, and measurements of exoplanet properties require accurate stellar classification. To construct a sample of exoplanet host parameters, we used the Exoplanet Orbit Database \citep[EOD][]{WFM+11}\footnote{http://exoplanets.org}, which was used in \citet{CMK16} to validate temperatures derived from the \emph{Hipparcos} sample of stars. From a selection of 5454 catalogued exoplanets, co-ordinates and $T_{\rm eff}$ were returned for 2616 unique hosts. Of these, 591 could be matched with stars in the Tycho--\emph{Gaia} catalogue. Of those, 150 have measureable parallaxes and are present in our final catalogue.

Among the 150 measured stars, the EOD quotes a literature stellar mass of 1.06 $\pm$ 0.43 M$_\odot$ (st.\ dev.) and a metallicity of [Fe/H] = 0.05 $\pm$ 0.24 dex (st.\ dev.). The average spectroscopic temperature was quoted as 5960 K. These parameters provide a good match to typical stars in our sample.

A comparison of the photometric and spectroscopic temperatures of these 150 stars is shown in Figure \ref{ExoFig}. The average photometric temperature is 73 $\pm$ 200 K (1.2 $\pm$ 3.4 per cent) lower than the spectroscopic temperature. For comparison, the median difference is slightly less, 52 K lower, and the 68th centile interval is --245 to 61 K, showing that the uncertainties are inflated by a number of poorly fit outliers.

Warmer stars have their temperature under-predicted more frequently, and the scatter is greater towards under-predicted temperatures (1$\sigma$ = 193 K) than over-predicted temperatures (1$\sigma$ = 113 K). Scatter on the under-predicted side of the median will still be affected by interstellar reddening. However, the scatter on the over-predicted side of the median (113 K) should approximate the 1-$\sigma$ uncertainty in the results.

The same comparison was performed against the \emph{Hipparcos} dataset, where 359 stars could be matched against stars present in our final catalogue. Among those stars, the average stellar mass (with standard deviation) is 1.19 $\pm$ 0.37 M$_\odot$, the average metallicity is [Fe/H] = 0.09 $\pm$ 0.28 dex, and the average spectroscopic temperature is 5396 $\pm$ 658 K. The \emph{Hipparcos} exoplanet hosts are typically cooler, yet very slightly more massive, due to the larger fraction of evolved stars. They lie at a much closer average distance ($<d> = 66$ pc, cf.\ $<d> = 270$ pc for the Tycho-2 hosts). The average photometric temperature is 64 $\pm$ 163 K (1.2 $\pm$ 3.1 per cent) lower than the spectroscopic temperature. The median difference is marginally greater, at 69 K lower, however the 68th centile interval is considerably smaller, at --153 to 37 K, providing a scatter of $^{+106}_{-84}$ K.

The magnitude of the systematic offsets and scatter for both datasets are typical: other studies have made previous comparisons of these methods on small fields, over which interstellar reddening is both known and constant \citep{MJZ11,JMP+15,CMK16}. Based on these studies, the global systematic offset of $\sim$50--70 K probably represents an artificial difference in modelling approach, either in the fine detail of the model atmospheres used, few-per-cent differences in the zero points and colour terms in the underlying photometric catalouges, or the effects of atmospheres which are out of local thermodynamic equilibrium (see, e.g., discussions in \citealt{LML+14,JMP+15}). Meanwhile, the scatter of $\sim$100 K likely contains contributions from the uncertainty in the spectroscopic temperature ($\sim$50 K), errors from the assumed stellar metallicity ($\sim$30 K; Figure \ref{FeHCorrFig}), remaining scatter from the interstellar reddening ($\sim$10 K, based on the difference between the median Tycho-2 and \emph{Hipparcos} temperature offsets), and errors from the assumed stellar gravity ($\sim$50 K). The remainder ($\sim$60 K for the \emph{Hipparcos} stars and $\sim$80 for the Tycho-2 stars, if added in quadrature) probably comes from random uncertainties in the input photometry. We stress, however, that these estimated uncertainties are meant for indicative purposes only. They are not derived from an unbiased, random sample of the data, and should not be applied directly to any single star without great care. Our final adopted uncertainties (Section \ref{HRErrorTempSect}, below) are slightly inflated from these values to be conservative, regarding these values as a lower limit.

\subsubsection{Adopted uncertainty on the derived temperature}
\label{HRErrorTempSect}

\begin{figure}
\centerline{\includegraphics[height=0.47\textwidth,angle=-90]{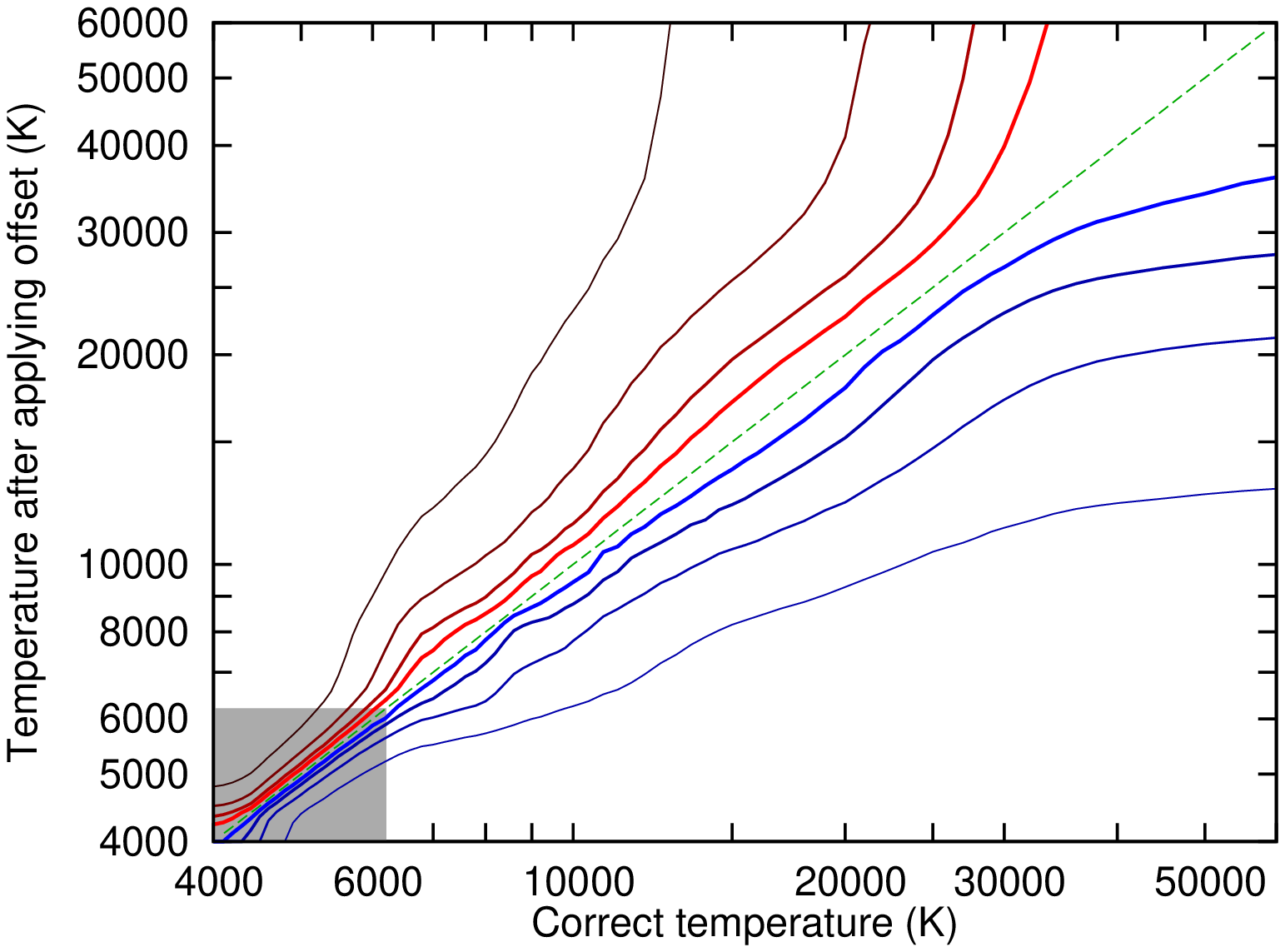}}
\centerline{\includegraphics[height=0.47\textwidth,angle=-90]{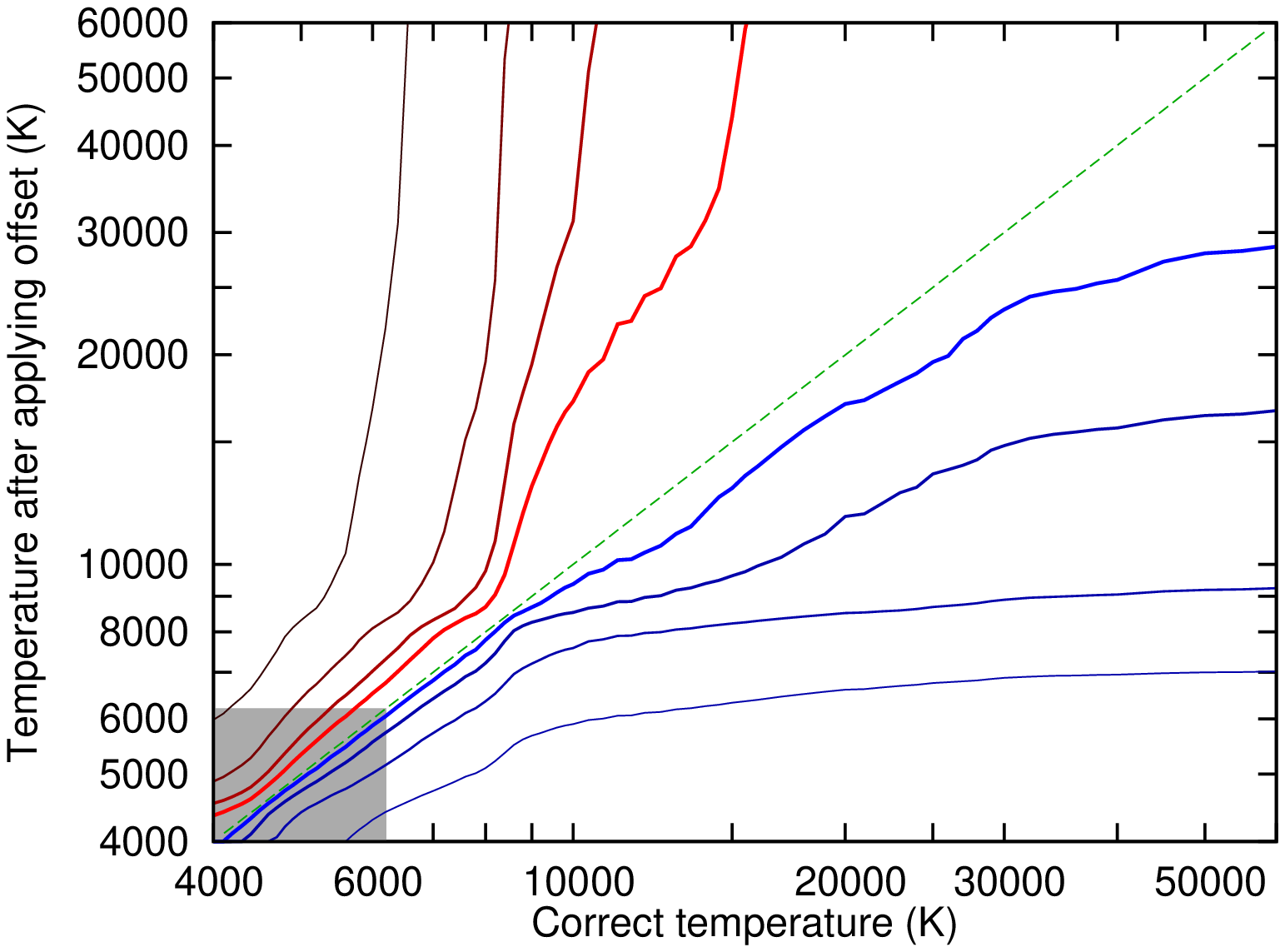}}
\caption{Temperature error arising from photometric inaccuracies in short-wavelength bands. {\it Top panel:} for data with $U$-band photometry. Red lines (above the dashed green line) show the effect of under-estimating the $U-B$, $U-B{\rm T}$, $B-V$ or $B_{\rm T}-V_{\rm T}$ colour (whichever is the most constraining) by 0.05, 0.1, 0.2 and 0.4 magnitudes. Blue lines (below the dashed green line) show the effects of over-estimating the colour by the same amount. {\it Bottom panel:} for data without $U$-band photometry, showing the effects of only the $B-V$ and $B_{\rm T}-V_{\rm T}$ colours. The grey box denotes temperatures below 6200 K, where redder colours provide better constraint on the temperature than those included here.}
\label{TempErrFig}
\end{figure}

\begin{figure*}
\centerline{\includegraphics[height=0.47\textwidth,angle=-90]{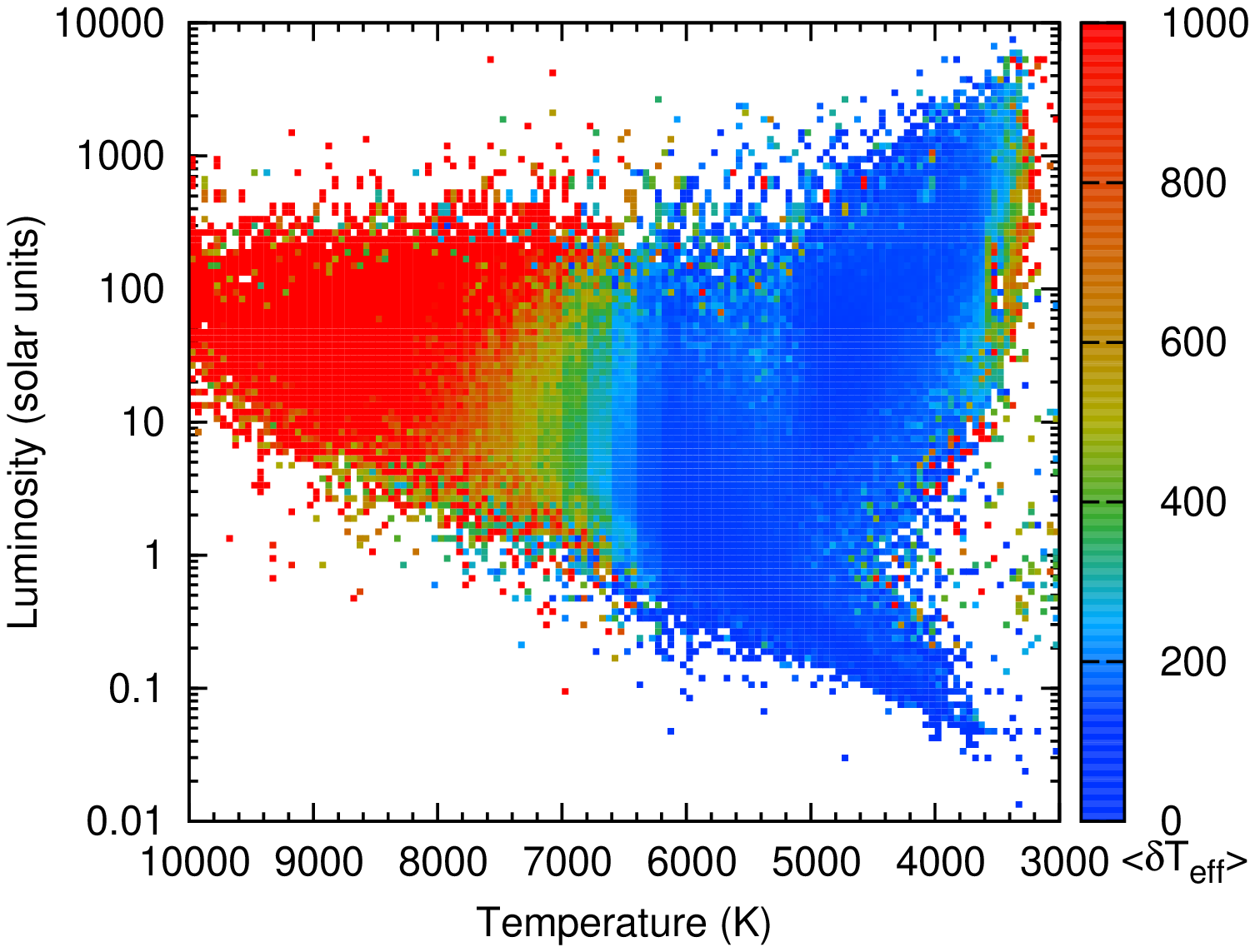}
            \includegraphics[height=0.47\textwidth,angle=-90]{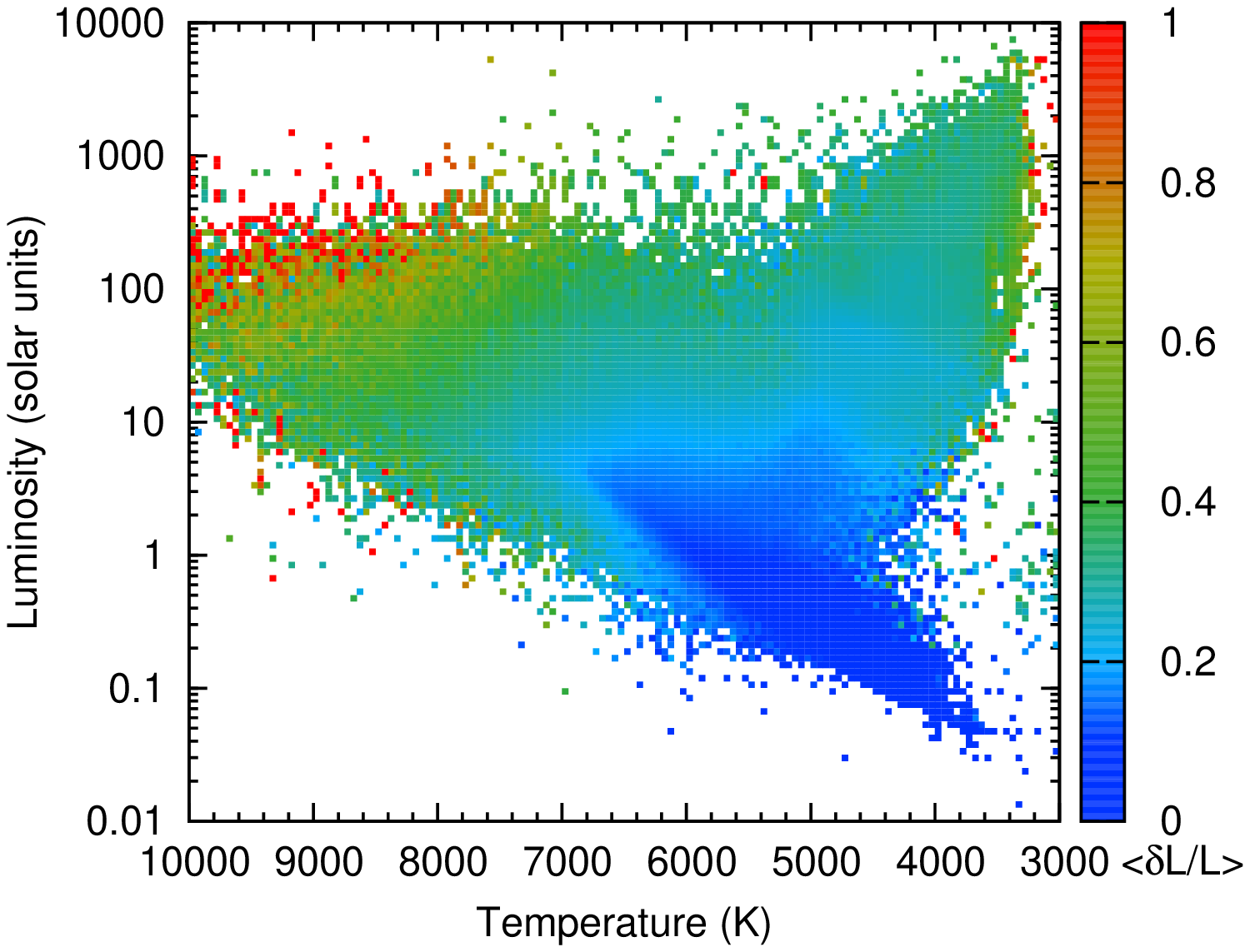}}
\centerline{\includegraphics[height=0.47\textwidth,angle=-90]{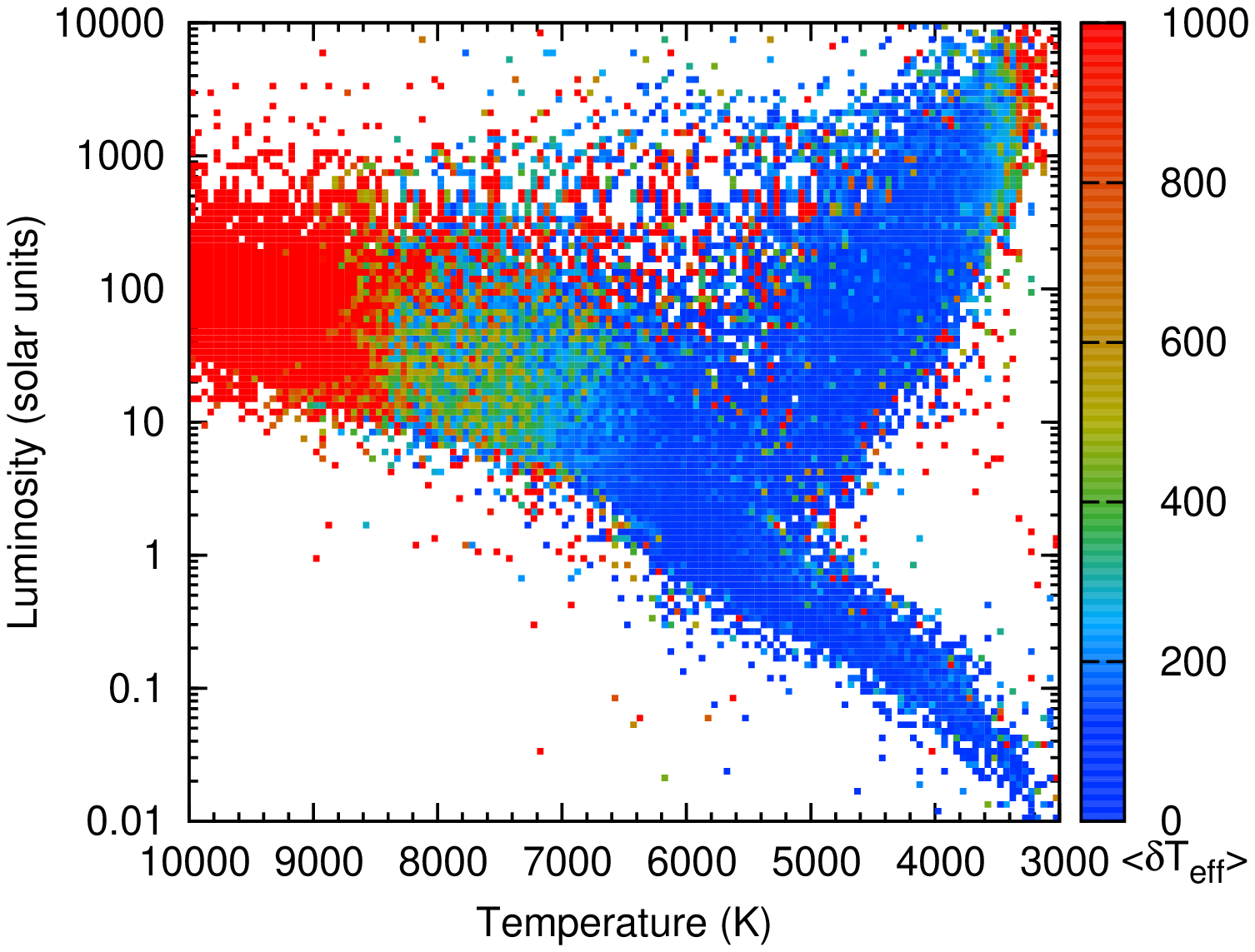}
            \includegraphics[height=0.47\textwidth,angle=-90]{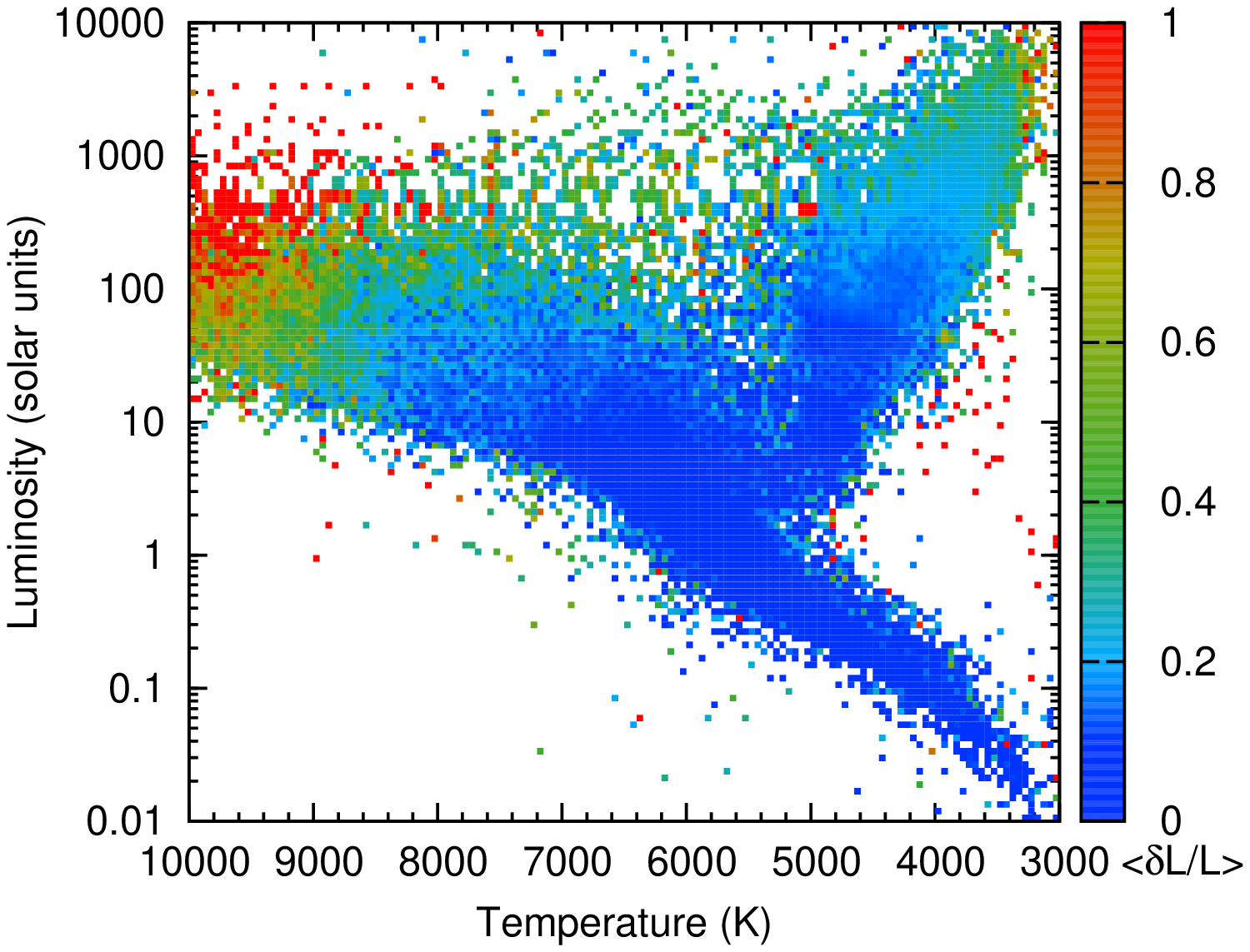}}
\caption{Uncertainties in temperature (left, absolute error) and luminosity (right, fractional error) averaged across the binned H--R diagram. The Tycho-2 sample is shown in the top panel, and the \emph{Hipparcos} sample in the bottom panels.}
\label{HRDErrFig}
\end{figure*}

To construct an error estimate that takes into account both the systematic offset and random scatter in Figure \ref{ExoFig}, we adopt the 68th centile of the distribution of \emph{absolute} deviations, as a measure that best reflects the uncertainty assigned to a typical star. For the Tycho-2 stars, this is $\sigma_T = 137$ K. For the \emph{Hipparcos} stars, $\sigma_T = 125$ K. These uncertainties should be appropriate for a star with typical fit uncertainties ($Q = 0.052$ and 0.042, respectively) providing the temperature is below $\sim$6200 K. The uncertainty should scale roughly with $Q$.

In hotter stars, there are no points sufficiently far down the Wien tail of the SED to accurately confine the stellar temperature. This limit is reached at $\sim$6200 K for photometry limited by the Johnson $B$, Tycho $B_{\rm T}$, or especially Sloan $g'$ filters (depending on the stellar gravity and metallicity). However, for some \emph{Hipparcos} stars, photometry extends to the Sloan $u'$ or Johnson $U$ filters. The magnitude of the Balmer jump, covered by these filters, can provide accurate temperatures up to a little over 10\,000 K.

Absolute flux calibration of the shortest wavelength bands are particularly important here. Figure \ref{TempErrFig} shows how the derived temperature departs from the mean for hot stars with and without $U$-band photometry, for a range of different photometric errors. For example, a 0.1 mag uncertainty in the $u' - B_{\rm T}$ colour of a 10\,000 K star can result in a temperature uncertainty of order $\pm$600 K, as will a 0.05 mag uncertainty in the $U - B$ colour. Equivalent uncertainties on a 12\,000 K star result in a range in temperatures from 11\,000 K to 19\,000 K, meaning stars with temperatures above 10\,000 K cannot be accurately placed on the H--R diagram via the SED method without UV photometry. In such cases, correctly accounting for interstellar extinction becomes extremely important (see Figure \ref{AVCorrFig}).

We assign an uncertainty on the derived temperature for \emph{Hipparcos} with $U$-band or $u'$-band photometry, given by the largest out of the following options:
\begin{itemize}
\item $\delta T = 125$ K;
\item $\delta T = 125 (Q / 0.051)$ K;
\item $\delta T = \Delta_Q$ K, as described below, if $T > 6250$ K (see note below);
\item $\delta T = \Delta_R$ K, as described below, if $T > 6250$ K (see note below).
\end{itemize}
The first option denotes a minimum standard error. The second option accounts for badly fit stars: roughly 68 per cent of stars have $Q < 0.051$, thus we can expect this to be the approximate threshold above which stars exceed the typical 125 K error calculated in the previous section\footnote{For comparison, the 68th centile for the planet hosts is comparable, at $Q = 0.053$.}.

The third option accounts for hot stars. Here, $\Delta_Q$ is the difference between the `correct' and `offset' temperatures in the top panel of Figure \ref{TempErrFig} for an offset of $\sqrt{2} Q$. For stars with $6250 < T < 10500$ K, this effect is brought in gradually, such that:
\begin{equation}
\delta T = \Delta_Q \frac{T - 6250}{10500 - 6250} {\rm K}.
\end{equation}
This accounts for the fact that some constraint is still applied by the longer-wavelength filters below 10\,500 K.

The fourth options accounts for hot stars that are otherwise well fit, but where the short-wavelength photometry is poorly fit. It subsitutes the offset of $\sqrt{2} Q$ for an offset of $R_{\rm U}$ or $R_{\rm u'}$ as appropriate. These options also account (to first order) for temperature uncertainties caused by circumstellar or interstellar reddening for both hot and cool stars. For \emph{Hipparcos} stars without $u'$-band or $U$-band photometry, we use the lower panel of Figure \ref{TempErrFig} for the third option, and $R_{\rm B}$ or $R_{\rm BT}$ for the fourth option. As with $\Delta_Q$, $\Delta_R$ is brought in gradually between 6250 and 10\,500 K for stars without $u'$-band or $U$-band photometry, and `instantaneously' at 10\,500 K for those with either of these bands observed.

Similarly, we assign an uncertainty for Tycho-2 stars as the largest out of the following options:
\begin{itemize}
\item $\delta T = 137$ K;
\item $\delta T = 137 (Q / 0.060)$ K;
\item $\delta T = \Delta_Q$ K, as described below, if $T > 6250$ K;
\item $\delta T = \Delta_R$ K, as described below, if $T > 6250$ K.
\end{itemize}
Since the Tycho-2 sample lacks reliably matched $U$-band or $u'$-band photometry, the lower panel of Figure \ref{TempErrFig} is always used for the third option, and $R_{\rm B}$ or $R_{\rm BT}$ is always used for the fourth option.

For both $\Delta_Q$ and $\Delta_R$, we round up to the nearest 0.01 mag in $Q$ and $R$, and round up to the temperature grid point above the derived temperature (this is almost universally more uncertain than the grid point below). This provides a fairly conservative estimate of the random uncertainty applied by both the photometry and fitting procedure to the temperature assigned to the star. It does not fully include uncertainties due to interstellar or circumstellar reddening, which are detailed in Section \ref{HRErrorEBVSect}. We stress that none of these uncertainties is a formal uncertainty measure, but instead simply an estimate of the 1-$\sigma$ uncertainty that can be assigned to the stellar temperature. These uncertainties are listed in Tables \ref{TychoTable} and \ref{HipparcosTable} and mapped onto the H--R diagram in Figure \ref{HRDErrFig}.

\subsubsection{Adopted uncertainty on the derived luminosity}
\label{HRErrorLumSect}

The contribution of photometric uncertainty to the uncertainty in derived luminosity is discussed with case studies in \citet{MBvL+11}. Photometric uncertainty affects temperature and luminosity in different ways, depending on the wavelength in question. Over-prediction of flux at wavelengths bluer than the SED peak leads to over-prediction in effective temperature and over-prediction in luminosity, while over-prediction of flux at redder wavelengths leads to \emph{under}-prediction of the effective temperature and \emph{under}-prediction of the luminosity. The greatest luminosity change that can normally be effected is $\delta L / L = 4 \delta T / T$, since (for a blackbody) $L \propto T^4$. The combination of the above effects means that the power law is shallower than this, but not normally by much. Therefore, $\delta L / L = 4 \delta T / T$ represents a fairly good estimate, yet also a conservative one. For example, an under-prediction of temperature of 137 K on a 4500 K star leads to an over-estimation of its luminosity by $\delta L / L$ = 12.1 per cent.

The uncertainty in luminosity has a reasonably strong correlation with the uncertainty in temperature, but that correlation and its direction depend on the photometric data causing the uncertainty. Optical data which is overly bright will lead to over-estimated temperature and luminosity; over-estimated infrared data will lead to \emph{under-}estimated temperature but still over-estimated luminosity. Photometric uncertainties are usually fractionally larger at longer wavelength (due to the thermal or astrophysical background, or sensitivity issues). Hence, there is more usually an anti-correlation between the photometric and luminosity uncertainties.

For hot stars, uncertainties in luminosity correlate with uncertainties in temperature, scaling as\footnote{In hot stars, the uncertainty is driven by the short-wavelength filters: the flux of the Rayleigh--Jeans tail is observationally well constrained. However, the flux at a wavelength on a blackbody's Rayleigh--Jeans tail varies linearly with temperature. If poor-quality optical photometry leads to an over-estimation in optical flux, the derived temperature increases. Accordingly, the derived surface area then decreases as $R \propto T^{-2}$. Thus, by $L \propto R^2 T^4$, the luminosity relation is to the third power, rather than the fourth.}: $\delta L / L = 3 \delta T / T$. The aforementioned $\sim$600 K uncertainty in the temperature of a 10\,000 K star results in a 24 per cent uncertainty in luminosity.

In most cases, the photometric contribution to the luminosity uncertainty is exceeded by the distance uncertainty to the star. The average parallax uncertainty on our Tycho--\emph{Gaia} sample is $\sigma_\varpi / \varpi = 16.4$ per cent, leading to an uncertainty in luminosity of $\sigma_L / L = 32.8$ per cent. For the \emph{Hipparcos} / \emph{Hipparcos--Gaia} sample, they are $\sigma_\varpi / \varpi = 7.6$ and $\sigma_L / L = 15.1$ per cent, respectively.

Our final luminosity uncertainty (see also Figure \ref{HRDErrFig}) is given as:
\begin{equation}
\delta L / L = \sqrt { \left( n \delta T / T \right)^2 + \left (\delta \varpi / \varpi \right)^2 } ,
\end{equation}
where $n = 4$ if $T < 6200$ K, $n = 3$ if $T > 10500$ K, and $n = 4 - (10500 - T) / (10500 - 6200)$ in between. These uncertainties are listed in Tables \ref{TychoTable} and \ref{HipparcosTable}. We again stress that these are not formal uncertainties.

\subsection{``Sanity checking'' of local population and interstellar extinction}
\label{DiscIsoSect}

\subsubsection{Galactic thick- and thin-disc populations}
\label{DiscIsoDiscSect}

\begin{figure}
\centerline{\includegraphics[height=0.47\textwidth,angle=-90]{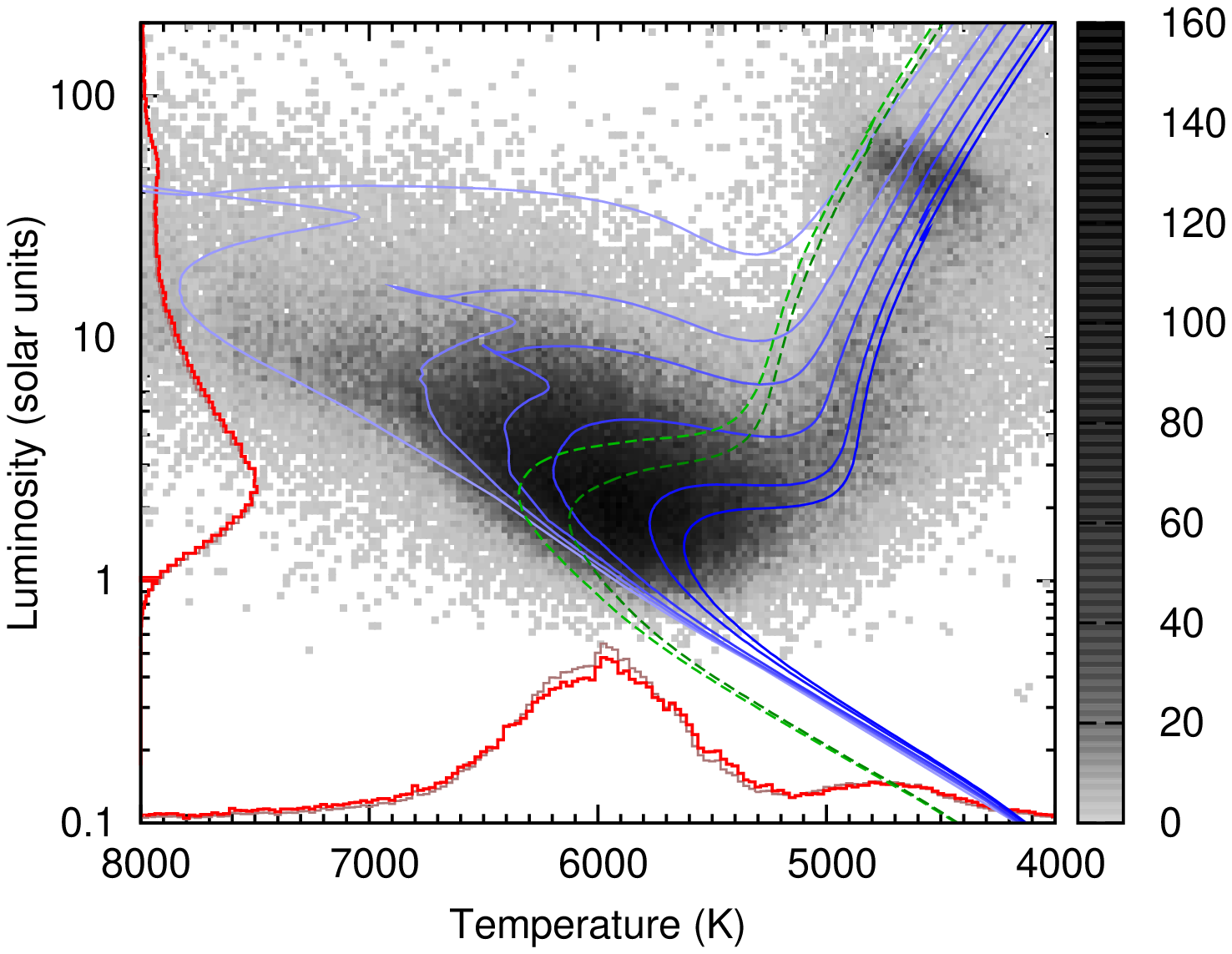}}
\centerline{\includegraphics[height=0.47\textwidth,angle=-90]{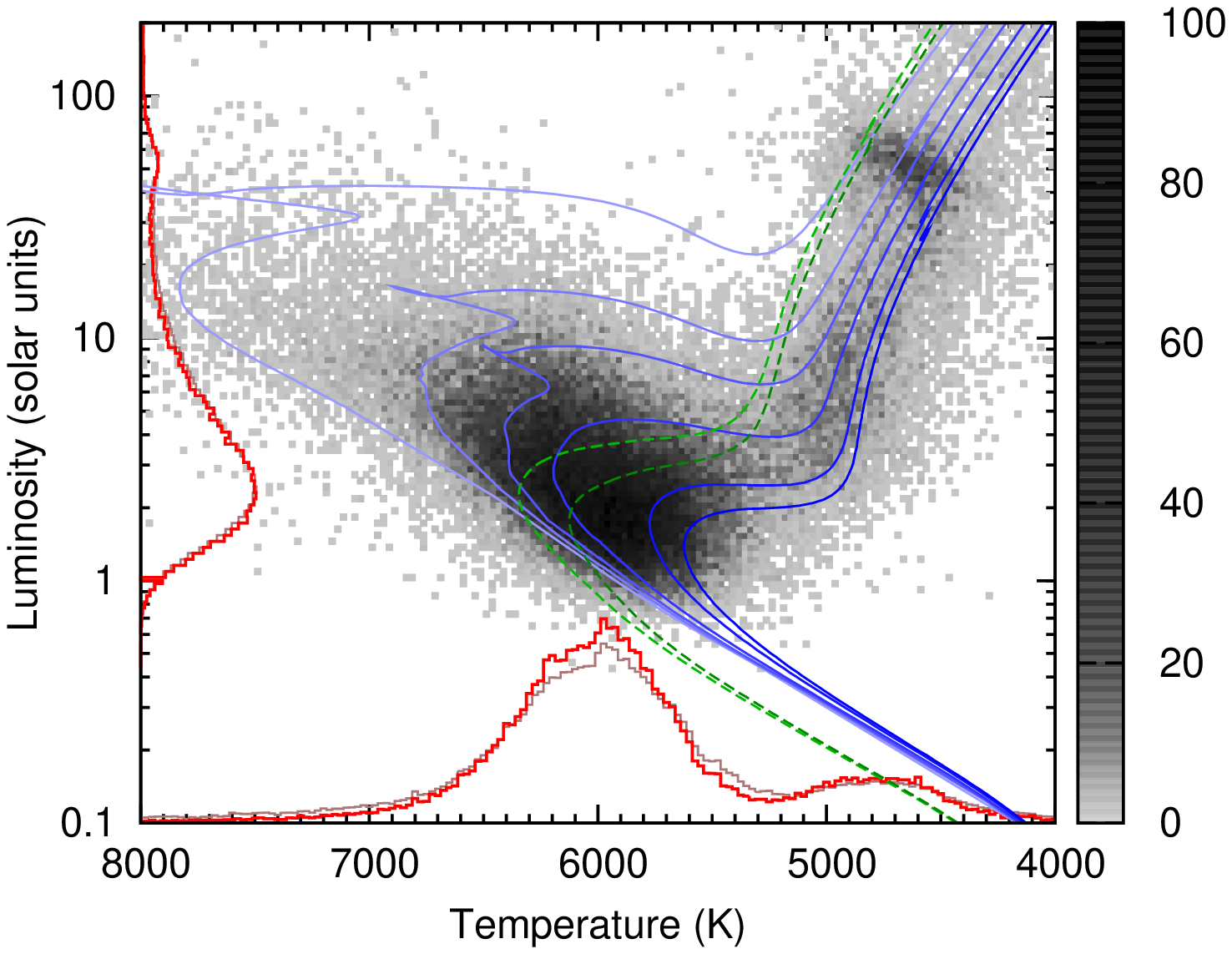}}
\centerline{\includegraphics[height=0.47\textwidth,angle=-90]{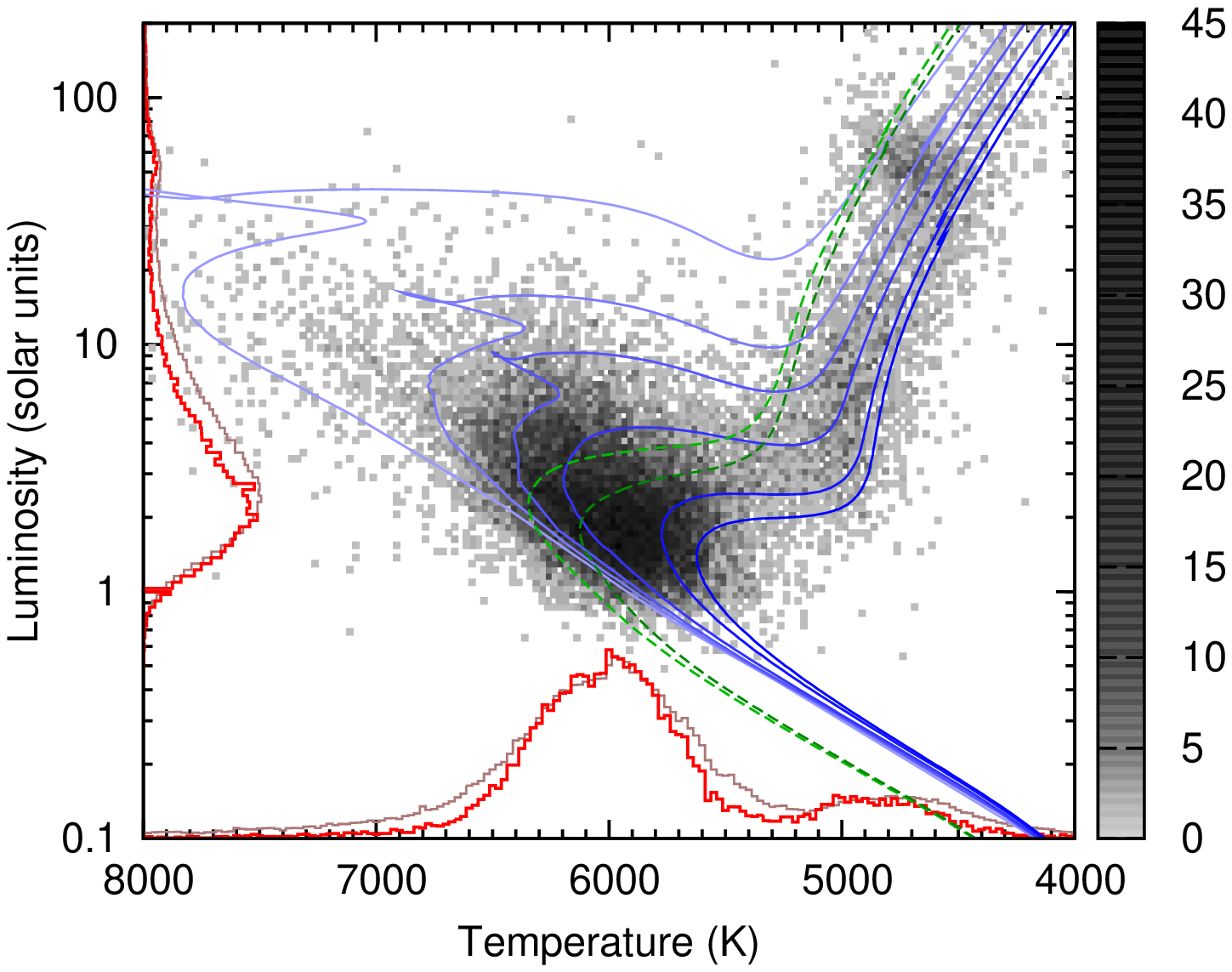}}
\caption{Density-coded (Hess) H--R diagram of stars between 300 and 400 pc from the Sun. The panels show (top to bottom) Galactic latitudes $\pm$0--30$^\circ$, 30--60$^\circ$ and 60--90$^\circ$, representing distances 0--200 pc, 150--350 pc and 260--400 pc from the Galactic Plane. Thick red lines show histograms of sources in that plot, compared to the lighter lines of sources at all latitudes. Overlain on the H--R diagrams are isochrones from \citet{MGB+08}, showing (in blue, top to bottom) isochrones for solar-composition stars at 1, 2, 3, 5, 10 and 13 Gyr. The dashed, green lines show 10 and 13 Gyr isochrones at [Fe/H] = --1 dex and [$\alpha$/Fe] = +0.2 dex.}
\label{GalLatFig}
\end{figure}

\begin{figure}
\centerline{\includegraphics[height=0.47\textwidth,angle=-90]{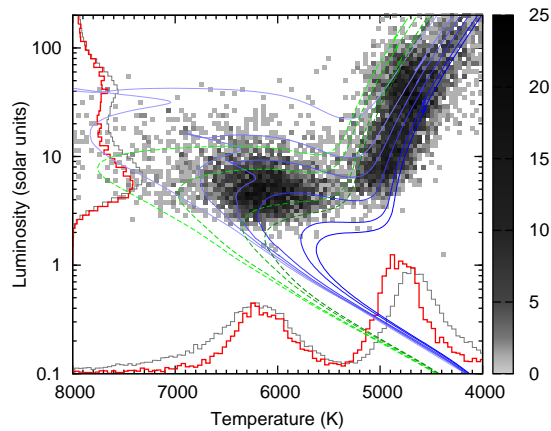}}
\caption{As the bottom panel of Figure \ref{GalLatFig}, but for stars between 600 and 800 pc from the Sun at at Galactic latitudes $\pm$60--90$^\circ$. The thinner, grey histogram shows stars in the range $\pm$10--50$^\circ$. Metal-poor isochrones are shown (green, dashed lines) for 3, 5, 10 and 13 Gyr, as well as the solar-metallicity isochrones from Figure \ref{GalLatFig}. Note the warmer giant branch.}
\label{GalLat2Fig}
\end{figure}

Figure \ref{GalLatFig} shows the H--R diagram for stars at a fixed range of distances (300--400 pc) at differing Galactic latitudes\footnote{A mild Lutz--Kelker bias exists at these distances, which is latitude dependent due to the changing density of objects.}. The solar-metallicity thin-disc population dominates at these scale heights. Stars are recovered down to the main-sequence turn-off in all cases, and extinction does not yet severely affect star counts in the Galactic Plane (however, see discussion on the Gould Belt, below). Without performing a detailed population model, it is still clear that completeness declines markedly below $\sim$3 L$_\odot$ at all latitudes.

At high latitudes, few stars at ages $<$3 Gyr are seen. The red clump appears both young and luminous if at solar metallicity\footnote{The metallicity correction in this region is typically $<$100 K per dex in metallicity (Figure \ref{FeHCorrFig}).}. \citet{MMN+16} determined a median age of $\sim$5 Gyr for red clump stars at scale heights of $\sim$300 pc. Even at high latitudes, we expects approximately solar abundances, as solar metallicity was reached by the time star formation ceased in the Galactic thick disc, $\sim$10 Gyr ago \citep{BFL04}. A significant component from the thick disc is not expected until scale heights of $>$500 pc \citep[e.g.][]{GR83,KZ08}. Along with our completeness limitations, this combination of factors explains the lack of stars lying below the solar-metallicity main sequence. However, the luminosity of the RGB bump is also strongly metallicity dependent \cite[cf.][]{BMvL+09,MBvL+11}, so including an old, metal-poor population which reduces the average abundance to slightly sub-solar metallicities ($\sim$0.2 dex), allows the RGB bump to be fit reasonably well.

Figure \ref{GalLat2Fig} shows the H--R diagram for high-latitude stars between 600 and 800 pc from the Sun (520--800 pc from the Plane). Sensitivity declines rapidly below $\sim$6 L$_\odot$, limiting inclusion to main-sequence turn-off stars $\lesssim$5 Gyr in age. Few stars are younger than $\sim$3 Gyr, or hotter than $>$6500 K. A significant shift in the temperature of the giant branch and red clump indicates stars are metal-poor: a crude estimate places them at [Fe/H] $\sim$ --0.5 dex, as expected from chemical studies \citep[e.g.][]{MG15}.

\subsubsection{The Galactic Plane and Gould Belt}
\label{DiscIsoGouldSect}

\begin{figure}
\centerline{\includegraphics[height=0.47\textwidth,angle=-90]{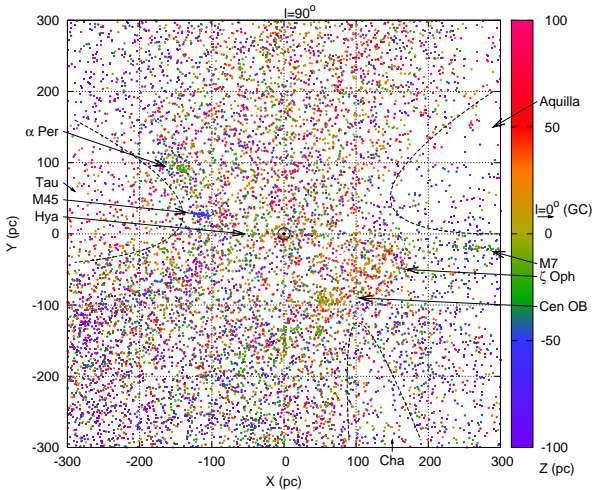}}
\caption{Hot stars ($>$8000 K) within 100 pc of the Galactic plane, colour-coded by height above/below the plane. Major known features are identified. The Galactic centre is to the right, and Galactic longitude ($l$) increases anti-clockwise.}
\label{GouldFig}
\end{figure}
\begin{figure}
\centerline{\includegraphics[height=0.47\textwidth,angle=-90]{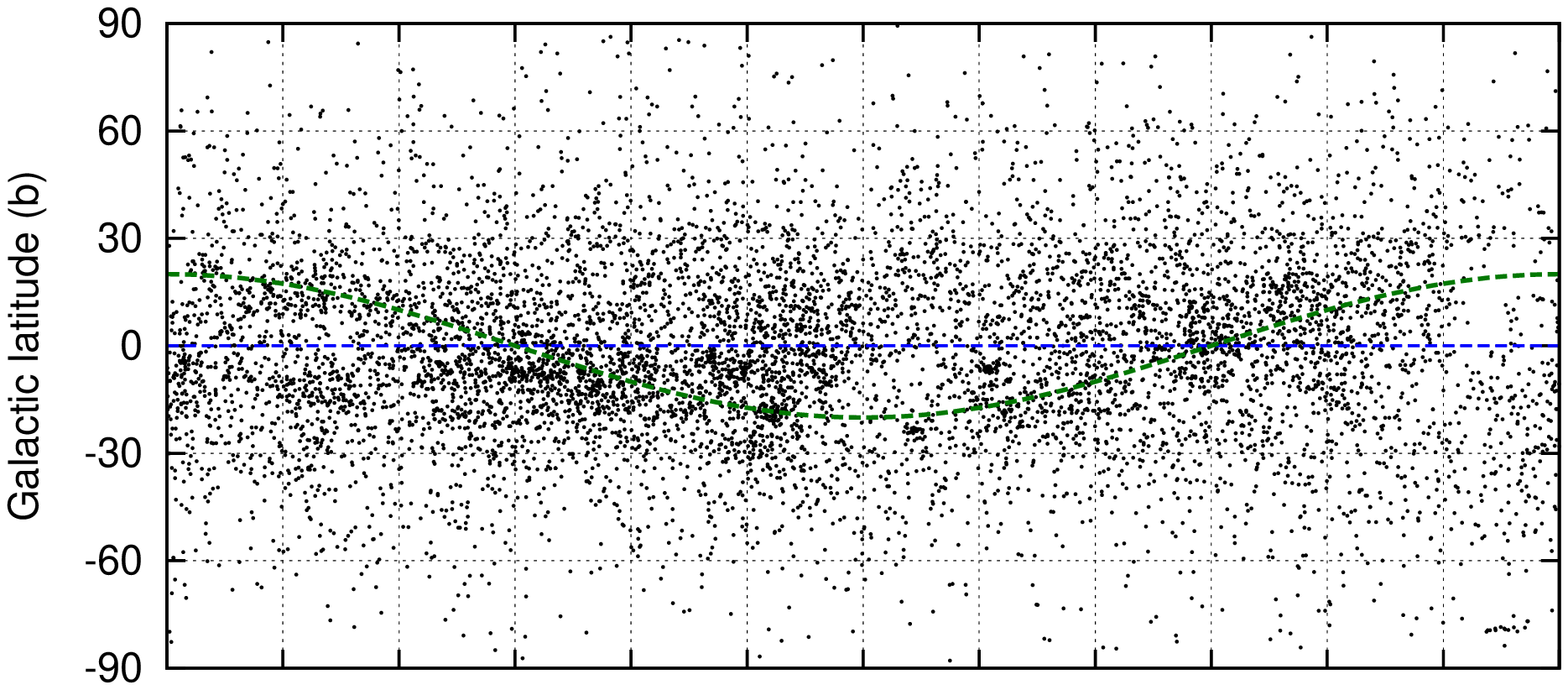}}
\centerline{\includegraphics[height=0.47\textwidth,angle=-90]{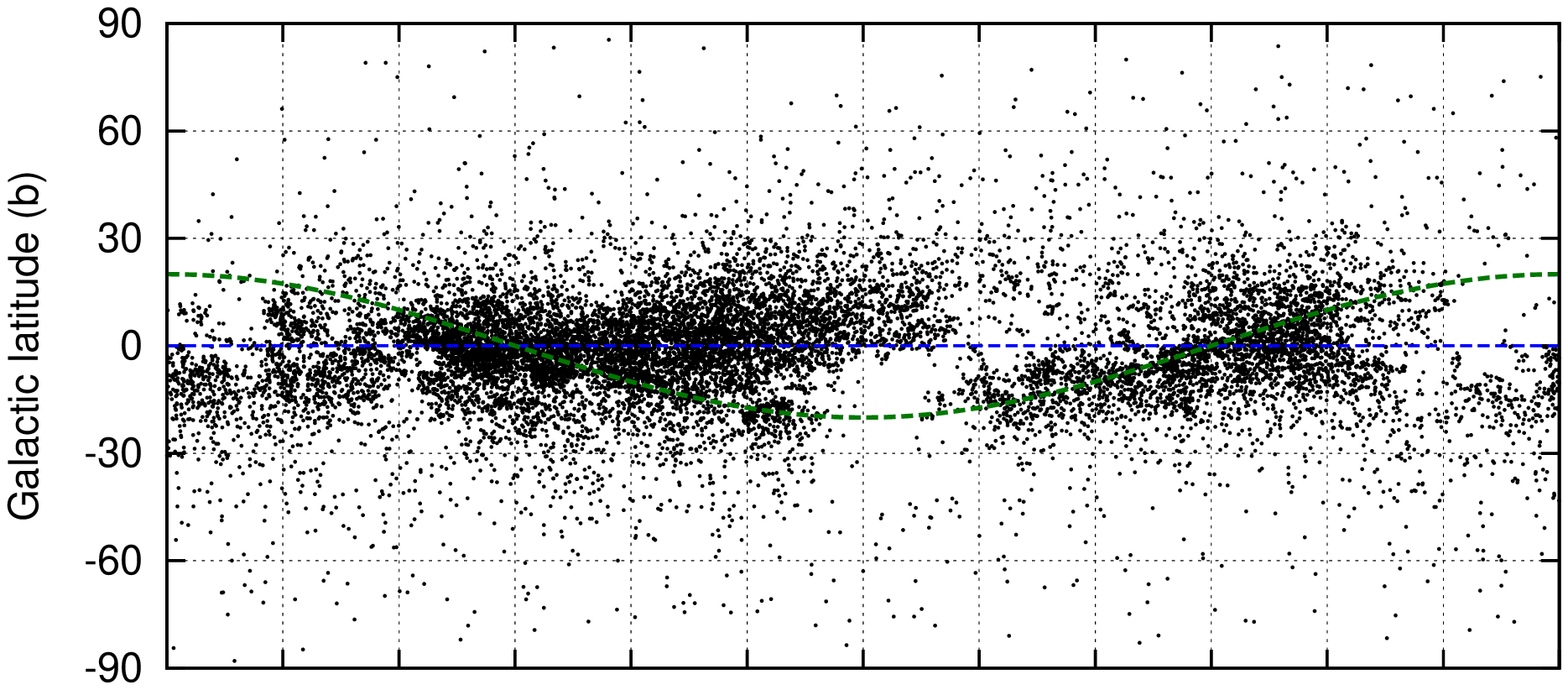}}
\centerline{\includegraphics[height=0.47\textwidth,angle=-90]{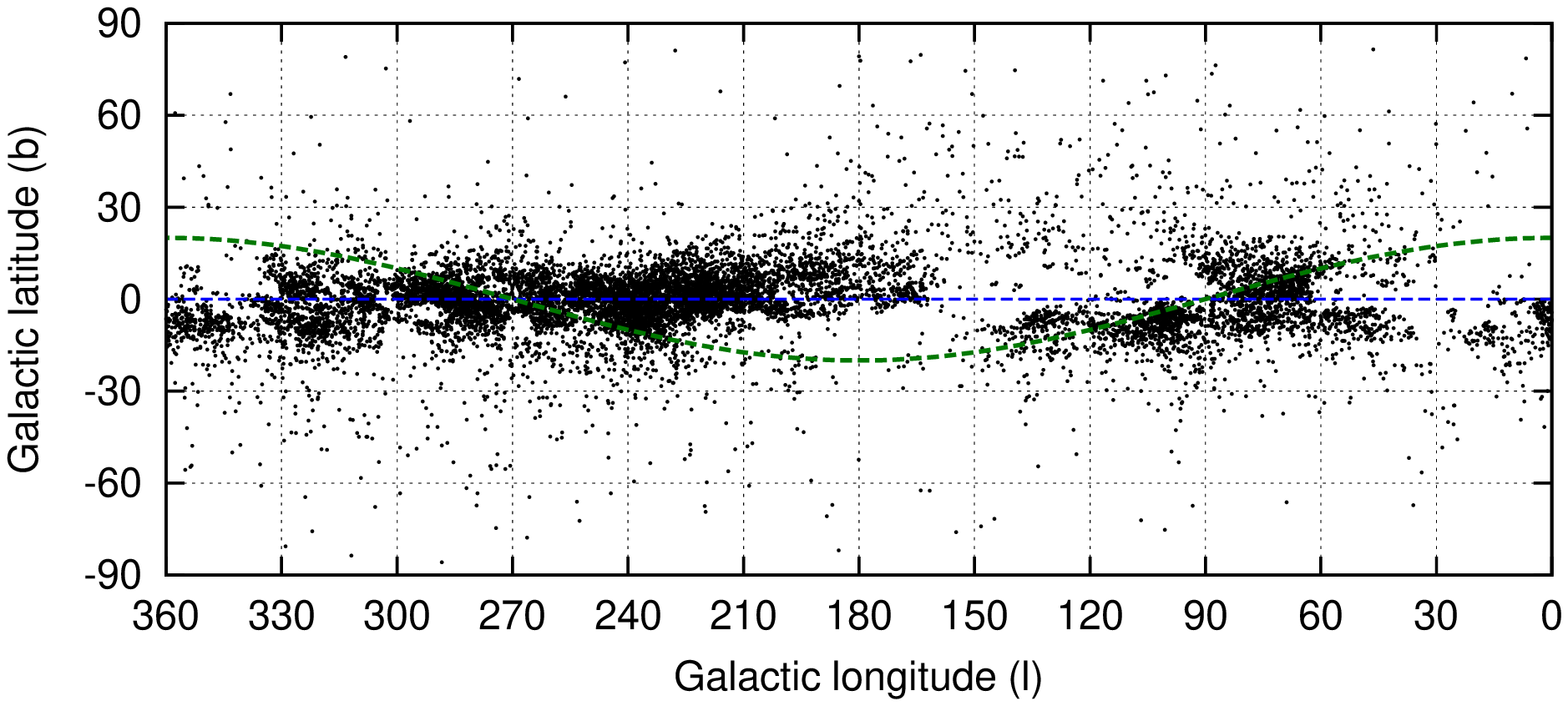}}
\caption{Hot stars ($>$8000 K) in the solar neighbourhood, showing the Gould Belt (dashed line). From top to bottom, the panels represent stars in the ranges 0--300, 300--600 and 600--900 pc from the Sun. Note that the bottom plot in particular will suffer from a strong Lutz--Kelker bias from stars at greater distances. Note the absence of stars along various Galactic Plane sightlines, indicating strong interstellar extinction.}
\label{Gould2Fig}
\end{figure}

The Gould Belt is an elliptical structure of young stars and star-formation regions, with major and minor axes roughly 400 $\times$ 300 pc. It is centred approximately on the $\alpha$ Per moving group, but presents on the terrestrial sky with a roughly constant 20$^\circ$ inclination with respect to the Galactic Plane. The Sun lies close to its inner edge, as traced by the Scorpius--Centaurus OB association \citep[e.g][]{Herschel1847,Olano82,Olano01,dZHdB+99,WTDIH07}. \emph{Gaia} DR1 records distances to individual stars with sufficient accuracy that membership of associations can be made within a few hundred pc of the Sun, covering roughly the nearer half of the Gould Belt. This region is presented in Figure \ref{GouldFig} and mapped onto the sky in Figure \ref{Gould2Fig}. In the further half of the Gould Belt, parallax uncertainties become large and smearing of associations in the radial direction and the associated Lutz--Kelker effects restrict detailed analysis of this region.

The majority of structures in the western part of the Gould belt ($150^\circ < l < 360^\circ$) are located within 300 pc, and the majority of the structures in the eastern part ($60^\circ < l < 150^\circ$) are between 300 and 600 pc, as in the studies cited above. However, at high resolution, the belt breaks up into the more discontinuous features of Figure \ref{GouldFig}. Figure \ref{Gould2Fig} also shows the regions affected by large interstellar dust clouds. The three primary offenders (Aquilla, Taurus and Chameleon) are shown in Figure \ref{GouldFig}. Stars in these regions suffer several magnitudes of visual extinction, so are either reddened sufficiently that they no longer appear to be above 8000 K (cf.\ Figure \ref{AVCorrFig}), or were otherwise rendered entirely invisible to the \emph{Hipparcos} and Tycho instruments. The presence of the Gould Belt is also traced by the distribution of stars with infrared excess in Figure \ref{XSMapFig}, indicating the large number of young stars (pre-main-sequence and Herbig Ae/Be stars) in this region.


\section{Infrared excess}
\label{DiscXSSect}

\subsection{Criteria for defining infrared excess}
\label{DiscXSCriteriaSect}

A definition of infrared excess must take into account all the above factors. We start with two assumptions:
\begin{enumerate}
\item The region $<$4.3 $\mu$m defines the stellar continuum. This region should be relatively free from circumstellar emission.
\item The region $\geq$4.3 $\mu$m defines the regime in which infrared excess occurs.
\end{enumerate}
The factors behind these assumptions are detailed in Appendix \ref{AppendixDust} (online only).

To help quantify infrared excess, we define the following statistics, using the individual observed/modelled flux ratios ($F_{\rm o}/F_{\rm m}$) and the overall quality of fit ($Q$) described in Section \ref{SEDMethodSect}:
\begin{itemize}
\item $\Re_{\rm opt}$ defines the average value of $F_{\rm o}/F_{\rm m}$ over the optical filters ($UBVR$, $ugr$).
\item Similarly, $\Re_{\rm NIR}$ defines the average of $F_{\rm o}/F_{\rm m}$ over the near-IR filters ($IJHK_{\rm s}L$, $iz$, and \emph{WISE} [3.4]).
\item Also, $\Re_{\rm MIR}$ defines the average of $F_{\rm o}/F_{\rm m}$ over the mid-IR filters (longward of $L$ and [3.4]).
\item $N_{\rm opt}$, $N_{\rm NIR}$ and $N_{\rm MIR}$ denote the number of near-IR and mid-IR datapoints, respectively, which contribute to the above.
\item The combined $\Re_{\rm opt+NIR}$ and $N_{\rm opt+NIR}$ represent the same quantities as $\Re_{\rm opt}$ and $N_{\rm opt}$, but computed over the full $U$ through [3.4] range.
\item $\Re^{\prime}_{\rm MIR}$ provides an alternative version of $\Re_{\rm MIR}$, removing the point with the maximum $R$ from the mid-IR data.
\item $X_{\rm MIR}$ provides a statistic of overall mid-infrared excess, calculated as:
	\begin{equation}
	X_{\rm MIR} = \Re_{\rm MIR} / \Re_{\rm opt+NIR} .
	\end{equation}
	This statistic should be most sensitive to faint mid-IR excess if the host star is unreddened. If it is substantially reddened, or contains a single bad mid-infrared datapoint, then:
	\begin{equation}
	X^\prime_{\rm MIR} = \Re^{\prime}_{\rm MIR} / \Re_{\rm NIR} 
	\end{equation}
	should provide a more accurate value. Robustness of the detection is therefore increased where both $X_{\rm MIR}$ and $X^\prime_{\rm MIR}$ are significantly above unity.
\item $S_{\rm MIR}$ provides a statistic of the significance of mid-infrared excess, calculated as:
	\begin{equation}
	S_{\rm MIR} = (\Re_{\rm MIR} - 1) \sqrt{N_{\rm MIR}} / {Q} .
	\end{equation}
	This approximates the signal-to-noise statistic of the infrared excess. Note that this will generally be an over-estimate for stars with little excess: scatter due to photometric errors will typically be much greater in the infrared than the optical and near-IR, meaning that the fit quality parameter, $Q$, will be an under-estimate for the `noise' component in this equation. For stars with significant excess, this will generally be an under-estimate, as the infrared excess artificially inflates the $Q$ parameter. We also note that this significance statistics does not exclude objects such as stars heavily reddened by interstellar extinction. This statistic is therefore presented for guidance only and should be used in combination with the others in this section to define whether a source has a significant excess.
\item To determine the amount of light emitted in the infrared excess, we construct a trapezoid integral, interpolated in the (log $F_\nu$)--(log $\lambda$) plane. This (respectively) provides the total luminosity and fraction of the stellar flux re-emitted into the infrared:
	\begin{equation}
	L_{\rm XS} = \int_{\nu = 0}^{7 \times 10^{13} {\rm Hz}} (F_{\nu} - F_\ast) \ {\rm d} \nu
	\end{equation}
and:
	\begin{equation}
	f_{\rm XS} = \frac { \int_{\nu = 0}^{7 \times 10^{13} {\rm Hz}} (F_{\nu} - F_\ast) \ {\rm d} \nu } { \int_{\nu = 0}^{\infty} F_\ast \ {\rm d} \nu } ,
	\end{equation}
	where we assume that the infrared excess beyond 1 mm is zero\footnote{Dust optical depth typically drops at longer wavelengths, as the emissivity of dust typically has a spectral slope steeper than a blackbody's (e.g.\ \citealt{SLO05}). For many objects, other emission mechanisms become important in the sub-millimetre and beyond (e.g.\ \citealt{RM97}).}, and that the stellar flux ($F_\ast$) is the modelled flux ($F_{\rm m}$) multiplied by $\Re_{\rm NIR}$. The cutoff of $7 \times 10^{13}$ Hz corresponds to 4.3 $\mu$m. This is a lower limit to the fraction of reprocessed light, since the SED fitting partially takes into account the optical absorption and infrared emission from this reprocessing.
\item Finally, we use this data to extract the wavelength at which the peak flux ($F_\nu$) of the infrared excess occurs, $\lambda_{\rm peak,XS}$, which is defined bythe point at which $(F_{\rm \nu} - F_\ast)$ reaches a maximum.
\end{itemize}

\subsection{A Hertzsprung--Russell diagram of infrared excess}
\label{DiscXSHRSect}

\begin{figure*}
\centerline{\includegraphics[height=0.97\textwidth,angle=-90]{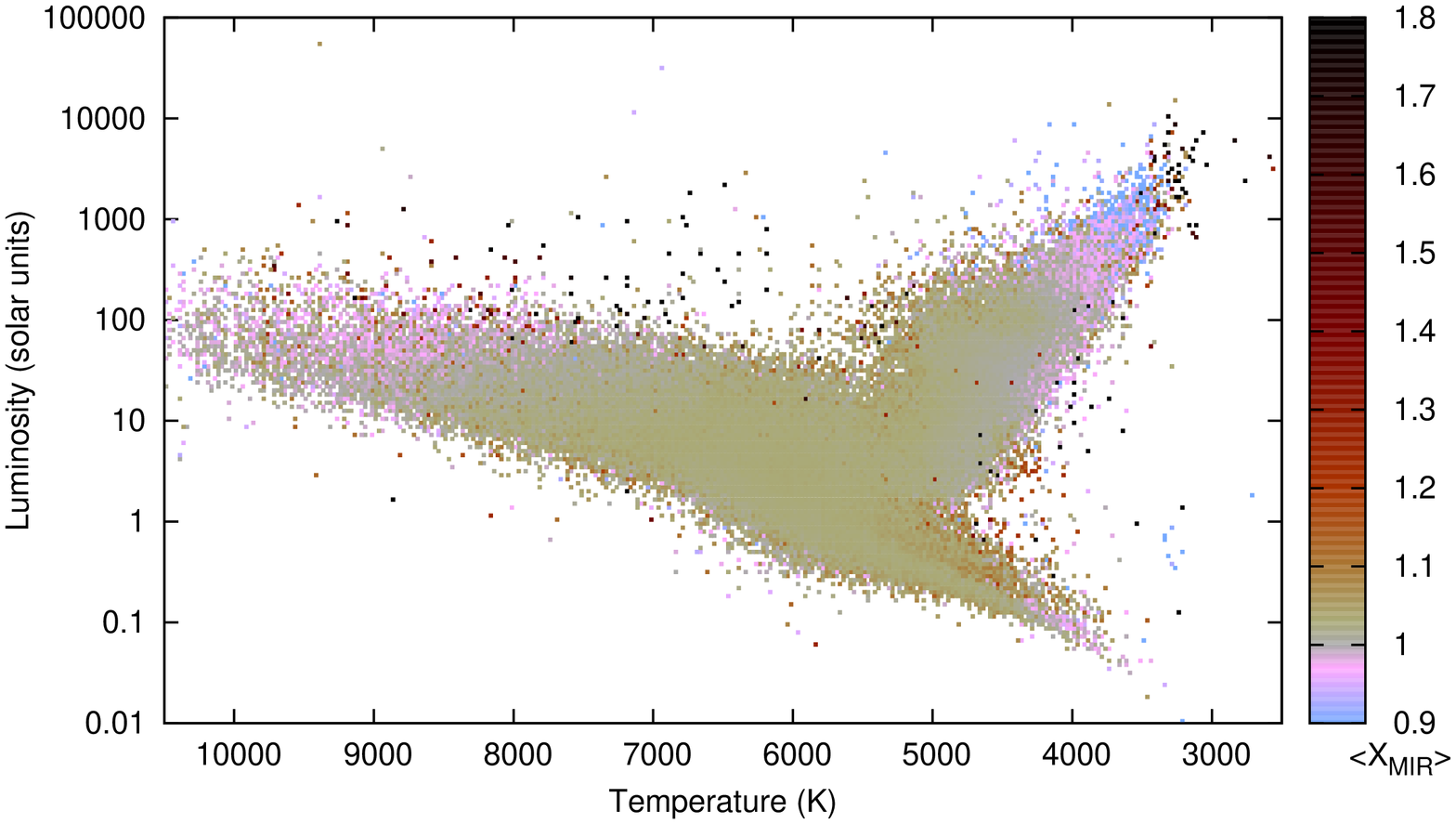}}
\centerline{\includegraphics[height=0.97\textwidth,angle=-90]{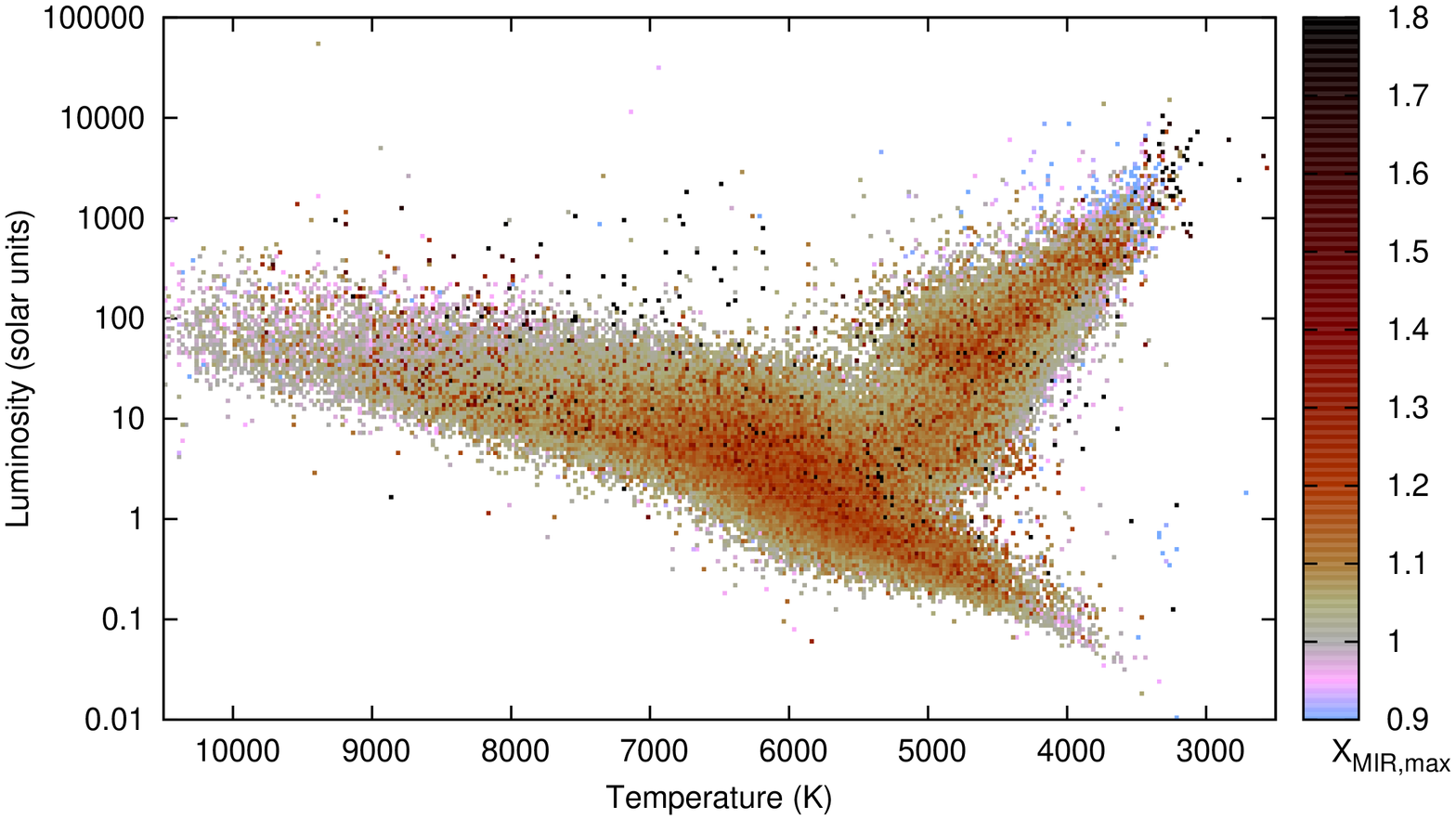}}
\caption{A binned Hertzsprung--Russell diagram, coloured to show the average mid-infrared excess ($X_{\rm MIR}$) in each bin. Stars are included if $N_{\rm opt} + N_{\rm NIR} > 0$, $N_{\rm MIR} > 1$, $\delta \varpi / \varpi < 0.2$ and $A_{\rm V} < 1.5$ mag. The top panel shows that average excess ($X_{\rm MIR}$) in each bin, with unity being no excess. The bottom panel shows the highest value of $X_{\rm MIR}$ in each bin, to show the most extreme sources.}
\label{HRDXSFig}
\end{figure*}

\begin{figure*}
\centerline{\includegraphics[height=0.97\textwidth,angle=-90]{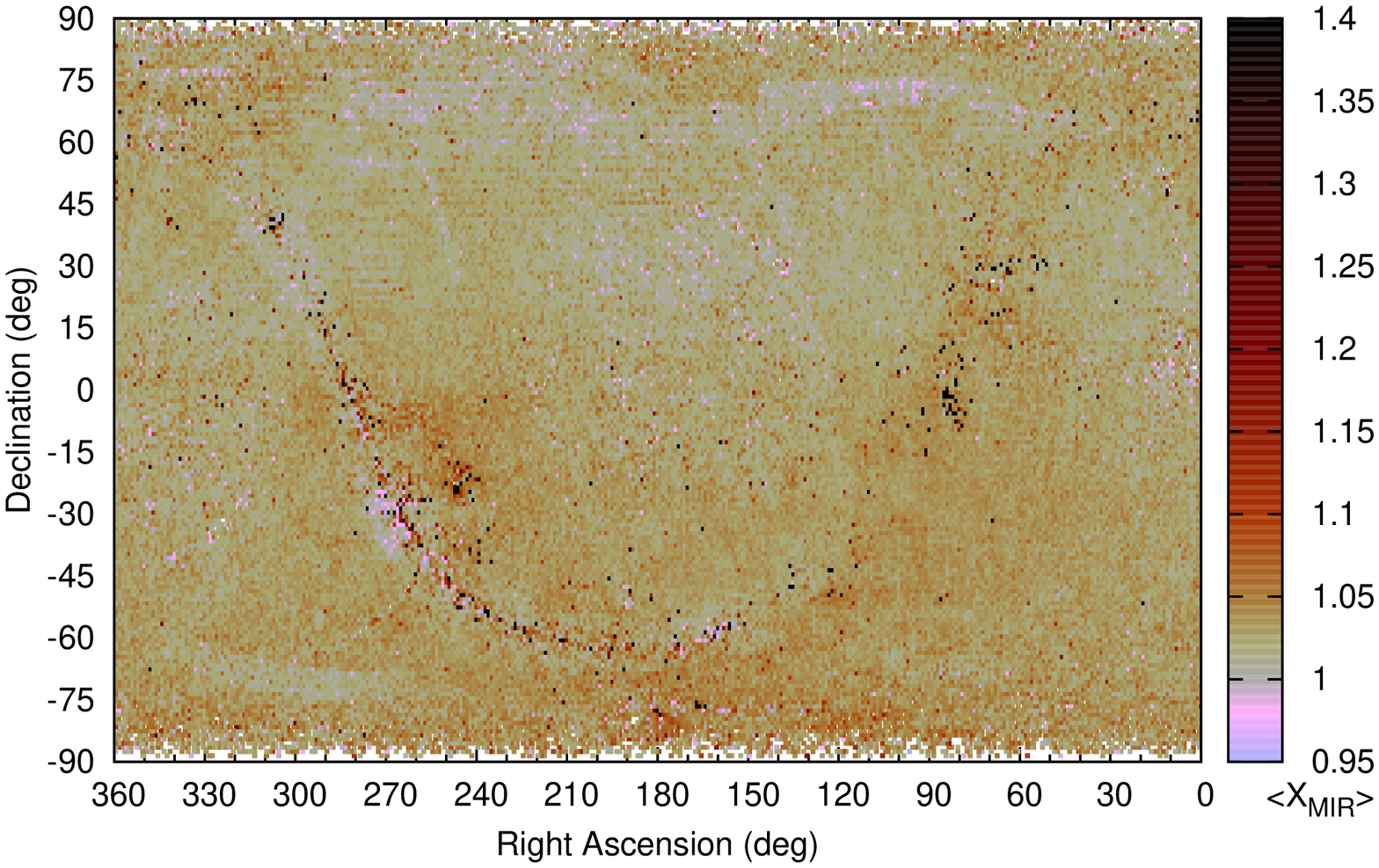}}
\centerline{\includegraphics[height=0.47\textwidth,angle=-90]{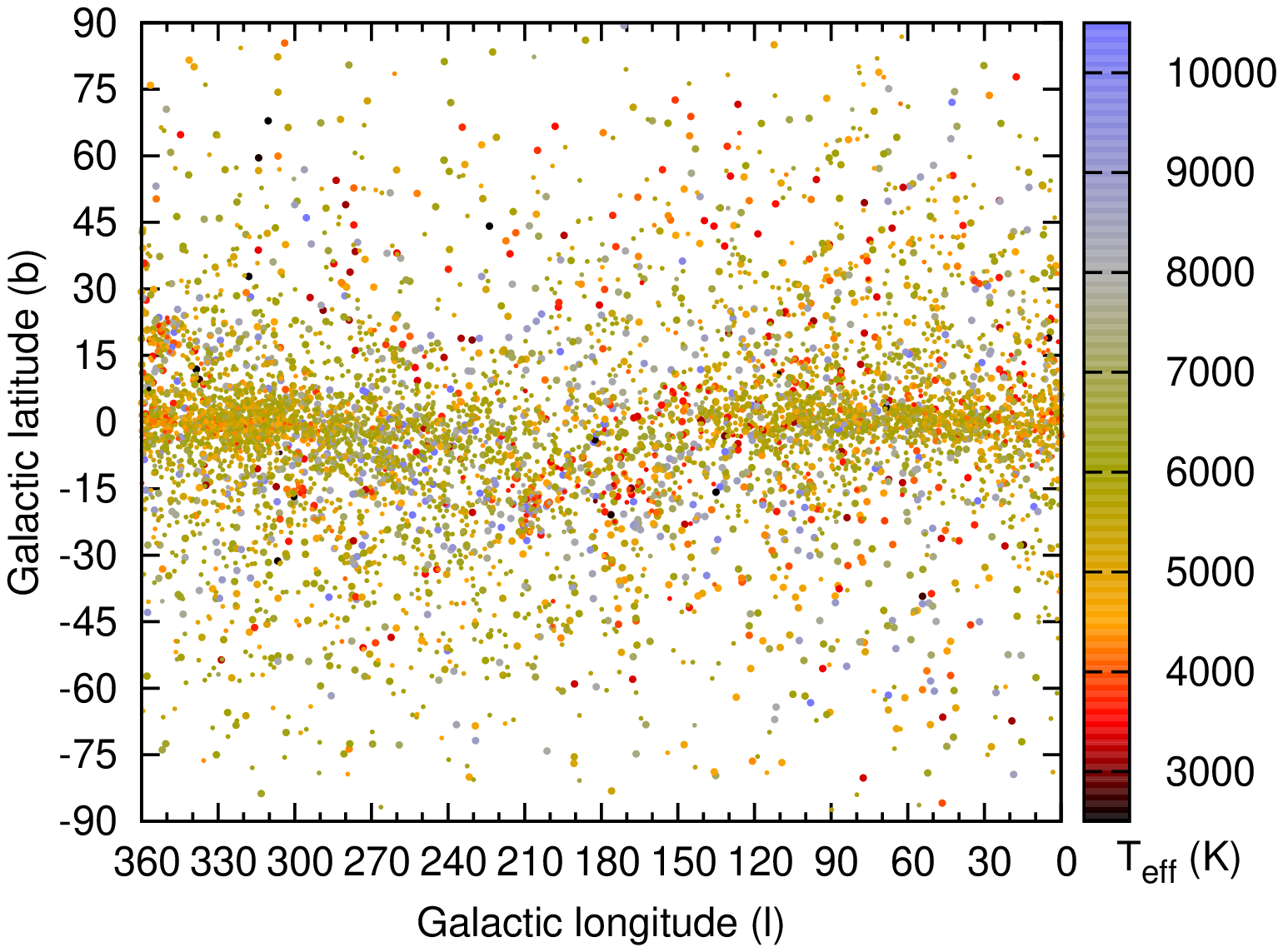}
            \includegraphics[height=0.47\textwidth,angle=-90]{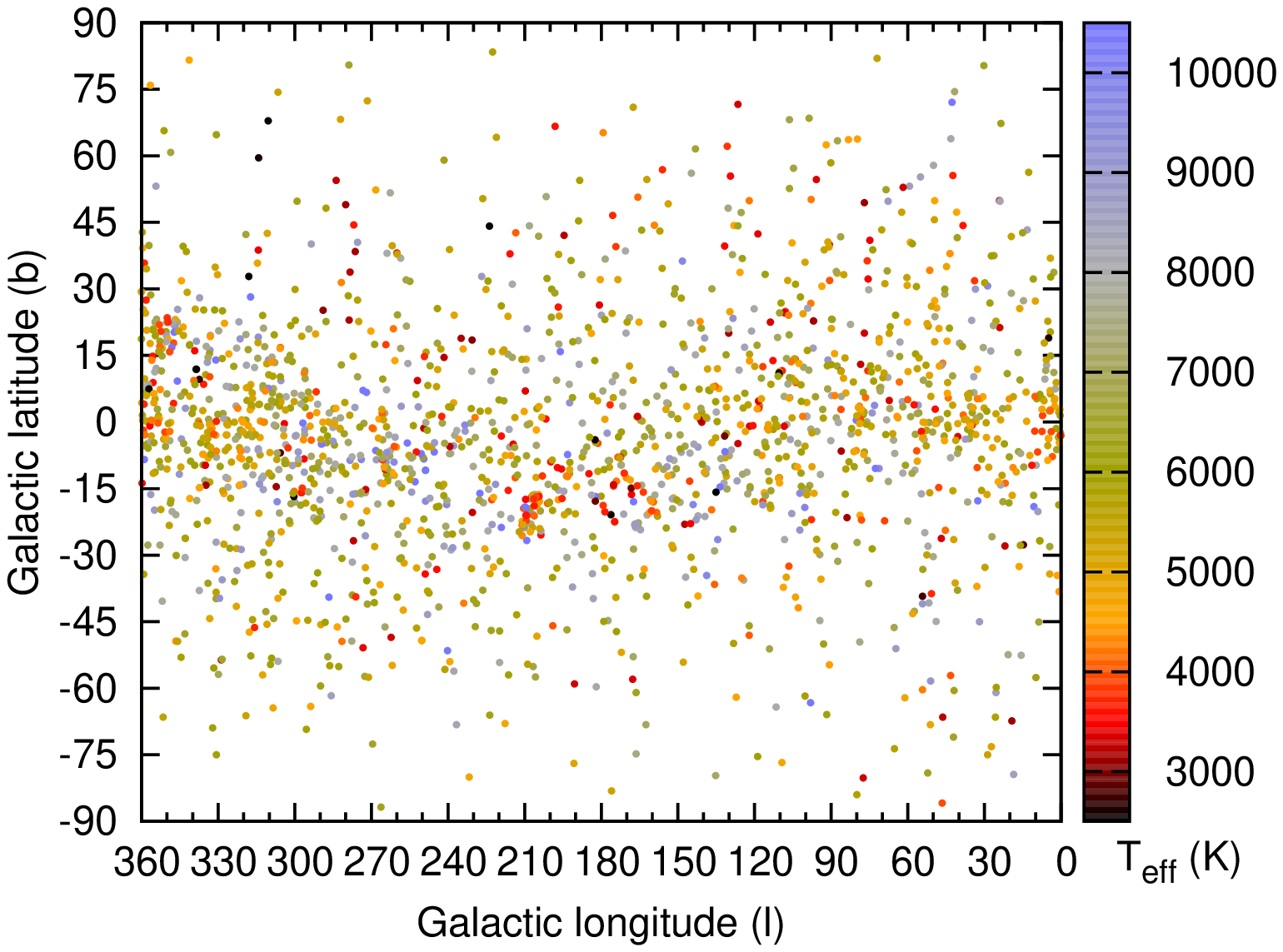}}
\caption{The spatial distribution of infrared excess for Tycho-2 and \emph{Hipparcos} catalogue stars. The bottom-left panel shows individual stars which are candidates for having infrared excess, colour-coded by temperature. The bottom-right panel shows the same plot for stars which are strong candidates (a score of more than three points).}
\label{XSMapFig}
\end{figure*}

\begin{figure}
\centerline{\includegraphics[height=0.47\textwidth,angle=-90]{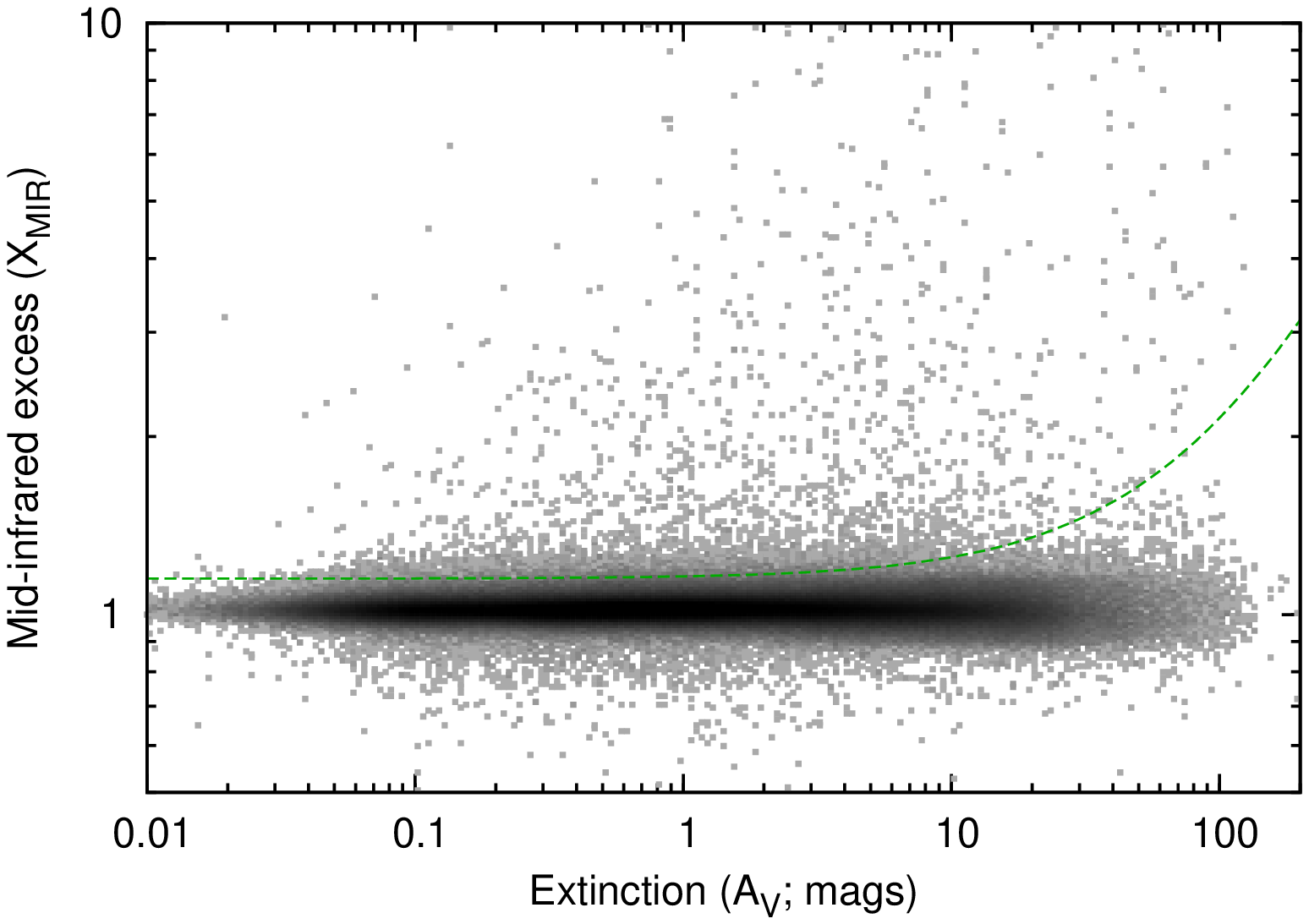}}
\caption{A Hess diagram showing the relationship between infrared excess ($X_{\rm MIR}$) and line-of-sight interstellar extinction ($A_{\rm V}$). The dashed line shows the cutoff used to determine candidacy for infrared excess.}
\label{XSEBVFig}
\end{figure}

Figure \ref{HRDXSFig} shows the H--R diagram of \emph{Hipparcos} and Tycho-2 stars, colour coded by infrared excess, while Figure \ref{XSMapFig} shows the distribution of sources across the sky. Sources are only included in these figures if $N_{\rm opt} + N_{\rm NIR} > 0$ (i.e.\ they have optical and infrared data), $N_{\rm MIR} > 1$ (i.e.\ they have more than one mid-IR datapoint), and if the parallax uncertainty $\delta \varpi / \varpi < 0.2$. Figure \ref{HRDXSFig} is also limited by $A_{\rm V} < 1.5$ mag.

The majority of these 600\,667 stars are well fit. The standard deviation of $X_{\rm MIR}$ is 0.185, however this is dominated by a small number of stars with large infrared excesses. If we take the central 68 per cent around the median of ${\rm Med}(X_{\rm MIR}) = 1.024$, the scatter is reduced to $\sigma_X = ^{+0.025}_{-0.027}$.

As a general trend, stars near the main sequence and lower giant branches tend to be well fit. Deviations become more apparent as one moves off these two sequences. Particularly noticeable are infrared deficits ($X_{\rm MIR} < 1$) among hot ($\gtrsim 8000$ K), luminous ($>$30 L$_\odot$) stars and cool ($\sim$3500--4500 K), luminous ($\sim$100--3000 L$_\odot$) stars.

Among hot stars, this deficit may be due to interstellar reddening. The opacity of interstellar dust has a steeper law than a blackbody's Wien tail in the optical, but a shallower law in the infrared. Reddened hot stars are modelled as cooler stars but, because of this opacity law, tend to be under-luminous in the optical and mid-infrared, and over-luminous in the near-infrared.

Reddened cool stars exhibit different qualities. Molecular opacity in the cool-star models has a strong temperature dependence. The opacity is mostly caused by TiO, and has a steeper wavelength dependence ($F \appropto \lambda^6$ over ($U-R$)) than interstellar extinction ($F \appropto \lambda^4$). Consequently, stars which are reddened by interstellar extinction and are fit by cooler stellar models tend to have a less sharp peak to their SEDs compared to stars intrinsically at that temperature, hence they tend to be over-luminous in the optical and mid-infrared, and under-luminous in the near-infrared, when compared to said models. This causes reddened giant branch stars to congregate around 3600--3700 K and exhibit mid-infrared excess (cf.\ the artefact at this temperature identified in Figure \ref{HRDFig}).

Instead, the mid-infrared deficit in giant stars seems to result from a combination of difficulties in accurately modelling the TiO absorption bands in the optical in cool stars, as well a under-estimation of flux in the $H$ band due to inaccurate modelling of the H$^{-}$ opacity peak (see the Appendix; Figure \ref{OETempFig}).

\subsection{Characteristics of infrared excess across the sky}
\label{DiscXSSkySect}

Small-scale variations of $X_{\rm MIR}$ can be seen across the sky (Figure \ref{XSMapFig}). Generally speaking, the regions of greatest deficit can be seen towards the Galactic Bulge and near the north Galactic pole (NGP). Towards the Bulge, crowding means that only optically brighter (typically hotter) stars are present in the \emph{Hipparcos}/Tycho-2 and \emph{Gaia} observations, which are then reddened. Towards the NGP, a large proportion of stars are old, cool stars. The previous section describes why these stars should be apparently under-luminous in the infrared.

Regions of moderate extinction, however, generally show a slight excess overall. This is most notable around the Musca interstellar clouds ($\alpha = 180^\circ$, $\delta = -80^\circ$), the $\rho$ Oph star-forming region ($\alpha = 250^\circ$, $\delta = -20^\circ$) and the Orion star-forming region ($\alpha = 90^\circ$, $\delta = 0^\circ$). Since these are regions of diffuse emission in the mid-IR, it is possible that background light affects some of the observations here at the level of a few percent. This background light may be from dust heated by the star in question (as seen in the Pleiades) or by other sources in the line of sight.

Stars with substantial infrared excess ($X_{\rm MIR} > 1.15$) also tend to occupy these regions, but are also more widely spread along the Galactic Plane.

\subsection{Defining criteria to flag infrared excess}

We define an infrared excess by two criteria. The first relates to the scatter calculated in Section \ref{DiscXSHRSect}. With 600\,667 stars, if our distribution of $X_{\rm MIR}$ was Gaussian in nature, we could expect a 5$\sigma$ threshold to remove random fluctuations in the data, hence sources with $X_{\rm MIR} > {\rm Med}(X_{\rm MIR}) + 5 \sigma_X = 1.15$ should be considered strong candidates for infrared excess. In practice, our distribution has a supra-Gaussian tail of badly fitting points on either side of the distribution, hence such a cutoff only removes the majority of badly fitting points.

The fraction of stars with $X_{\rm MIR} > 1.15$ is marginally larger towards lines of sight with higher extinction (Figure \ref{XSEBVFig}). Hence, we modify our criterion to remove stars with marginal infrared excess along high-extinction lines of sight. To qualify as a candidate for infrared excess, stars must have $X_{\rm MIR} > 1.15 + A_{\rm V} / 100$. This criterion is shown as the dashed line shown in Figure \ref{XSEBVFig}.

There are 1879 sources from the \emph{Hipparcos} sample which meet these criteria (0.18 per cent), and 2377 sources from the Tycho-2 sample (0.016 per cent). The much lower fraction from the Tycho-2 catalogue is caused primarily by the comparatively poor quality of the infrared photometry available for the Tycho-2 stars, due to their faintness and (in high-extinction lines of sight) the consequent difficulty of extracting them from the diffuse infrared background. Secondary effects include the less certain parallax measurements for the Tycho-2 sample and the propensity for bright (\emph{Hipparcos}) stars to display infrared excess (e.g.\ Herbig Ae/Be stars, Cepheids, giant branch stars). Improvements in the resolution and depth of the available infrared databases would substantially improve our ability to extract infrared excess.

We define these 4256 stars as having candidate infrared excess associated with them. We strongly advise users of this data to inspect the associated mid-infrared imagery of each object, and cross-check the relevant values of $Q$, $S_{\rm MIR}$ and $X^\prime_{\rm MIR}$, to help confirm or refute its presence.

\subsection{A catalogue of stars with infrared excess}
\label{DiscXSSubSect}

\begin{center}
\begin{table*}
\caption{Catalogue of stars with candidacy for hosting infrared excess. A portion of the online table is shown here, where table columns are numbered for clarity. The columns are described in full in the text, but can briefly be described as: (1) Tycho-2 or \emph{Hipparcos} identifier; (2--18) as Table \ref{TychoTable}; (19) mid-infrared excess; (20) mid-infrared excess, calculated with the point with the strongest excess removed; (21) uncalibrated significance of the excess; (22) {\sc simbad} primary name; (23) {\sc simbad} primary object type; (24) full list of {\sc simbad} object types; (25) {\sc simbad} spectral class; (26) points-based quality criterion. Complete tables are to be found at CDS.}
\label{XSTable}
\begin{tabular}{cccccccccccc}
    \hline \hline
\multicolumn{1}{c}{(1)} & \multicolumn{1}{c}{\nodata} & \multicolumn{1}{c}{(18)} & \multicolumn{1}{c}{(19)} & \multicolumn{1}{c}{(20)} & \multicolumn{1}{c}{(21)} & \multicolumn{1}{c}{(22)} & \multicolumn{1}{c}{(23)} & \multicolumn{1}{c}{(24)} & \multicolumn{1}{c}{(25)} & \multicolumn{1}{c}{(26)} \\
\multicolumn{1}{c}{Name} & \multicolumn{1}{c}{\nodata} & \multicolumn{1}{c}{$Q$} & \multicolumn{1}{c}{$X_{\rm MIR}$} & \multicolumn{1}{c}{$X_{\rm MIR}^\prime$} & \multicolumn{1}{c}{$S_{\rm MIR}$} & \multicolumn{1}{c}{{\sc simbad}} & \multicolumn{1}{c}{{\sc simbad}} & \multicolumn{1}{c}{{\sc simbad}} & \multicolumn{1}{c}{{\sc simbad}} & \multicolumn{1}{c}{Quality} \\
\multicolumn{1}{c}{\ } & \multicolumn{1}{c}{\nodata} & \multicolumn{1}{c}{\ } & \multicolumn{1}{c}{\ } & \multicolumn{1}{c}{\ } & \multicolumn{1}{c}{\ } & \multicolumn{1}{c}{Name} & \multicolumn{1}{c}{\texttt{otype}} & \multicolumn{1}{c}{\texttt{otypes}} & \multicolumn{1}{c}{Class} & \multicolumn{1}{c}{(points)}\\
    \hline
HIP 66          & \nodata & 0.107 & 1.211 & 1.278 & 3.245 & HD 224790 & * & *,IR & F2V & 4\\ 
HIP 75          & \nodata & 0.108 & 1.153 & 1.215 & 3.256 & HD 224821 & * & *,IR & K4III & 3\\ 
HIP 122         & \nodata & 1.324 & 1.909 & 2.461 & 1.257 & * tet Oct & * & *,IR & K3III & 5\\ 
\nodata         & \nodata &\nodata&\nodata&\nodata&\nodata&\nodata&\nodata&\nodata&\nodata\\
TYC 9529-1698-2 & \nodata & 0.522 & 1.940 & 2.092 & 2.855 & CPD-85 549B & * & **,*,IR & G5 & 6\\
\hline
\end{tabular}
\end{table*}
\end{center}

\begin{center}
\begin{table}
\caption{Summary of spectral types among stars with mid-infrared excess. The first count column gives all candidate sources; the second column gives sources with $>$3 points.}
\label{XSSpecTypeTable}
\begin{tabular}{cccl}
    \hline \hline
\multicolumn{1}{c}{Spectral} & \multicolumn{2}{c}{Count} & \multicolumn{1}{c}{Notes}\\
\multicolumn{1}{c}{Type} & \multicolumn{1}{c}{\ } & \multicolumn{1}{c}{\ } & \multicolumn{1}{c}{\ } \\
    \hline
O &  13 &  11 & \\
B & 565 & 475 & Including two DB stars \\
A & 549 & 355 & Including five DA stars \\
F & 382 & 186 & \\
G & 302 & 132 & \\
K & 410 & 133 & \\
M & 124 &  74 & \\
C &  15 &  11 & \\
S &   5 &   2 & \\
\hline
\end{tabular}
\end{table}
\end{center}

\begin{center}
\begin{table}
\caption{Summary of common {\sc simbad} object types among stars with mid-infrared excess. Objects may appear more than once in the list. Only those types with $\geq$3 entries are shown. Purely observational characteristics (e.g.\ infrared source) are excluded.}
\label{XSObjTypeTable}
\begin{tabular}{ccl}
    \hline \hline
\multicolumn{1}{c}{Object} & \multicolumn{1}{c}{Count} & \multicolumn{1}{c}{Notes}\\
\multicolumn{1}{c}{Type} & \multicolumn{1}{c}{\ } & \multicolumn{1}{c}{\ } \\
    \hline
\multicolumn{3}{c}{\it Young stellar types \& hot stars} \\
    \hline
Be* & 199 & Herbig Be star \\
Y*O &  38 & Young stellar object (YSO) \\
TT* &  37 & T Tauri star\\
Ae* &  30 & Herbig Ae star \\
Ae? &  30 & Candidate Ae star \\
pr* &  28 & Pre-main-sequence star \\
Y*? &   8 & Candidate YSO \\
HH  &   5 & Herbig--Haro object \\
bC* &   4 & $\beta$ Cephei variable \\
    \hline
\multicolumn{3}{c}{\it Evolved stellar types} \\
    \hline
C*  &  19 & Carbon star \\
Mi* &   8 & Mira variable \\
S*  &   6 & S-type star \\
AB* &   7 & AGB star \\
WD* &   5 & White dwarf \\
pA? &   7 & (Candidate) post-AGB star \\
    \hline
\multicolumn{3}{c}{\it Variable star types} \\
    \hline
V*  & 437 & Variable star \\
LP* &  56 & Long-period variable (LPV) \\
Ro* &  15 & Rotational variable stars \\
Or* &  25 & ``Orion type'' variable stars \\
dS* &  12 & $\delta$ Scu star \\
Pu* &  10 & Pulsating variable \\
a2* &   9 & Rotational ($\alpha_2$ CVn) variable \\
LP? &   9 & Candidate LPV \\
Ir* &   7 & Irregular variable \\
No* &   6 & Nova \\
BY* &   5 & Rotational (BY Dra) variable \\
V*? &   5 & Candidate variable \\
El* &   4 & Ellipsoidal variable \\
Ce* &   3 & Cepheid variable \\
RI* &   3 & Rapid, irregular variable \\
NL* &   3 & Nova-like star \\
Fl* &   3 & Flare star \\
    \hline
\multicolumn{3}{c}{\it Binary star types} \\
    \hline
**  & 425 & Binary star \\
SB* &  85 & Spectroscopic binary star \\
*i* &  20 & In multiple star system \\
Al* &  33 & Detatched (Algol) eclipsing binary \\
WU* &  17 & Contact binary (W UMa) stars \\
bL* &  13 & Semi-detached ($\beta$ Lyr) system \\
RS* &  12 & RS CVn close binary stars \\
EB* &  11 & Eclipsing binary stars \\
EB? &   8 & Candidate eclipsing binary \\
blu &   5 & Blue straggler \\
HXB &   3 & High-mass X-ray binary \\
    \hline
\multicolumn{3}{c}{\it Other types of object} \\
    \hline
Em* & 290 & Emission-line star \\
*iC &  72 & Star in cluster \\
*iN &  44 & Star in nebula \\
EmO &   7 & Emission object (ISM) \\
As* &   7 & Stellar associations \\
*iA &   7 & Star in association \\
Pe* &   3 & Peculiar stars \\
\hline
\end{tabular}
\end{table}
\end{center}

\begin{figure*}
\centerline{\includegraphics[height=0.97\textwidth,angle=-90]{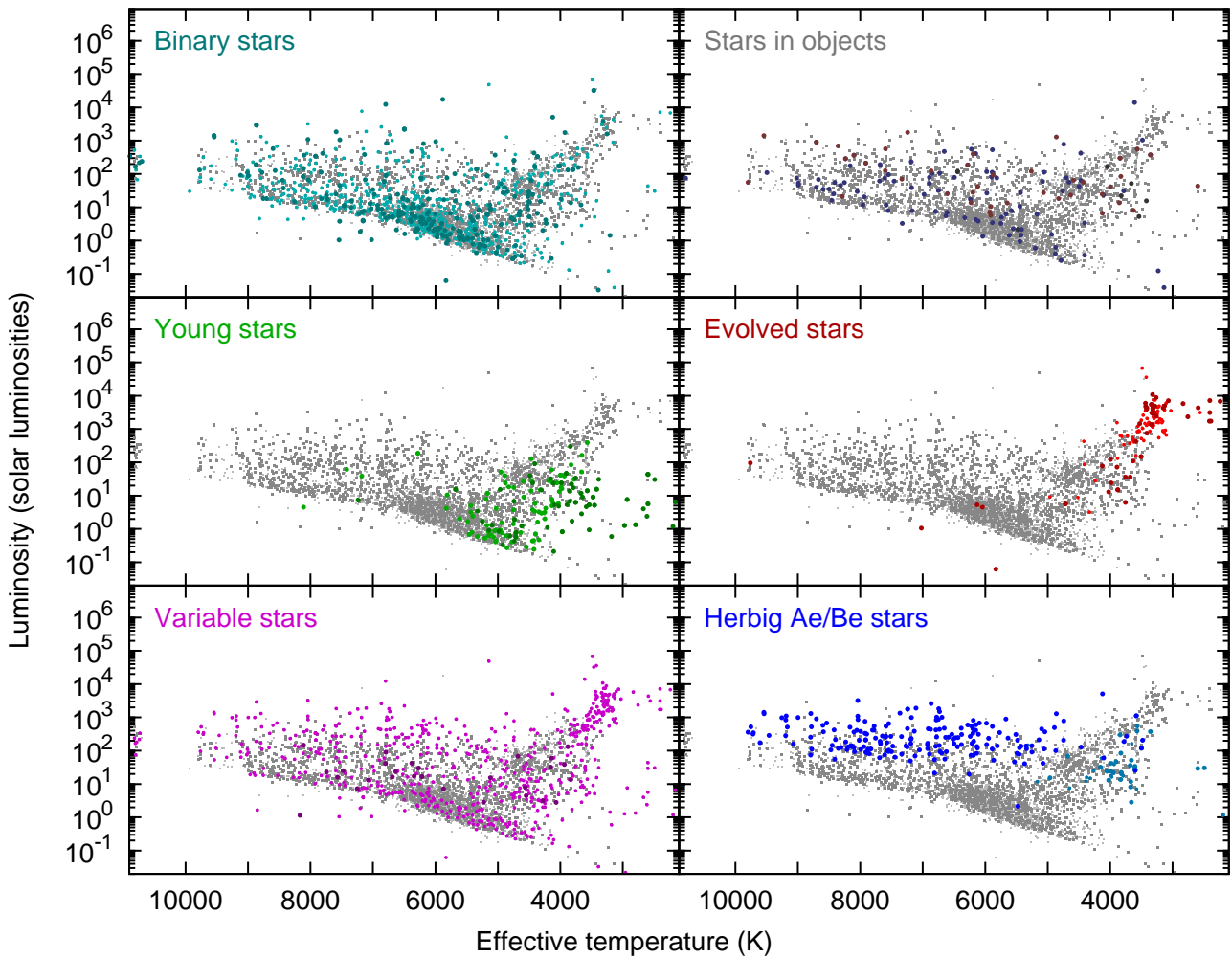}}
\caption{H--R diagrams, showing the locations of different classifications of stars. In each case, the light grey dots show all candidate stars in Table \ref{XSTable}, with the slightly darker grey dots showing stars with high confidence ($>$3 points). Binary stars with no further designator are shown as smaller points with lighter colour. Stars within objects are shaded red to denote in nebulae and blue to denote in clusters. Young testers are coloured lighter for pre-main-sequence stars and YSOs, and darker for T Tauri stars and Herbig--Haro objects. Evolved stars are coloured light for long-period variables, and dark if their designator provides further information (e.g.\ Mira variable, carbon star, etc.). Variable stars are shown in larger, darker points if they are known instability strip variables (e.g.\ Cepheids). Herbig Ae/Be stars are shown in cyan for Ae and blue for Be stars: smaller symbols denote questionable designations ({\sc simbad}'s Ae? and Be?).}
\label{XSOTypeFig}
\end{figure*}

\subsubsection{The catalogue and its contents}
\label{DiscXSSubCatSect}

Table \ref{XSTable} catalogues the objects defined as having infared excess. The {\sc simbad} spectral types are listed (Table \ref{XSSpecTypeTable}), as well as \texttt{otype} parameters\footnote{http://simbad.u-strasbg.fr/simbad/sim-display?data=otypes} (Table \ref{XSObjTypeTable}), providing a basic categorisation of each source. There are 95 entries which did not receive a {\sc simbad} match. The location of different categories of source on the H--R diagram is shown in Figure \ref{XSOTypeFig}.

The statistics in Tables \ref{XSSpecTypeTable} \& \ref{XSObjTypeTable} are not complete, and each list is not exhaustive. Of the 4161 sources with {\sc simbad} entries, 3049 are have a primary classification of `star'. Examination of individual records indicates that many of these are known objects of interest (e.g.\ emission-line stars, late-type giants, etc.) which have not yet been correctly designated as such by {\sc simbad}. Yet, may of these 3049 objects appear to be new candidates for hosting infrared excess.

The inhomegeneity of our input data quality means that the confidence on the detection of infrared excess varies. We therefore introduce a point-based quality criterion to judge the likelihood of excess being present. Points are awarded successively if $X_{\rm MIR} > 1.2 + A_{\rm V} / 80$, $X_{\rm MIR} > 1.3 + A_{\rm V} / 40$, and $X_{\rm MIR} > 1.5 + A_{\rm V} / 3.1$; if $X_{\rm MIR} > Q + 1$; if $X^\prime_{\rm MIR} > Q + 1$; or if $S_{\rm MIR} > 1$, giving a maximum possible six points. Examination of individual sources shows that, typically, more than three points are needed to show a high-quality detection of infrared excess: there are 1883 objects with more than three points, 1156 of which have either no {\sc simbad} classification, or a primary classification of `star'.

\subsubsection{Types of object with infrared excess}
\label{DiscXSSubTypeSect}

The statistics in Table \ref{XSObjTypeTable} show we detect a variety of stellar types that are expected to host infrared excess. These include Herbig Ae/Be stars, and a variety of young and pre-main-sequence stars, evolved (post-)AGB stars and stars experiencing third dredge-up (S-type stars and carbon stars; see, e.g., \citealt{KL14}), and a variety of variable stars which are known to exhibit dust. Also included are a variety of binary stars. Some of these are expected to host circumstellar or circumbinary material, and some are not. In many cases, the infrared excess may simply arise from problems caused by fitting two superimposed stellar SEDs with a single stellar atmosphere model.

There are a variety of other types of object which are not {\it a priori} expected to host infrared excess. These are stars in clusters, nebulae and stellar associations. Several of these stars are in regions of known nebulosity, such as the Pleiades and various parts of the Orion star-forming complex. It also includes stars in nearby clusters, but clearly not associated with them, such as HIP 81894. Other causes of infrared excess in such objects may be attributable to stellar blending (e.g.\ \citealt{MBvL+11}).

A number of objects are identified by {\sc simbad} as extra-galactic, but are unlikely to be so. These include TYC 273-677-1 and TYC 705-746-1, where \emph{Gaia} has measured parallaxes of 5.99 $\pm$ 0.95 mas and 2.42 $\pm$ 0.31 mas, respectively, and TYC 7415-696-1, which is the T Tauri object Hen 3-1722 \cite{Wray66,SW72,Henize76}.

\subsubsection{Properties of infrared-excess stars on the H--R diagram}
\label{DiscXSSubHRSect}

Figure \ref{XSOTypeFig} places various categories of infrared-excess stars in the H--R diagram. Stars with infrared excess at high confidence are typically found away from the main sequence and giant branch, mostly above the main sequence. Variable stars are found all over the H--R diagram, with no clear sign of the bounds of the instability strip. Likewise, binary stars are found in many locations, although they do not frequent the giant branch due to observational biases against their detection.

Stars associated with clusters or nebulosity scatter above the main sequence, suggesting source confusion or incorporation of background light into the SED may have occurred. In some cases, these may also be young stars that have yet to descend to the main sequence.

Young stars in the cool end of the H--R diagram tend to lie at varying distances above the main sequence. The majority of the T Tauri stars and Herbig--Haro objects lie in the Hayashi forbidden zone \citep{Hayashi61}, commensurate with their young age.

By contrast, evolved stars are logically found predominantly near the top of the giant branch. However, a large number of `evolved' stars are well down the giant branch ($<$200 L$_\odot$), and there are even some on the main sequence. Such objects include:
\begin{itemize}
\item The carbon star HIP 56551 (HD 100764), which may be an extrinsic carbon star.
\item HIP 91260 (CE Lyr), which is a Mira variable, but which suffers from contamination by a nearby star.
\item A number of post-AGB objects also fall into this category. They include the post-AGB star HM Aqr, and the candidate post-AGB stars / proto-planetary nebulae TYC 2858-542-1 (IRAS 02529+4350) and TYC 718-517-1 (HD 246299).The remainder appear to either be mis-classified Herbig Ae/Be stars or T Tauri stars: HIP 78092 (HD 142527), HIP 78943 (HD 144432), TYC 6679-305-1 (HD 143006) and TYC 6856-876-1 (HD 169142).
\end{itemize}

Finally, Herbig Ae/Be stars scatter to cooler temperatures than expected for their spectral classifications, as a result of the circumstellar material that surrounds them. Ae stars cluster around 4000 K and 2 L$_\odot$, while Be stars occupy a broader range, between 7000 and 10000 K, and 100 and 3000 L$_\odot$. Generally speaking, they lie well above the main sequence. Many of the undesignated objects in the same region of the H--R diagram may also be Be stars in their own right.


\subsection{Application to mass-losing stars on the giant branch}
\label{DiscObjSect}

\begin{figure}
\centerline{\includegraphics[height=0.47\textwidth,angle=-90]{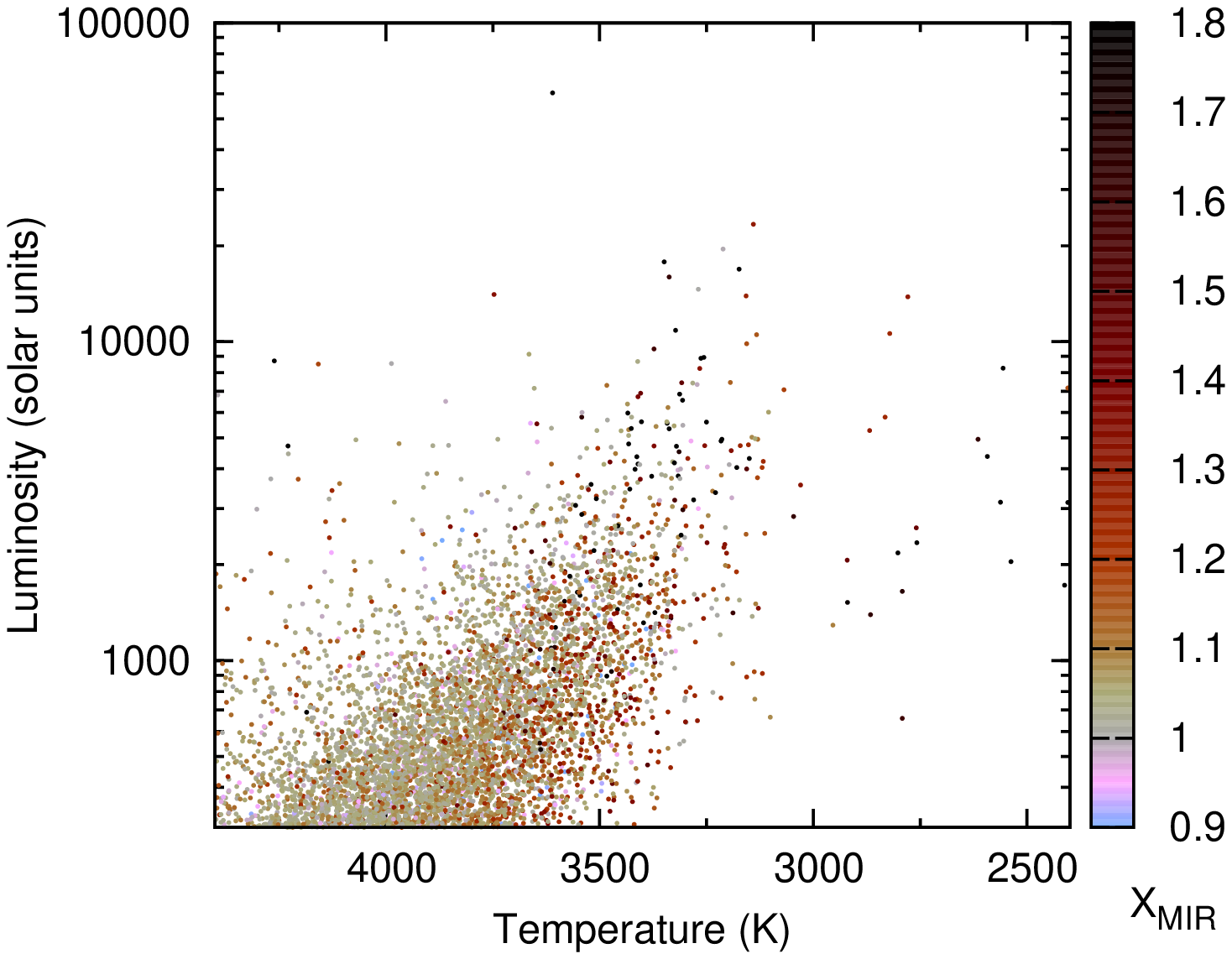}}
\centerline{\includegraphics[height=0.47\textwidth,angle=-90]{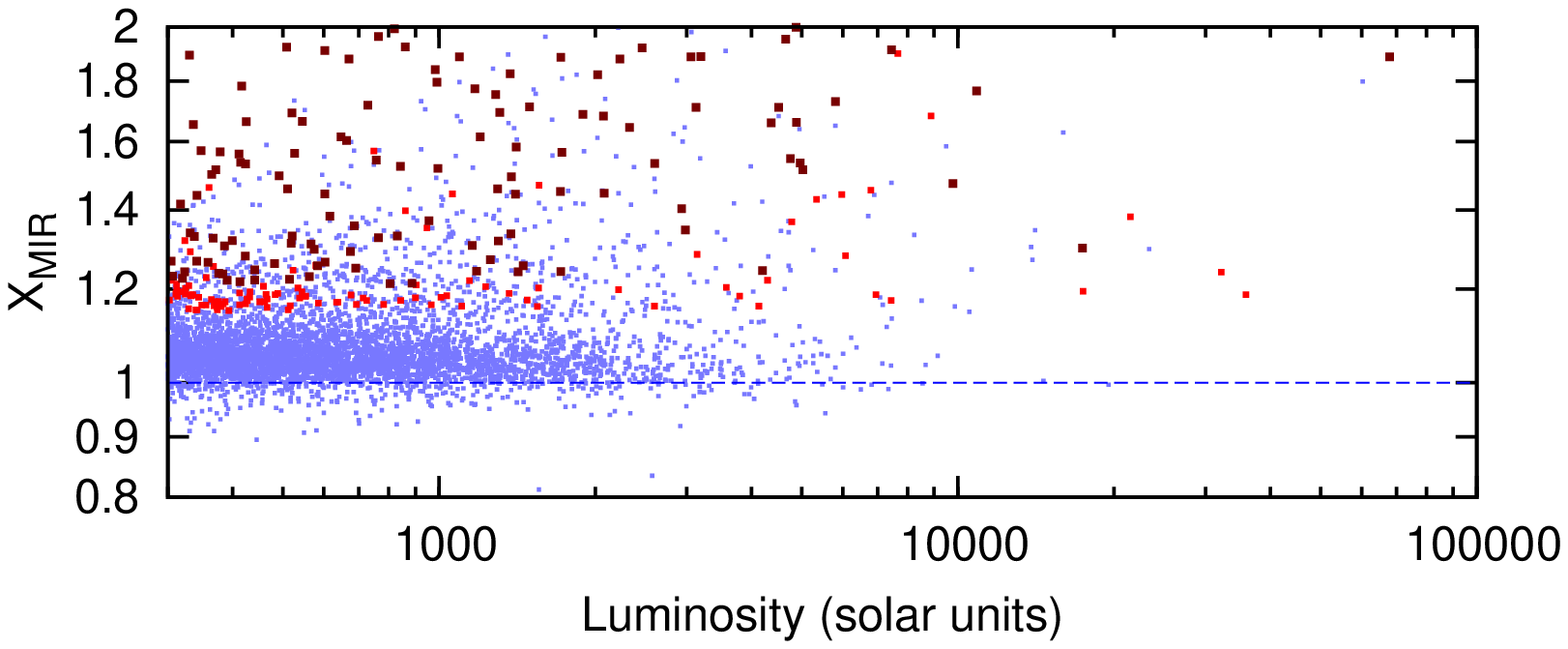}}
\centerline{\includegraphics[height=0.47\textwidth,angle=-90]{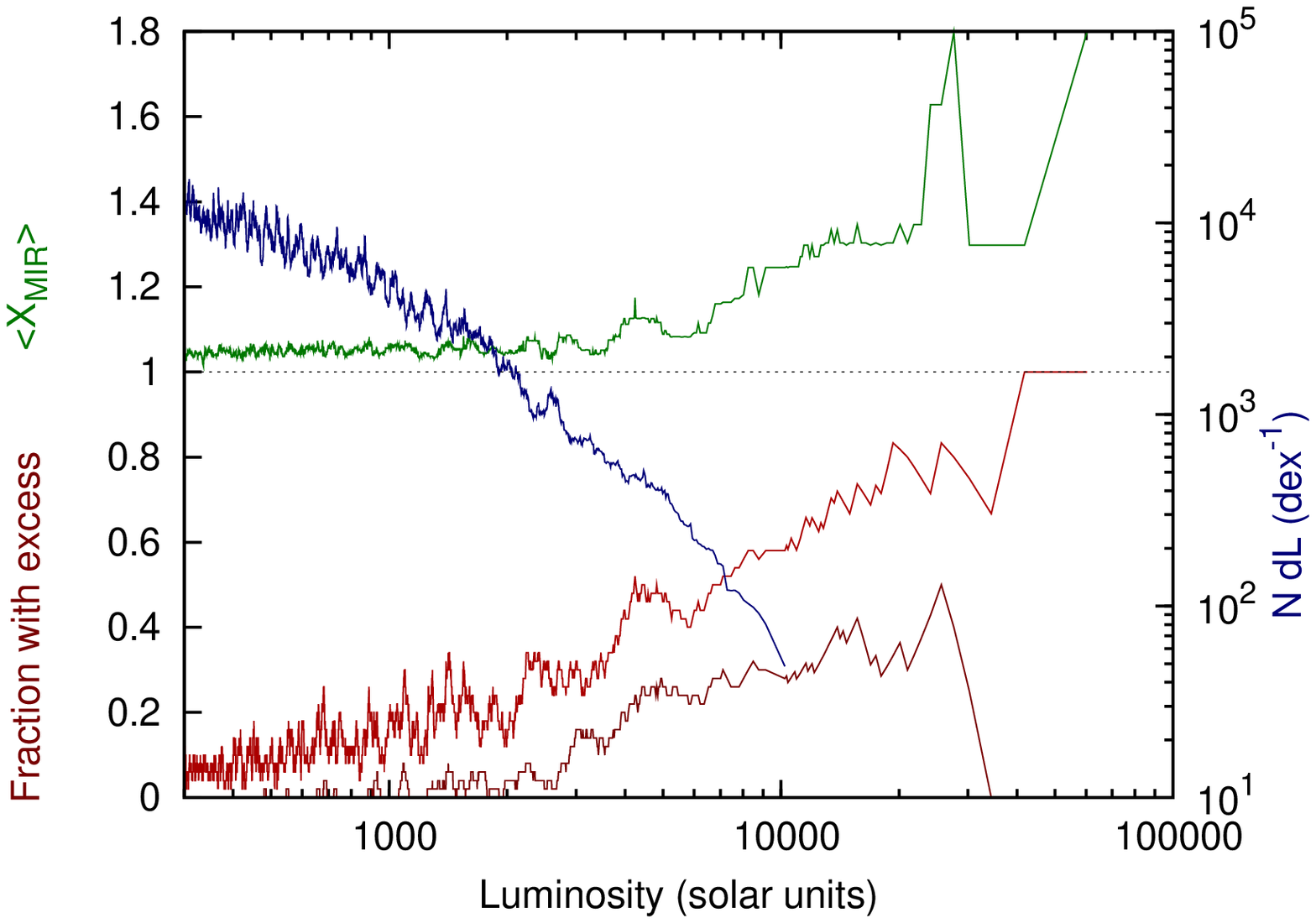}}
\caption{\emph{Top panel:} The upper giant branches, with infrared excess colour-coded as in the top panel of Figure \ref{XSMapFig}. Only stars with $A_{\rm V} < 1.5$ mag and $\delta \varpi / \varpi < 0.2$ are considered. \emph{Middle panel:} Mid-infrared excess of individual stars as a function of luminosity (light blue points). Red points show stars with candidate infrared excess; dark red points show those stars with a score of more than three points. \emph{Bottom panel:} The evolution of stellar properties with luminosity, smoothed by a running mean of 50 stars. From top to bottom (as viewed from the left-hand side of the plot), the lines represent: (1) The blue line shows the number of sources per dex in luminosity. This is shown on the right-hand (logarithmic) scale. (2) The green line shows the median $X_{\rm MIR}$ stars at that luminosity have. (3) The red line shows the fraction of stars meeting our infrared excess criterion ($X_{\rm MIR} > 1.15 + A_{\rm V} / 100$). (4) The darker red line shows the same plot for $X_{\rm MIR} > 1.5 + A_{\rm V} / 3.1$. The RGB tip lies between around 2400 L$_\odot$ to 2500 L$_\odot$, for most solar-metallicity stars. A dotted line is placed at unity to guide the eye.}
\label{GiantFig}
\end{figure}

A useful application of this research is into the minimum luminosity of dusty giant branch stars. This is one of the few places on the H--R diagram where dust production is expected to be confined to a specific region. Figure \ref{GiantFig} shows the upper giant branches of the H--R diagram. Below $\sim$300 L$_\odot$, source densities are affected by our temperature cutoff at 4400 K. Above $\sim$300 L$_\odot$, our parallax uncertainty criteria of $<$20 per cent limits us to nearby sources. This closely matches the bright limit of \emph{Gaia} DR1, so parallaxes of giant stars above 300 L$_\odot$ largely come from the \emph{Hipparcos} mission, and are within $\sim$1 kpc of Earth. At these distances, all stars will be easily detectable by either \emph{Hipparcos} or \emph{Gaia}, so the source density is not strongly influenced by the easier detectability of luminous stars.

The precise conditions needed to initiate dust production around evolved stars remain unknown. Circumstellar dust around RGB stars is thought to be very rare, though not necessarily impossible \citep[e.g.][]{Groenewegen12,MZB12,MZS+14,MZ16b}. In (metal-poor) globular clusters and the Magellanic Clouds, the onset appears between 700 and 1500 L$_\odot$ \citep{BMvL+09,MBvL+11,MvLS+11,BMS+15}. While the total mass-loss rate (at least in older stars) does not appear to be strongly linked to metallicity \citep{vLBM08,MZ15b}, the onset luminosity is likely to have some metallicity dependence \citep[e.g][]{MvLDB10}, as the dust column density should scale approximately with metallicity \citep{vL06,GVM+16}. However, the onset is hard to trace in solar metallicity populations due to distance or contamination. Based on the above studies, we can expect the onset of dust production to be traced by a gradual increase in the fraction of stars with infrared excess, starting at some point below the RGB tip.

The RGB tip is present in the upper panel at $\sim$2000 L$_\odot$. However, it is poorly defined due to a variety of observation and astrophysical factors: primarily the distance uncertainty, which can alter the luminosity by up to $\pm$40 per cent, and the stellar mass and metallicity, which can alter the luminosity by $\pm$20 per cent \citep[e.g.][]{MGB+08}. For intermediate-age and older populations, the evolutionary speed on the AGB is $\sim$3--5$\times$ faster than on the RGB, hence density declines above the RGB tip by a factor of $\sim$4--6. The inexact position of the RGB tip obfuscates its presence in the source density plot (the blue line in the bottom panel of Figure \ref{GiantFig}), but it can be seen as a small discontinuity between 2000 and 3000 L$_\odot$. Beyond the RGB tip, source density declines sharply as one ascends the upper AGB (the thermally pulsating, or TP-AGB).

The limitations in modelling these cool stellar atmospheres become problematic here, however. The median $X_{\rm MIR}$ ratio starts at just above unity near the middle of the giant branch and rises slowly (the offset being largely due to the poor $H^{-}$ modelling). Beyond the RGB tip, the median $X_{\rm MIR}$ rises more rapidly, until the value becomes stochastic among the most luminous AGB stars.

Simultaneously, the fraction of stars with identified excess rises slowly towards the RGB tip. However, the number of stars with clear-cut excess remains negligible until $\sim$890 L$_\odot$. Only a handful of giant stars with excess fall below this luminosity: l Vir, Z Peg, FW Vir, HD 68425, SU And (carbon star), RT Boo, AU Peg (W Vir variable), HM Aqr (post-AGB star), HD 100764 (carbon star), DY Boo and RU Crt. With the possible exceptions of l Vir (686 L$_\odot$), RU Crt (664 L$_\odot$) and HD 68425 (483 L$_\odot$), these objects all have very strong infrared excess, are not well modelled by a simple stellar photosphere, and do not fall on the giant branch in the H--R diagram. It is likely that the luminosity has been under-estimated for these stars. Circumstellar material has been detected from RU Crt (McDonald et al., in prep.), identifying it as the lowest luminosity giant where a dusty outflow has been convincingly detected.

As one progresses above 890 L$_\odot$, there comes a steady list of sources with infrared excess. The fraction of sources is fairly low at first, but increases significantly at the RGB tip (Figure \ref{GiantFig}, bottom panel). The luminosity function of sources with strong infrared excess does not change appreciably across the RGB tip, arguing that few (if any) RGB stars exhibit circumstellar dust. All the giant stars which have infrared excess and are near the RGB tip are therefore expected to be AGB stars. The fraction of stars with infrared excess, and the amount of infrared excess they have, both increase with luminosity as stars ascend the AGB.


\section{Conclusions}
\label{ConcSect}

In this paper, we have photometrically matched numerous public databases of stellar photometry against parallactic measurements of stellar distances from the \emph{Gaia} satellite's first data release. Modelling of the resulting SEDs have allowed us to derive the temperature and luminosity for 1\,583\,066 unique objects, placing them on the H--R diagram. We report on the goodness-of-fit of each best-fit model, and quantify the presence of infrared excess around each star.

We list 4256 stars which are candidates for infrared excess, of which 1883 are qualified as having strong evidence of infrared excess. These objects have been categorised by their literature classifications. A large number of previously identified binary, variable and emission-line stars are recovered, along with a substantial number of potentially new detections.

We briefly explore some of the facets of this dataset:
\begin{itemize}
\item We identify that the vast majority of the \emph{Gaia} DR1 dataset exhibits relatively little extinction, although a small but significant number of stars (mainly giant stars) are still considerably affected.
\item We explore dust production among nearby giant stars, confirming that little or no dust condensation takes place around RGB stars, but becomes prevalent in AGB stars at an evolution point close to the RGB tip.
\item We explore populations at different Galactic scale heights, identifying that stars with ages $<$3 Gyr have a strong tendency to be located within $\sim$200 pc of the Galactic plane, and that the metallicity of nearby stars remains close to the solar value until one exceeds $\sim$600 pc from the plane.
\item We identify hot stars within a few hundred parsecs of the Sun, and use these to map out sites of recent star formation in the solar neighbourhood. Dust clouds and hot stars are presented in three dimensions and basic inferences drawn on their relation to the Gould Belt.
\end{itemize}
Our closing recommendations for repeating this study on a larger data set, following future \emph{Gaia} data releases, are presented in Appendix \ref{AppendixGaia} (online-only).


\section*{Acknowledgements}

The authors acknowledge support from the UK Science and Technology Facility Council under grant ST/L000768/1. This paper could not have been possible without data from a large of surveys and facilities. Their standard requested acknowledgements are listed below:
\begin{itemize}
\item This research made use of the cross-match service provided by CDS, Strasbourg.
\item This research has made use of the SIMBAD database, operated at CDS, Strasbourg, France. 
\item This work has made use of data from the European Space Agency (ESA) mission {\it Gaia} ({http://www.cosmos.esa.int/gaia}), processed by the {\it Gaia} Data Processing and Analysis Consortium (DPAC, {http://www.cosmos.esa.int/web/gaia/dpac/consortium}). Funding for the DPAC has been provided by national institutions, in particular the institutions participating in the {\it Gaia} Multilateral Agreement.
\item Funding for the Sloan Digital Sky Survey IV has been provided by the Alfred P. Sloan Foundation, the U.S. Department of Energy Office of Science, and the Participating Institutions. SDSS acknowledges support and resources from the Center for High-Performance Computing at the University of Utah. The SDSS web site is www.sdss.org.

SDSS is managed by the Astrophysical Research Consortium for the Participating Institutions of the SDSS Collaboration including the Brazilian Participation Group, the Carnegie Institution for Science, Carnegie Mellon University, the Chilean Participation Group, the French Participation Group, Harvard-Smithsonian Center for Astrophysics, Instituto de Astrofísica de Canarias, The Johns Hopkins University, Kavli Institute for the Physics and Mathematics of the Universe (IPMU) / University of Tokyo, Lawrence Berkeley National Laboratory, Leibniz Institut f\"ur Astrophysik Potsdam (AIP), Max-Planck-Institut f\"ur Astronomie (MPIA Heidelberg), Max-Planck-Institut für Astrophysik (MPA Garching), Max-Planck-Institut f\"ur Extraterrestrische Physik (MPE), National Astronomical Observatories of China, New Mexico State University, New York University, University of Notre Dame, Observatório Nacional / MCTI, The Ohio State University, Pennsylvania State University, Shanghai Astronomical Observatory, United Kingdom Participation Group, Universidad Nacional Aut\'onoma de M\'exico, University of Arizona, University of Colorado Boulder, University of Oxford, University of Portsmouth, University of Utah, University of Virginia, University of Washington, University of Wisconsin, Vanderbilt University, and Yale University.
\item The DENIS project has been partly funded by the SCIENCE and the HCM plans of the European Commission under grants CT920791 and CT940627. It is supported by INSU, MEN and CNRS in France, by the State of Baden-W\"urttemberg in Germany, by DGICYT in Spain, by CNR in Italy, by FFwFBWF in Austria, by FAPESP in Brazil, by OTKA grants F-4239 and F-013990 in Hungary, and by the ESO C\&EE grant A-04-046.

Jean Claude Renault from IAP was the Project manager.  Observations were carried out thanks to the contribution of numerous students and young scientists from all involved institutes, under the supervision of  P. Fouqu\'e,  survey astronomer resident in Chile.  
\item Based on data obtained as part of the INT Photometric H-Alpha Survey of the Northern Galactic Plane.
\item This work is based in part on data obtained as part of the UKIRT Infrared Deep Sky Survey.
\item This publication makes use of data products from the Two Micron All Sky Survey, which is a joint project of the University of Massachusetts and the Infrared Processing and Analysis Center/California Institute of Technology, funded by the National Aeronautics and Space Administration and the National Science Foundation.
\item Based on observations with \emph{AKARI}, a JAXA project with the participation of ESA.
\item This publication makes use of data products from the Wide-field Infrared Survey Explorer, which is a joint project of the University of California, Los Angeles, and the Jet Propulsion Laboratory/California Institute of Technology, funded by the National Aeronautics and Space Administration.
\item The \emph{InfraRed Astronomical Satellite} was developed and operated by the Netherlands Agency for Aerospace Programmes (NIVR), the U.S.\ National Aeronautics and Space Administration (NASA), and the U.K.\ Science and Engineering Research Council (SERC).
\item This research used the \emph{DIRBE} Point Source Photometry Research Tool, a service provided by the Legacy Archive for Microwave Background Data at NASA's Goddard Space Flight Center.
\item This research made use of data products from the \emph{Midcourse Space Experiment}. Processing of the data was funded by the Ballistic Missile Defense Organization with additional support from NASA Office of Space Science. This research has also made use of the NASA/IPAC Infrared Science Archive, which is operated by the Jet Propulsion Laboratory, California Institute of Technology, under contract with the National Aeronautics and Space Administration.
\item Based on observations obtained with Planck (http://www.esa.int/Planck), an ESA science mission with instruments and contributions directly funded by ESA Member States, NASA, and Canada.
\end{itemize}
The authors also wish to thank the anonymous referee for their careful scrutiny of the paper and the enlightening comments raised during the ensuing discussion.


\newpage
\newpage

\appendix
\section{Data flagging in Tycho-2 data}
\label{Appendix}




This Appendix describes the process for removing bad data from the Tycho-2 data. Data reduction took place in a series of `runs'. During each run, a portion of the data were fed through the SED-fitting routine and the output inspected using a number of metrics for obvious signs of bad data. The primary criterion used was the ratio of observed to modelled flux ($F_{\rm o} / F_{\rm m}$).

This ratio can be plotted against a number of different input and output parameters to identify the presence of any bad data and determine its origin. The primary comparisons are observed flux, stellar luminosity, stellar effective temperature, and line-of-sight interstellar reddening. In all cases, accurately modelled stars should have $F_{\rm o} / F_{\rm m}$ close to unity. The final versions of each of these plots are shown at the end of this Appendix.

Within each band, deviations from unity which are correlated with observed flux are useful in identifying bad data in the input catalogues. Examples of this are stars scattered to spuriously high fluxes when they are close to the detection limit, or stars with unphysically low fluxes, which may be experiencing saturation problems. Deviations which are correlated with luminosity identify problems arising from the accuracy of the model atmospheres in certain regimes (e.g.\ pulsating, luminous giants out of thermodynamic equilibrium). Deviations which are correlated with temperature are useful in determining the effects of the model atmospheres in other regimes (e.g.\ very cool stars with high molecular opacity), inaccuracies in filter transmission curves and the effects of interstellar reddening. Deviations which are correlated with interstellar reddening are useful identifying how the SED-fitting process behaves under such conditions.

\subsection{Run 1: identification of strong saturation and poor detections in the initial catalogue}

\begin{figure*}
\centerline{\includegraphics[height=0.97\textwidth,angle=-90]{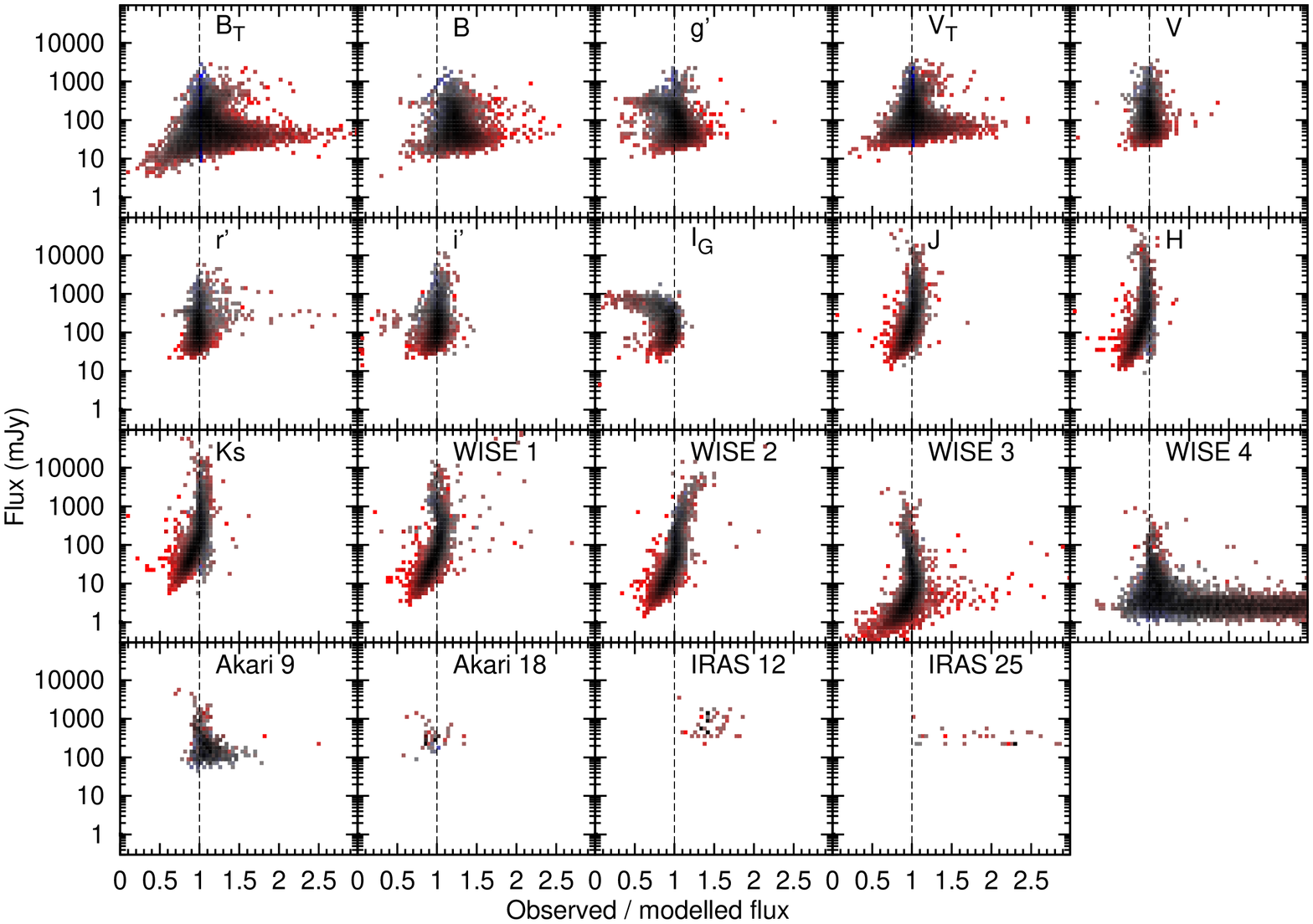}}
\caption{Run 1. The ratio of observed flux to that of the best-fit model atmosphere as a function of input flux in the indicated photometric band, binned in both dimensions for clarity. A line drawn at unity shows the target fit. Point colours are as in Figure \ref{HRDFig}. One per cent of the sample was analysed for this run.}
\label{OEFluxRun1Fig}
\end{figure*}

Figure \ref{OEFluxRun1Fig} shows the results of a preliminary fitting analysis, during which every 100th star from the matched Tycho--\emph{Gaia} set was modelled. This `selection by number' ensures a representative distribution of stars across the sample and across the sky. 

From Figure \ref{OEFluxRun1Fig} it is clear that there are some substantial systematic deviations from unity:
\begin{itemize}
\item At bright magnitudes, the DENIS $I$-band data suffers from significant saturation problems. Stars were restricted to magnitudes of $I > 9.7$ mag.
\item The IPHAS photometry also suffers from saturation problems. Stars were restricted to magnitudes of $r^\prime > 11.5$ mag and $i^\prime > 11.5$ mag.
\item At faint magnitudes, the \emph{IRAS} data suffers from spurious matches to objects near the noise limits. A limit was placed restricting \emph{IRAS} [25] $>$ 360 mJy.
\item The \emph{AllWISE} data suffers the same issue. A limit was placed restricting \emph{WISE} [22] $<$ 6.5 mag.
\item The \emph{AllWISE} data also suffers from issues near the saturation point. While this has improved markedly since early \emph{WISE} releases (cf.\ \citealt{MZB12}), this is still an issue for some stars. A limit was placed restricting \emph{WISE} [4.6] $>$ 6.0 mag.
\end{itemize}


\subsection{Run 2: removal of poor cross-correlations across catalogues}

\begin{figure*}
\centerline{\includegraphics[height=0.97\textwidth,angle=-90]{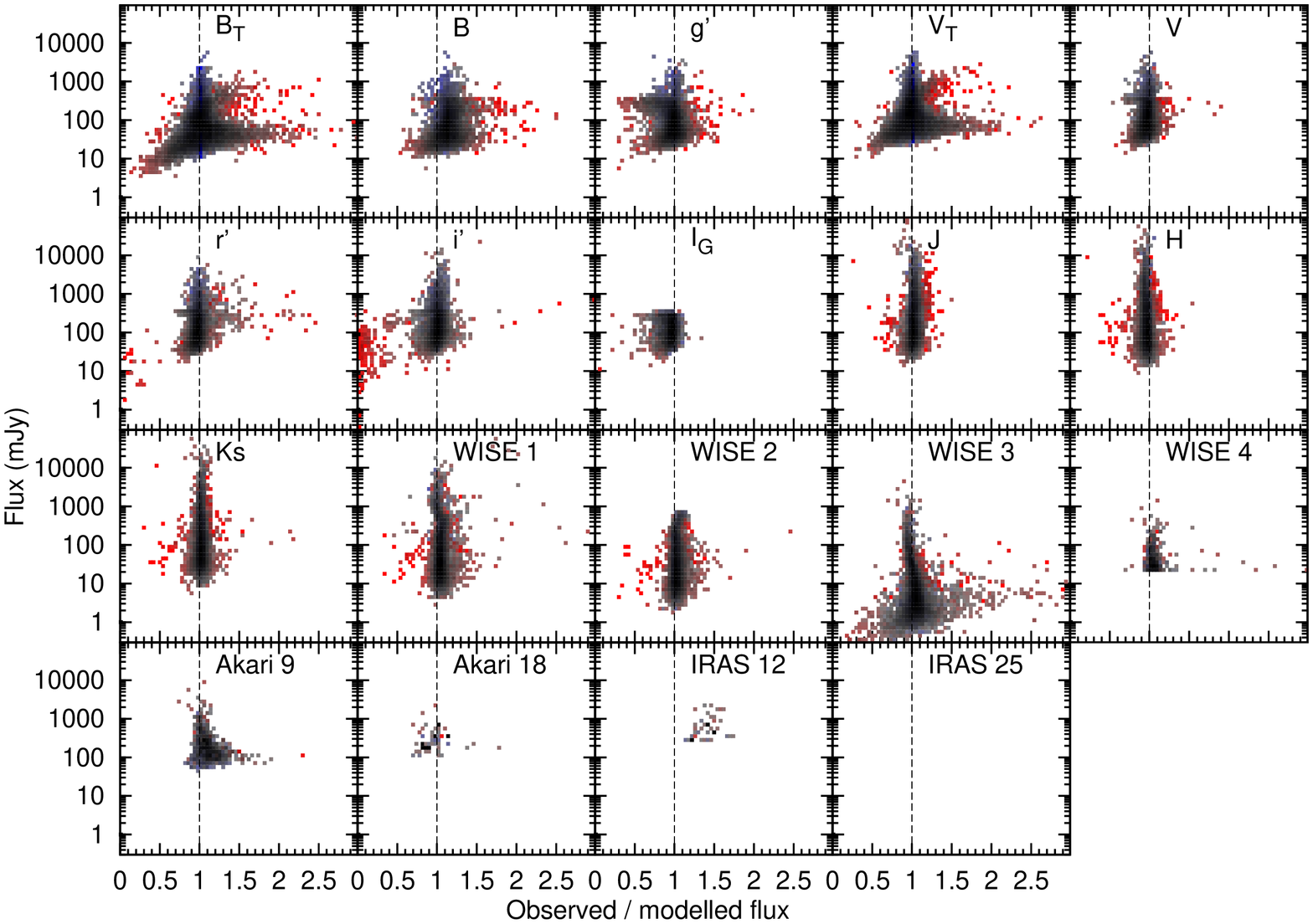}}
\caption{Run 2. See Figure \ref{OEFluxRun1Fig} for description.}
\label{OEFluxRun2Fig}
\end{figure*}

Figure \ref{OEFluxRun2Fig} shows the flux ratios for this run. There are still clear problems in a number of bands. These are caused by two factors: saturation problems in APASS, and poor flux measurements around stars which are saturated in SDSS. In the latter case, it is clear that significant issues affected a small fraction of the SDSS photometry we imported into our database. 

TYC 5281-1870-1 is an example of this. It is a nondescript, 11th magnitude star, recorded as $B_T = 11.718$ mag and $V_T = 10.974$ mag by Tycho-2 in 1980 and $J,\ H,\ K_{\rm s}$ = 9.764, 9.456, 9.364 mag by 2MASS (1998). APASS records (in 2013): $g',\ r',\ i'$ = 11.257, 10.788, 10.664 mag. However, corresponding magnitudes from SDSS DR7 (epoch 2000) are $g',\ r',\ i'$ = 14.231, 10.848, 10.699 mag, making the $g'$ and $i'$ magnitudes each discrepant from both Tycho-2 and APASS by $\sim$8 mag. These come from object 587727178999595066, which is a child object of 587727178999595063. In this case, a saturated star has been classified as a galaxy and split among a number of child objects.

Since the Vizier XMatch service does not incorporate the flagging data for SDSS sources, there is currently no trivial way to remove this photometry from the cross-matched source list\footnote{We thank the CDS for acknowledging and resolving this issue during preparation of this manuscript.}. Instead, the following manual cuts to remove the photometry have been implemented for the SDSS photometry:
\begin{itemize}
\item $g'$ is removed if $B_{\rm T}-g > B_{\rm T}-V_{\rm T} - 2.2$ and $B_{\rm T}-g < 0$ and $B_{\rm T} - V_{\rm T} > -1$;
\item $r'$ is removed if $V_{\rm T}-r > V_{\rm T}-J - 1.5$ and $V_{\rm T}-r < -0.9$ and $V_{\rm T} - J > -1$;
\item $i'$ is removed if $V_{\rm T}-i > V_{\rm T}-J - 1.5$ and $V_{\rm T}-r < -0.5$ and $V_{\rm T} - J > -1$;
\item $z'$ is removed if $V_{\rm T}-z > V_{\rm T}-J - 2.8$ and $V_{\rm T}-r < 0$ and $V_{\rm T} - J > -1$;
\item $g'$ is removed if $g - J > (V - J) / 0.8 + 0.7$;
\item $r'$ is removed if $r - J > (V - J) / 1.2 + 0.7$;
\item $i'$ is removed if $i - J > (V - J) / 2.0 + 0.7$;
\item $g'$ is removed if $g - J > (V_{\rm T} - J) / 0.8 + 1.3$;
\item $r'$ is removed if $r - J > (V_{\rm T} - J) / 1.2 + 1.0$;
\item $i'$ is removed if $i - J > (V_{\rm T} - J) / 2.0 + 1.0$.
\end{itemize}
These cuts have been designed to remove the vast majority of suspect photometry, while avoiding sources which have correct photometry but where the source is not well-fit by the stellar models (e.g.\ due to strong interstellar reddening).

For the APASS photometry, we introduce the following cuts to reduce saturation issues:
\begin{itemize}
\item $g'$ is removed if $g - V > (B - V) / 1.8 + 0.5$;
\item $g'$ is removed if $g - V_{\rm T} > (B_{\rm T} - V_{\rm T}) / 2.2 + 0.7$;
\item $i'$ is removed if $i - J > (V - J) / 2.0 + 0.7$.
\end{itemize}


\subsection{Run 3: more saturation flagging}

\begin{figure*}
\centerline{\includegraphics[height=0.97\textwidth,angle=-90]{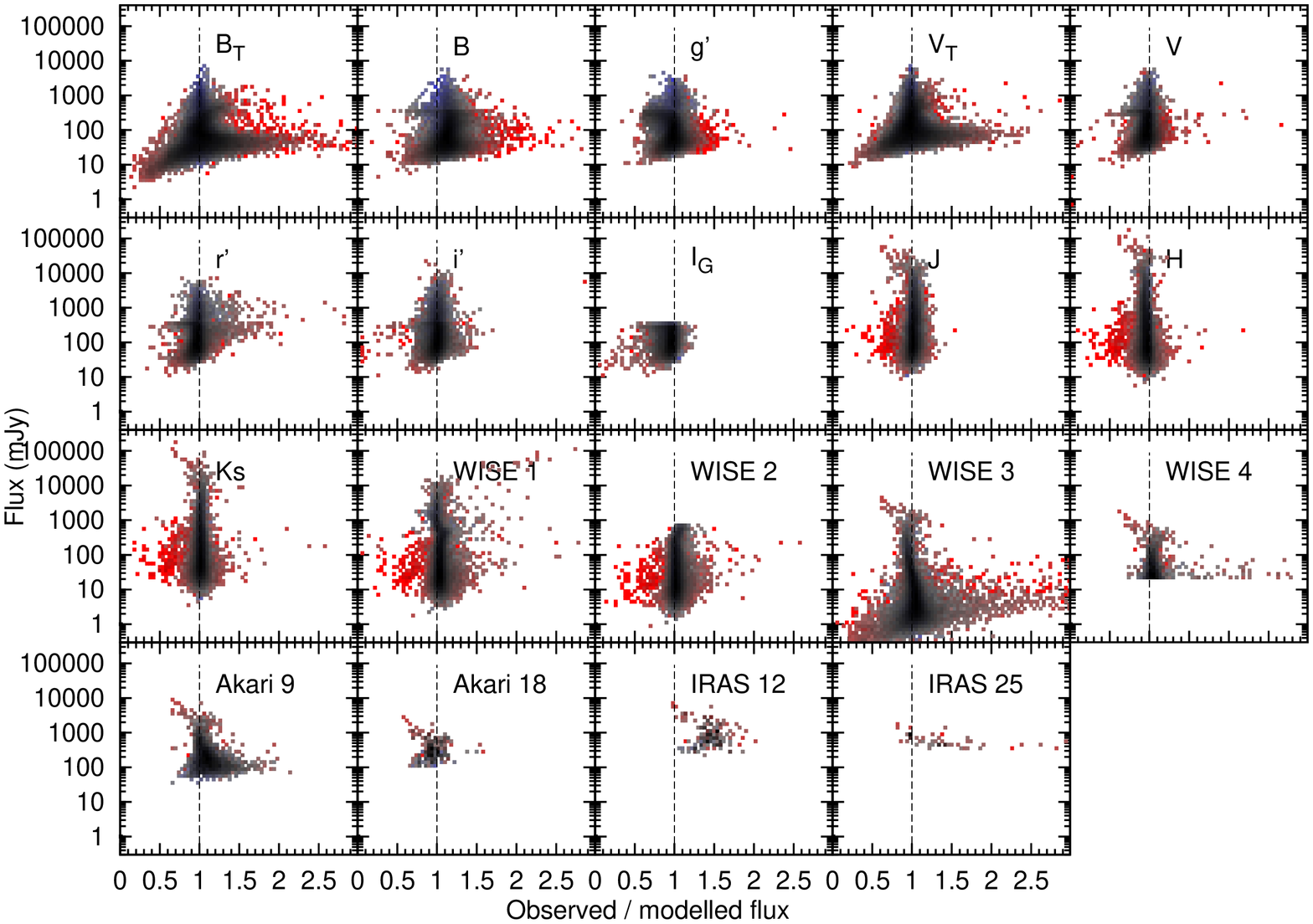}}
\caption{Run 3. See Figure \ref{OEFluxRun1Fig} for description. One 40th of the sample was computed in this run.}
\label{OEFluxRun3Fig}
\end{figure*}

\begin{figure*}
\centerline{\includegraphics[height=0.97\textwidth,angle=-90]{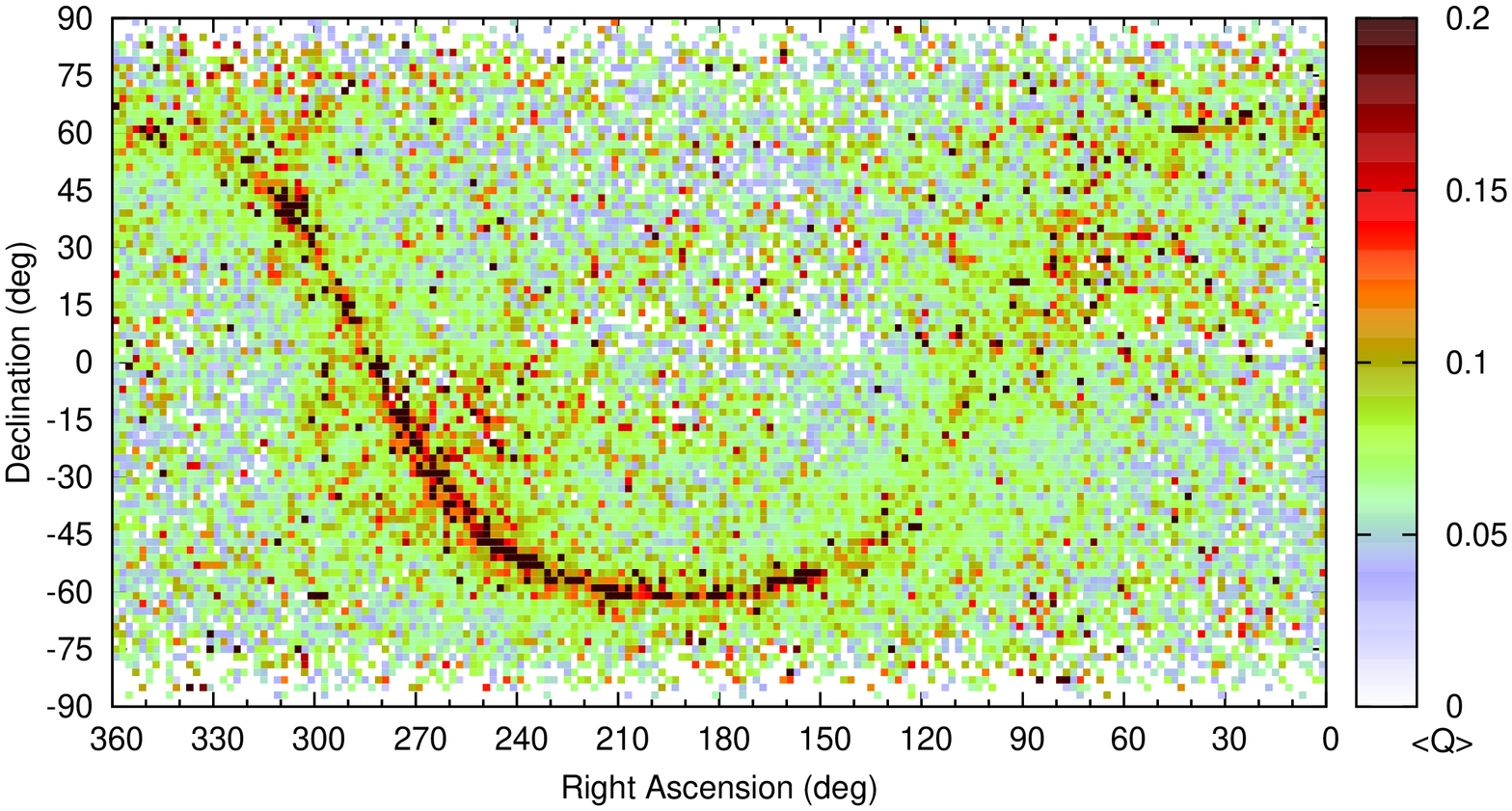}}
\caption{Run 3. The the average fit quality ($Q$) in each sky pixel is shown on the colour scale. The effects caused by stellar blending and extinction within a few degrees of the Galactic Plane are clear. Problems on a subset of sources can be seen in the northern hemisphere, away from the Plane. The distribution of these sources matches the footprint of the SDSS survey.}
\label{OESkyRun3Fig}
\end{figure*}

During this run a larger number of stars were fitted (every 40th star, or 54\,725 in total) to identify rarer effects in the data. Figure \ref{OEFluxRun3Fig} shows the same ratio of observed to model flux as previously, while Figure \ref{OESkyRun3Fig} shows the goodness-of-fit metric, $Q$, averaged over the entire sky in $1^\circ \times 1^\circ$ regions.

The fraction of poorly fitting data in each band has decreased, although there are still some significant effects. Most of the poorly fitting sources are located along the Galactic Plane. Two effects become important here: stellar blending and interstellar extinction. In dense environments, sources may appear as a single entry in low-resolution catalogues (e.g.\ \emph{IRAS}) but multiple entries in high-resolution catalogues (e.g.\ SDSS). This can lead to very poor fitting of the SED and a scattering of points. In cases of nearby stars in the Plane, this is compounded by stars' proper motions. Meanwhile, interstellar extinction has a progressive effect on the optical SED, but has a different wavelength dependence than the stellar atmosphere models. Extinction typically leads to the pattern of over-estimated flux in blue filters and under-estimated flux in near-IR filters that can account for most of the scatter of red points in Figure \ref{OEFluxRun3Fig} (see further explanation in Section \ref{DiscXSHRSect}).

Stars with the brightest infrared fluxes have poorly-fitted data in these bands. This is a combination of saturation issues in the \emph{WISE} data and sensitivity limits in the \emph{IRAS} data.

To combat all these effects, we adopt the following cuts to the APASS photometry:
\begin{itemize}
\item $B$ is removed of $9.7 < B < 10.7$ and $B_{\rm T} - B < -0.3$;
\item $V$ is removed of $9.7 < V < 10.7$ and $B_{\rm T} - V < -0.5$;
\item $V$ is removed of $9.7 < V < 10.7$ and $B_{\rm T} - V < -0.25$ and $B_{\rm T} - V_{\rm T} < 1.5$;
\item $g'$ is removed of $9.7 < g' < 10.7$ and $B_{\rm T} - g < (B_{\rm T} - V_{\rm T}) / 1.8 - 0.3$;
\end{itemize}
and the following cuts to the \emph{WISE} photometry:
\begin{itemize}
\item $W_1$ is deleted if $-10 < K - W_1 < -1$ and $-10 < K - [12] < -10$;
\item $W_1$ is deleted if $W_1 < 2$;
\item $W_3$ is deleted if $W_3 - [12] > -4$ and $[12] > 4.8$;
\item $W_4$ is deleted if $W_4 - [25] > -6$ and $[25] > 5.0$;
\item $W_3$ is deleted if $[12] \leq 4.8$;
\item $W_4$ is deleted if $[25] \leq 5.0$.
\end{itemize}
The optical cuts are chosen to exclude the range that are not covered by any stellar model atmospheres. A significant scatter (roughly $\sim$0.3 to $\sim$0.5 mag) beyond this range is allowed to account for the effects of normal photometric errors. The nature of these cuts is such that they tend to avoid wrongly excluding photometry affected by interstellar reddening.


\subsection{Run 4: a first complete run}

\begin{figure*}
\centerline{\includegraphics[height=0.97\textwidth,angle=-90]{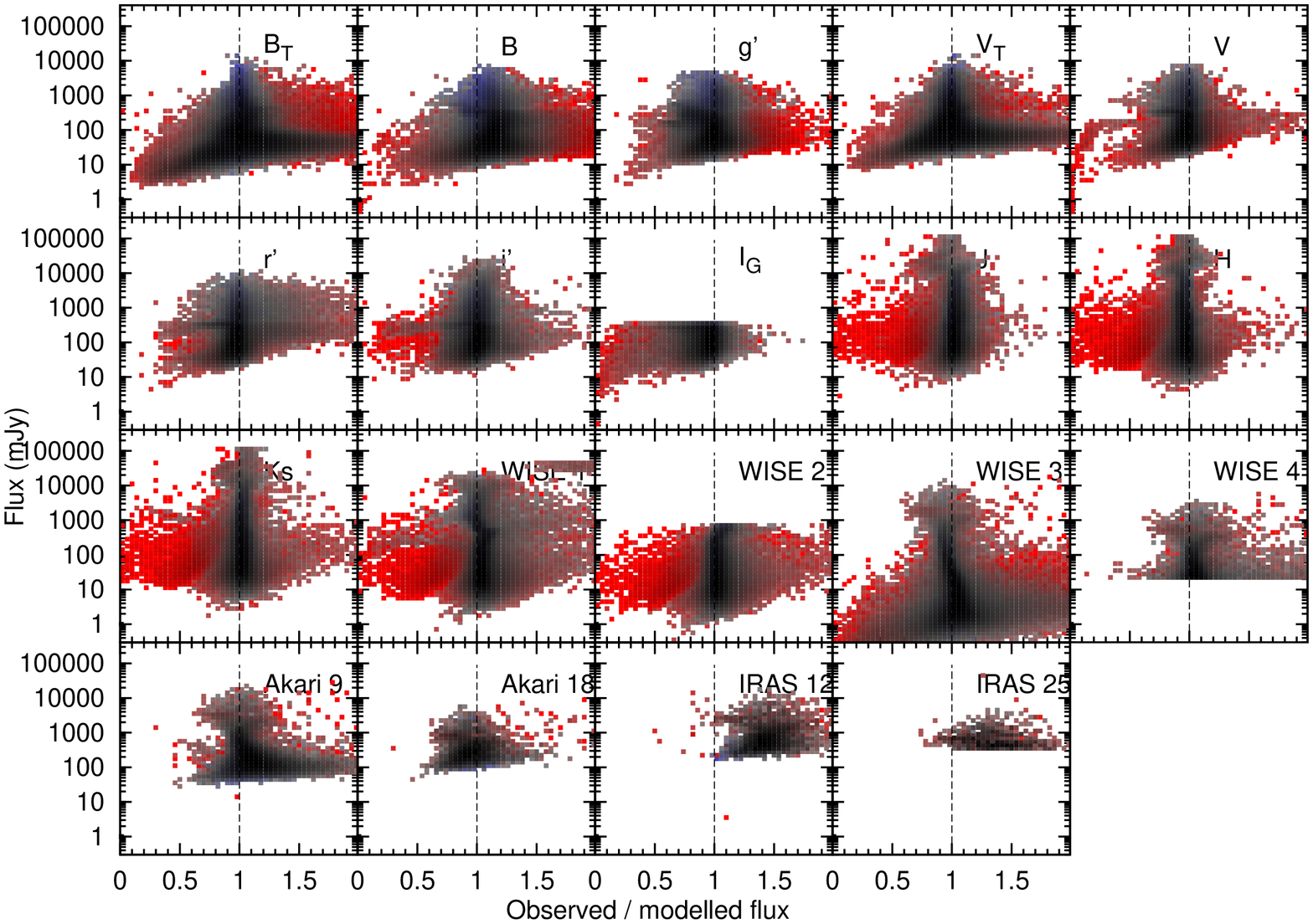}}
\caption{Run 4. See Figure \ref{OEFluxRun1Fig} for description.}
\label{OEFluxRun4Fig}
\end{figure*}

This run represents the first run where every star is analysed, and where the majority of the bad data has been taken out. This allows us to identify individual photometric points that are not well fit by the SED fitter, which can be individually removed from the input database. The ratio of observed to modelled flux for each object in the dataset is shown in Figure \ref{OEFluxRun4Fig}. It is clear that some systematic effects are still present, including:
\begin{itemize}
\item saturation issues in the APASS data ($B$, $V$, $r'$ and $i'$ filters), 
\item saturation effects in the $W_1$ filter, which have a knock-on effect in $J$, $H$ and $K_{\rm s}$,
\item systematic offsets in the zero point of the $B$ filter with respect to $B_{\rm T}$, 
\item systematic offsets in the mean \emph{IRAS} [12] and [25] fits, partly due to proximity to the sensitivity limit, partly due to beam size effects, and partly due to colour correction problems, and
\item sensitivity issues at the faint end of the Tycho-2 photometry.
\end{itemize}

The following cuts were performed to the photometry to alleviate these problems:
\begin{itemize}
\item Correct the \emph{IRAS} colour offset by reducing the flux by 47 per cent for \emph{IRAS} [12] and 41 per cent for \emph{IRAS} [25] \citep{BHW88}\footnote{See also: https://lambda.gsfc.nasa.gov/product/iras/colorcorr.cfm}.
\item Reduce the \emph{WISE} [3.4] saturation point to remove points if $W_1<3$ mag.
\item Delete APASS $B$ magnitudes fainter than the nominal detection limit $B>15$ mag.
\item Delete APASS $V$ magnitudes fainter than the nominal detection limit $V>14$ mag.
\item Delete APASS $r'$ magnitudes in the range $10<r'<10.2$ mag, if $V_{\rm T}-r'<-0.1$ mag.
\item Delete APASS $V$ magnitudes if $B-V<-0.3$ mag.
\item Delete APASS $r'$ magnitudes if the following criteria are met: $r_{\rm SDSS}<10$ and $r_{\rm APASS}>10$ and $r_{\rm APASS}-r_{\rm SDSS}>0.1$ mag.
\item Delete APASS $i'$ magnitudes if the following criteria are met: $i_{\rm SDSS}<10$ and $i_{\rm APASS}>10$ and $i_{\rm APASS}-i_{\rm SDSS}>0.04$ mag.
\end{itemize}

The offset in the $B$ filter is colour dependent, and appears to represent a slight offset of the filter transmission curve with respect to the Johnson $B$ band. Excess flux is found at cooler stellar temperatures, suggesting the filter profile includes more red flux than the standard. The offset has only a small impact on our results, primarily on the effective temperatures of the stars (typically increasing them by $\ll$2.5 per cent). Given the complexity of the required correction, it was therefore decided not to change the zero point of these data.

\subsection{Run 5: a second complete run}

\begin{figure*}
\centerline{\includegraphics[height=0.97\textwidth,angle=-90]{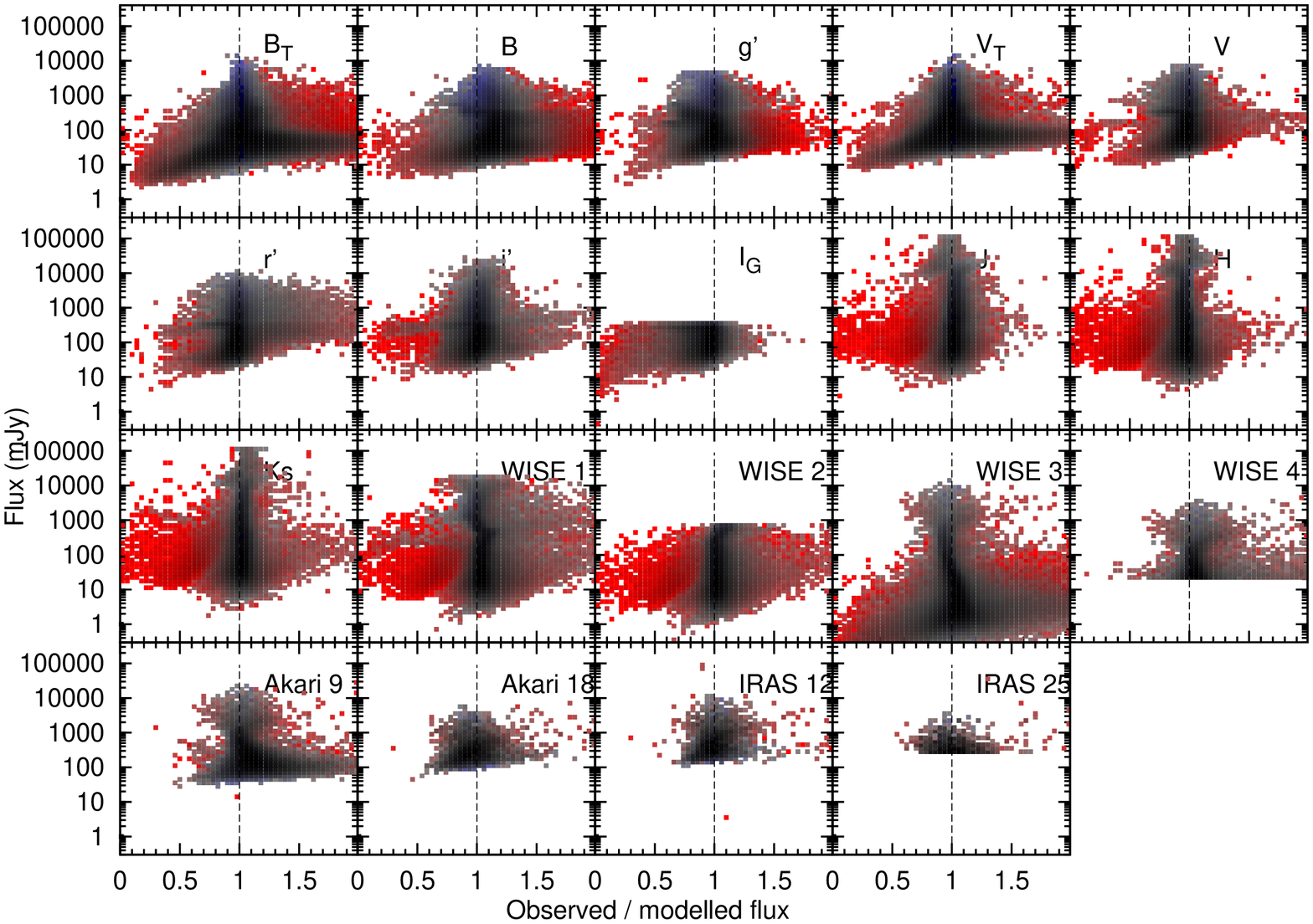}}
\caption{Run 5. See Figure \ref{OEFluxRun1Fig} for description.}
\label{OEFluxRun5Fig}
\end{figure*}

\begin{figure*}
\centerline{\includegraphics[height=0.97\textwidth,angle=-90]{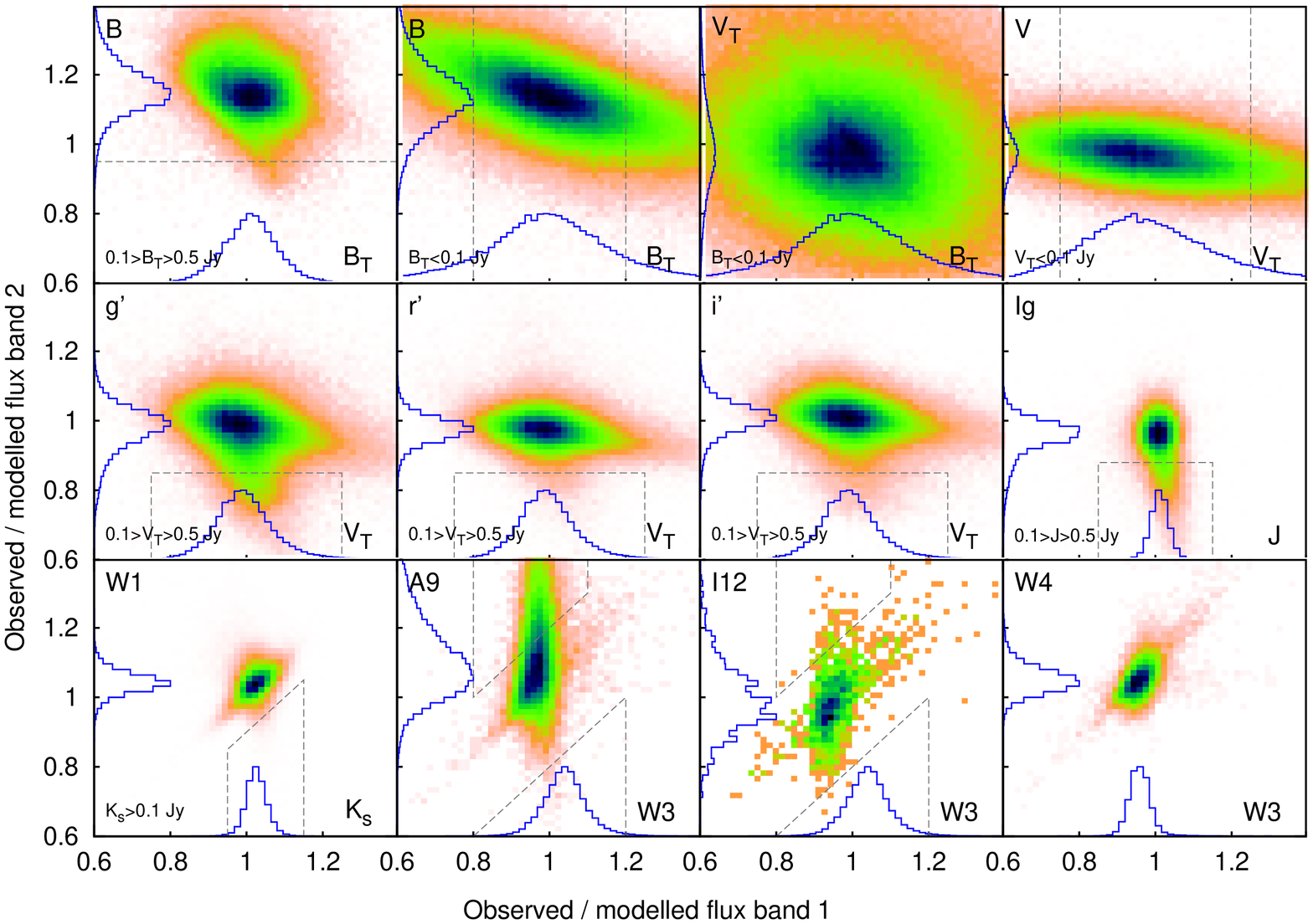}}
\caption{Run 5. The ratios of observed to model flux ($R$) for pairs of filters. Darker/bluer colours indicate a greater density of points. Histograms are displayed along each axis. In some cases (labelled) only a certain range of fluxes is shown to highlight particular sets of badly fitting data. Dashed grey lines show the regions which were cut from the final analysis.}
\label{Run5CutsFig}
\end{figure*}

At this point, the majority of bad data that could be cut out by simple colour--magnitude cuts had been removed. We refocussed our attention on data points which were badly fit. Figure \ref{Run5CutsFig} shows how badly fitting data points were selected based on the ratio of observed data to the best-fit stellar atmosphere models ($R_{\rm n}$). The following cuts were applied to the dataset for stars where $A_{\rm V} < 3.1$ mag:
\begin{itemize}
\item $B$ photometry was removed if $0.1 < B_{\rm T} < 0.5$ Jy and $R_{\rm B} < 0.95$. This removes saturation effects in the APASS $B$ data.
\item $B_{\rm T}$ photometry was removed if $B_{\rm T} < 0.1$ Jy and $R_{\rm BT} < 0.8$ or $R_{\rm BT} > 1.2$. This removes scattered, sensitivity-limited data in Tycho $B$.
\item $V_{\rm T}$ photometry was removed if $V_{\rm T} < 0.1$ Jy and $R_{\rm VT} < 0.8$ or $R_{\rm VT} > 1.2$. This removes scattered, sensitivity-limited data in Tycho $V$.
\item $g'$, $r'$ or $i'$ photometry was removed if $0.1 < V_{\rm T} < 0.5$ Jy and $0.75 < R_{\rm VT} < 1.25$ and $R_{\rm g',r',i'} < 0.85$. This removes saturation effects in the APASS data.
\item $i'$ photometry was also removed if $0.1 < J < 0.5$ Jy and $0.85 < R_{\rm J} < 1.15$ and $R_{\rm i'} < 0.85$. This removes saturation effects in the APASS $i'$ data.
\item $I_{\rm Gunn}$ photometry was removed if $0.1 < J < 0.5$ Jy and $0.85 < R_{\rm J} < 1.15$ and $R_{\rm I} < 0.88$. This removes saturation effects in the DENIS $I$ data.
\item $WISE$ 1 photometry was removed if $K_{\rm s} > 0.1$ Jy and $0.95 < R_{\rm Ks} < 1.15$ and $R_{W1} < R_{\rm Ks} - 0.1$. This removes saturation effects in the WISE [3.4] data.
\item $Akari$ [9] photometry was removed if $0.8 < R_{W3} < 1.2$ and $|R_{A9} - R_{W3}| > 0.2$. This removes scattered, sensitivity-limited data in $Akari$ [9].
\item $IRAS$ [12] photometry was similarly removed if $0.8 < R_{W3} < 1.2$ and $|R_{I12} - R_{W3}| > 0.2$. This removes scattered, sensitivity-limited data in $IRAS$ [12].
\item $IRAS$ [25] photometry was removed if $0.6 < R_{W3} < 1.4$ and $R_{I25} > R_{W3} + 0.15$. This removes scattered, sensitivity-limited data in $IRAS$ [25].
\item $WISE$ 4 photometry was taken out if $R_{W4} / R_{W3} > 16$ (or $> A_{\rm V}$ (in mag) if $A_{\rm V} > 16$ mag). This removes spurious matches in $WISE$ [22] photometry near the detection limit. This was also applied to stars with $A_{\rm V} \geq 3.1$ mag.
\item Any data was removed if $R > 20 Q$ (or $> A_{\rm V}Q$ (in mag) if $A_{\rm V} > 20$ mag) and either $R<0.5$ or $R>2$. This was also applied to stars with $A_{\rm V} \geq 3.1$ mag.
\end{itemize}
The design of the these cuts removes individual outliers (e.g.\ unmasked cosmic rays, poorly subtracted backgrounds, or artefacts from differing telescope beam sizes). At the same time, it allows stars which are broadly discrepant from stellar models over several filters to remain in the dataset, such as binary stars, dust-enshrouded and heavily extincted stars. Such stars exhibit SEDs less strongly peaked than an equivalent-temperature, unextincted blackbody.

These cuts resulted in 489\,792 datapoints being removed from 395\,166 stars. The majority of these datapoints (230\,232 and 147\,240, respectively) were to remove faint sources in $B_{\rm T}$ and $V_{\rm T}$ Tycho-2 data with poor data quality. A further 42\,167 $I$-band points were removed from the DENIS catalogue, and smaller numbers from other catalogues. These stars were re-run through the fitter and merged back into the catalogue.

\subsection{Run 6: selective removal of bad data}

\begin{figure*}
\centerline{\includegraphics[height=0.97\textwidth,angle=-90]{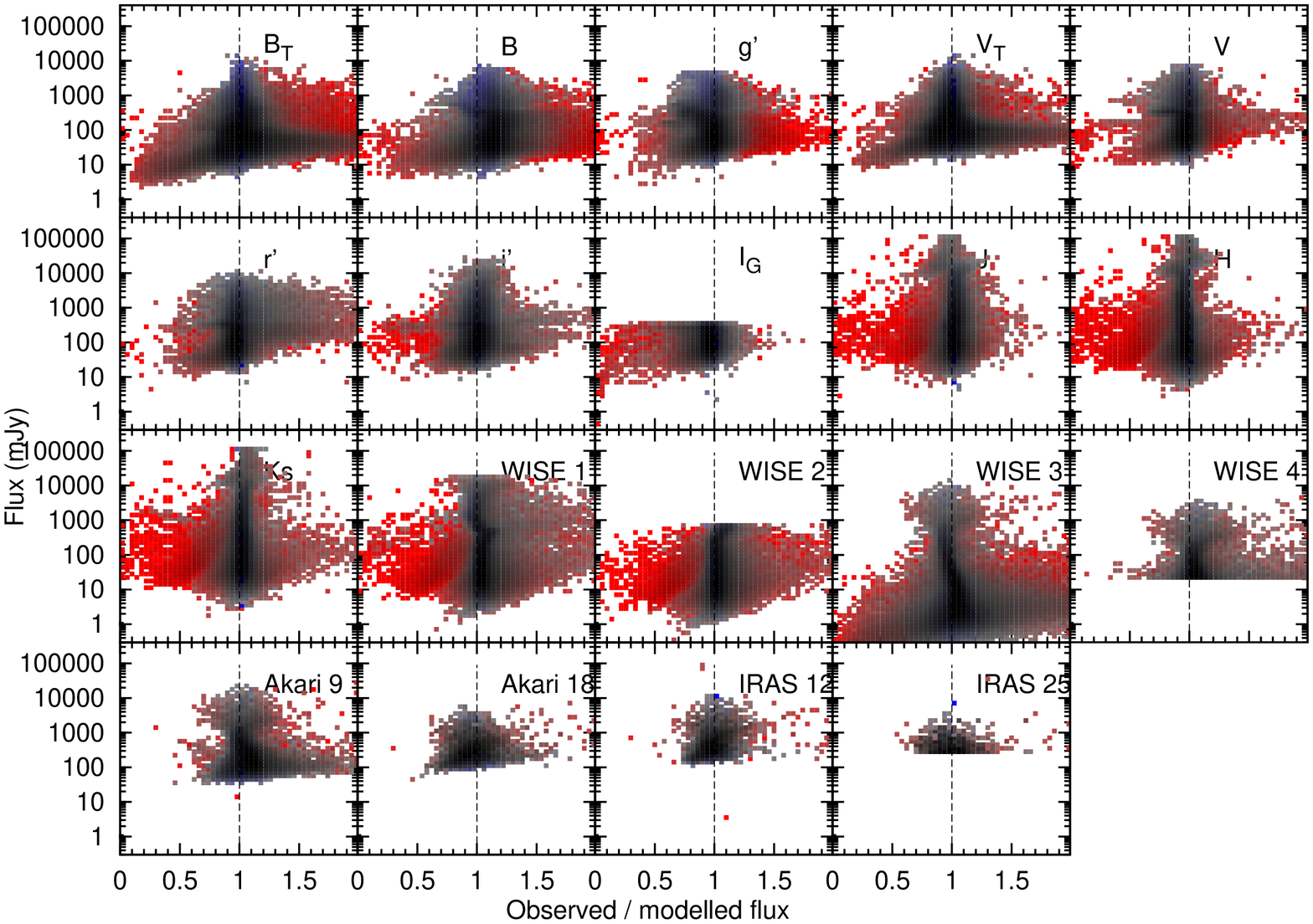}}
\caption{Run 6. See Figure \ref{OEFluxRun1Fig} for description.}
\label{OEFluxRun6Fig}
\end{figure*}

Substantial improvement in the quality of fits can be seen following this run. Several changes were made to the criteria used to remove bad data:
\begin{itemize}
\item Flux limits on $B_{\rm T}$ and $V_{\rm T}$ were changed from $<$0.1 Jy and $0.1 < (B_{\rm T} | V_{\rm T}) < 0.5$ Jy to $<$0.2 Jy and $0.2 < (B_{\rm T} | V_{\rm T}) < 0.5$ Jy, to reflect the significant scatter remaining in these bands, compared to the now-more-accurate APASS, IPHAS and SDSS photometry.
\item Cuts from run 5 applied to stars with $A_{\rm V} < 3.1$ were now also applied to stars within 400 pc which have $A_{\rm V} > 3.1$ mag. At this distance, the Lutz--Kelker bias is relatively small, but we caution that this does not imply a fixed boundary.
\item For the final cut, data was removed if $R - 1 > 20 Q$ (or $> A_{\rm V}Q$ (in mag) if $A_{\rm V} > 20$ mag), i.e.\ if the point was discrepant by more than 20 times the average discrepancy.
\end{itemize}

These cuts resulted in 132\,260 datapoints being removed from 129\,676 stars. The majority of these datapoints (109\,298) were to remove faint, poor-quality $V_{\rm T}$ Tycho-2 data once the corresponding $B_{\rm T}$ data had been removed. These stars were again re-run through the fitter and ingested back into the catalogue.

\subsection{Run 7: more selective removal of bad data}

Further improvement of the fits was seen. The same set of cuts was processed to provide an eighth run: 168\,222 datapoints were removed from 147\,925 stars. The majority (109\,298) were removal of Tycho $V_{\rm T}$ data.

\subsection{Run 8: selective removal of WISE [11.3] data}

At this stage, the largest source of bad data is close to the sensitivity limit of $WISE$ 3, where there is a large scatter of points. Most stars with fluxes of $<$20 mJy at 11.3 $\mu$m are well fit. However, $\sim$4 per cent have substantial offset from a perfect fit. These tend to correlate with areas of high extinction, where emission from interstellar dust (and potentially unresolved stars) can contribute significantly to the sky background.

A small fraction of these 4 per cent of sources could be stars with genuine infrared excess, which we would ideally like to keep in the database, making the criterion for removing bad data quite important. Objects with infrared excess will typically be extincted in the optical, but have flux excess in other infrared bands. These include both young and evolved objects with strong infrared excess \citep{WOK+11,RKJ+15}, and evolved carbon stars \citep{MWZ+12,SML+12,SKM+16}. Evolved stars with silicate emission typically do not show much infrared excess shortward of the silicate emission peak at $\sim$10 $\mu$m. However, even these stars do show measureable excess at $\sim$4.6 $\mu$m where the $WISE$ 2 band is located (e.g.\ \citealt{vLBM08,MvLS+11,BMS+15}). A flux limit of 20 mJy retains most sources on (or cooler than) the giant branch, which are within 1 kpc and more luminous than 680 L$_\odot$. Sources with genuine infrared excess are more likely to scatter above the 20 mJy limit, so these are more likely to be retained. Around 90 per cent of sources in this sample are within 1 kpc, and 680 L$_\odot$ represents the expected lower luminosity limit for giant branch stars producing significant quantities of dust \citep{MZB12}. An additional criterion was therefore established whereby $WISE$ [11.3] data was removed if $R_{W3} < 0.75$, or both $R_{W3} > 1.33$ and $R_{W2} < 1$. 

The full selection of cuts were applied to the catalogue, which was run again. A total of 113\,956 datapoints were removed from 100\,789 stars. The majority (88\,053) were to remove $WISE$ 3 data.

\subsection{Run 9: selective removal of other bad data}

At this stage, bad data from other infrared bands dominates the remaining bad data in the sample. These were dealt with using the following cuts, which apply the principles that: (1) sources with strongly rising infrared SEDs are likely to either be sufficiently obscured that they are optically invisible, or be associated with line-of-sight sources that are not directly tied to the observed star. This can be applied as a general rule, although harsh application of it does risk removing certain kinds of sources (e.g.\ near-face-on disc sources). The following cuts were applied, in addition to repeats of those previously mentioned:
\begin{itemize}
\item If $R_{W3} < 0.75$ and $R_{W3} < R_{W2} - 0.3$ then delete $WISE$ 3, in order to remove negative scatter caused by low signal-to-noise $WISE$ 3 photometry.
\item If $R_{W2} < 2$ and $R_{W3} > 1.33$ and $R_{W3} > 2 (R_{W2} - 1) + 1.5$ then delete $WISE$ 3, in order to positive scatter caused by low signal-to-noise $WISE$ 3 photometry.
\item If $R_{W3} < 2$ and $R_{W4} > 1.33$ and $R_{W4} > 4 (R_{W2} - 1) + 3$ then delete $WISE$ 4, in order to positive scatter caused by low signal-to-noise $WISE$ 4 photometry.
\item If $R_{W3} < 2$ and $R_{A9} > 1.33$ and $R_{A9} > R_{W3} + 0.5$ then delete $AKARI$ [9], in order to positive scatter caused by low signal-to-noise $AKARI$ [9] photometry.
\end{itemize}

A few bad datapoints from optical bands still remain, mostly arising from saturation issues in the APASS photometry. These were dealt with using the following cuts:
\begin{itemize}
\item If $R_{g'} < 0.5$ and $R_{g'} < R_{\rm VT} - 0.3$ then delete $g'$.
\item If $R_{\rm V} < 0.5$ and $R_{\rm V} < R_{\rm VT} - 0.3$ then delete $V$.
\item If $R_{r'} < 0.5$ and $R_{r'} < R_{\rm VT} - 0.3$ then delete $r'$.
\item If $R_{i'} < 0.45$ and $R_{i'} < R_{\rm J} - 0.33$ then delete $i'$.
\end{itemize}

These cuts resulted in the removal of 15\,508 datapoints from 12\,974 objects.

\subsection{Run 10: selective removal of high-background WISE 4 data}

The most problematic bad data at this stage is sources with unexpectedly large excess in $WISE$ [22]. Examination of the location of these objects on the sky shows that they are predominantly associated with regions of high extinction or nebulosity, such as the Galactic Plane, Orion complex and NGC 7000. A cut was included to remove $WISE$ 4 data from stars with $R_{W4} > 1.33$ and $A_{\rm V} > 5$ mag. Re-running all of the cuts removed 11\,845 points from 8100 stars. Of these, 3575 were from $WISE$ 4.

\subsection{Run 11: selective removal of other high-background infrared data}

This removes most of the remaining outliers in the Galactic Plane, however other infrared data are also affected to a lesser extent, particularly in regions such as Orion. A cut was included to remove \emph{any} infrared data longward of 8 $\mu$m which have $R > 1.33$, $A_{\rm V} > 10$ mag and $Q > 0.3$. Re-running all of the cuts removed 7\,517 points from 5333 stars. Of these, 1657 were from $WISE$ 3. These cuts effectively blind us to sources with intrinsic excess infrared emission in very high extinction regions. However, in the majority of cases these would not be confidently discernable anyway.

\subsection{Run 12: final run}

\begin{figure*}
\centerline{\includegraphics[height=0.97\textwidth,angle=-90]{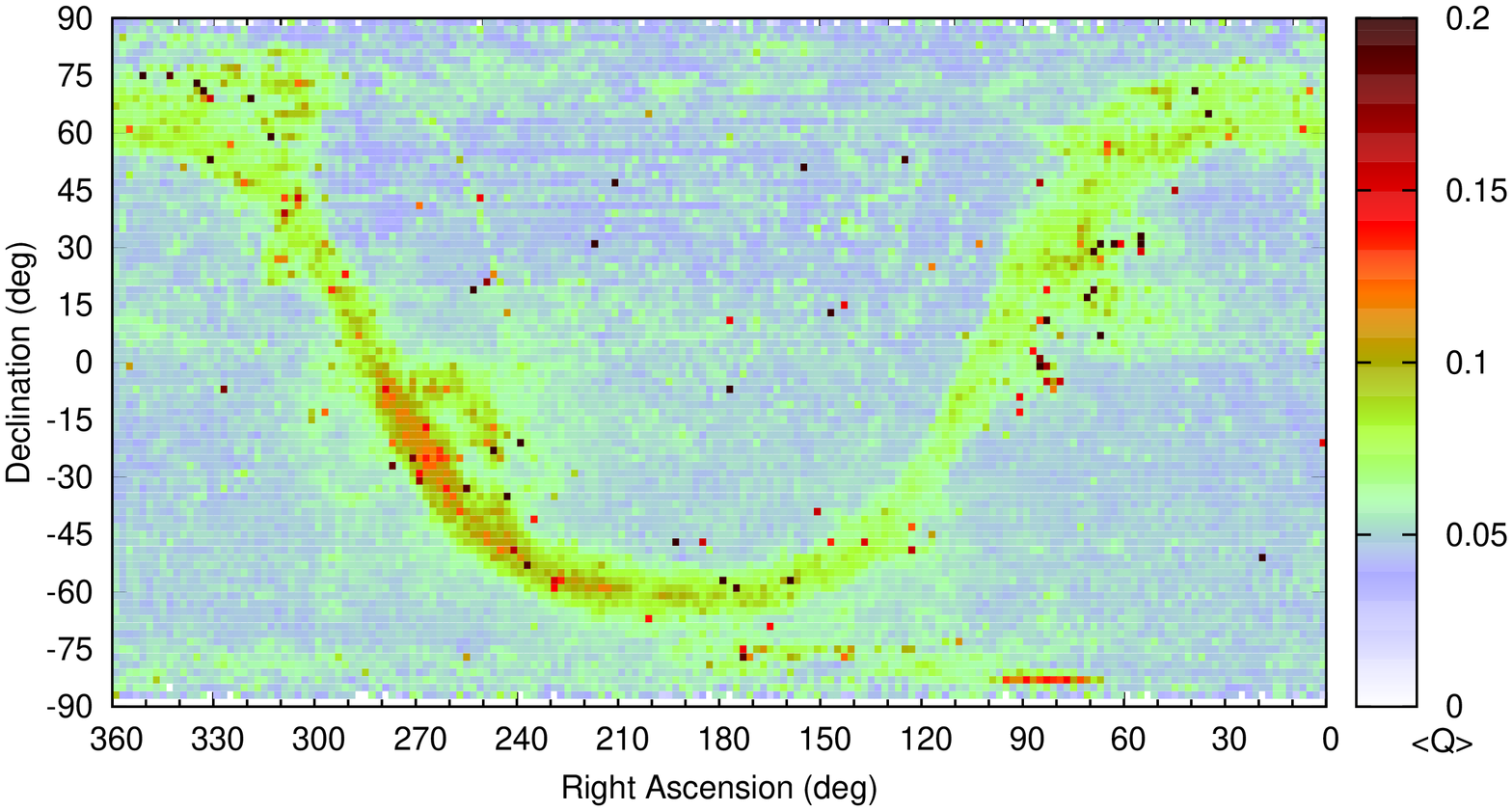}}
\caption{Final run. As Figure \ref{OESkyRun3Fig}.}
\label{OESkyFig}
\end{figure*}

\begin{figure*}
\centerline{\includegraphics[height=0.97\textwidth,angle=-90]{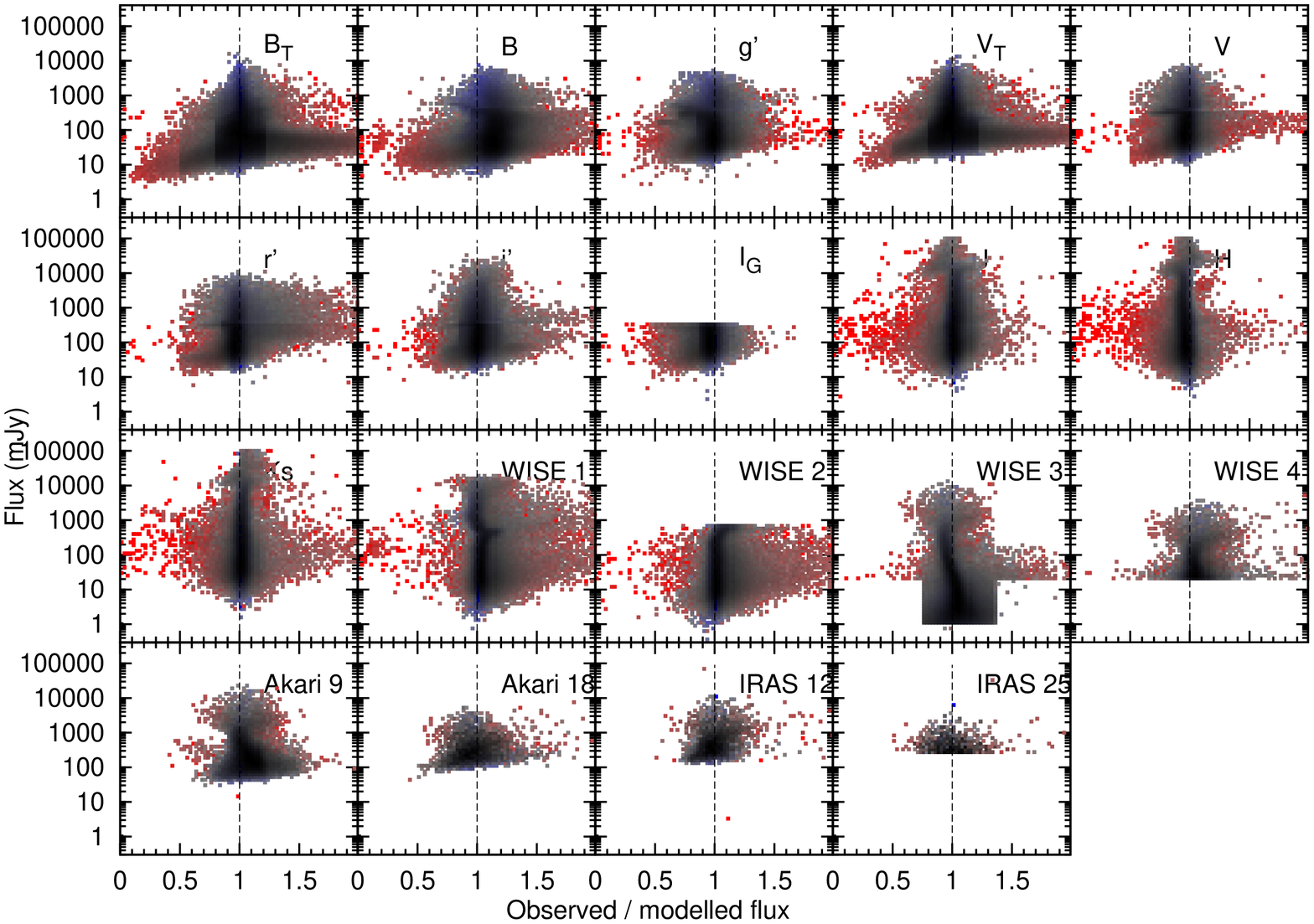}}
\caption{Final run. See Figure \ref{OEFluxRun1Fig} for description.}
\label{OEFluxFig}
\end{figure*}

\begin{figure*}
\centerline{\includegraphics[height=0.97\textwidth,angle=-90]{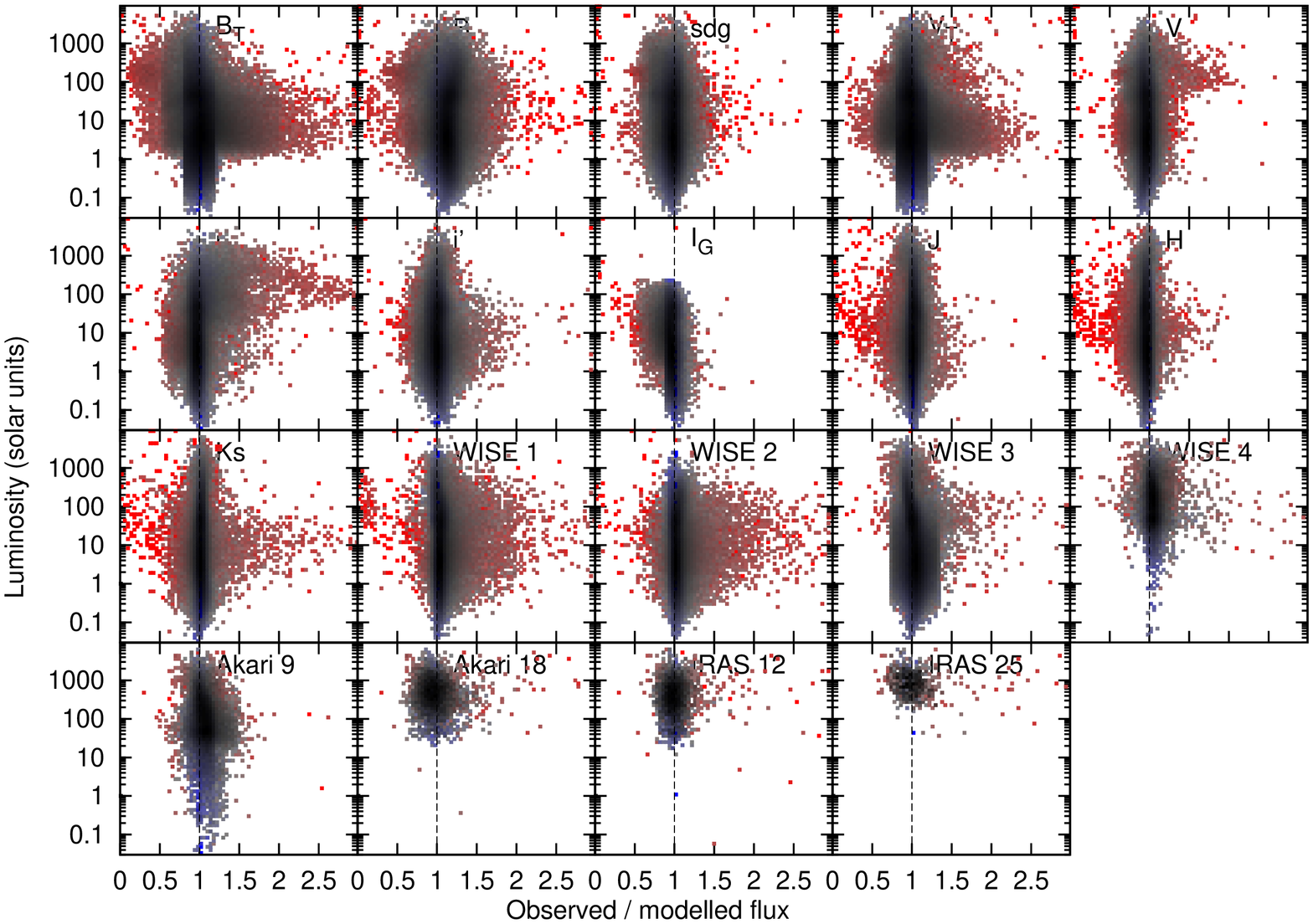}}
\caption{Final run. As Figure \ref{OEFluxFig}, but showing deviation from the stellar model versus luminosity.}
\label{OELumFig}
\end{figure*}

\begin{figure*}
\centerline{\includegraphics[height=0.97\textwidth,angle=-90]{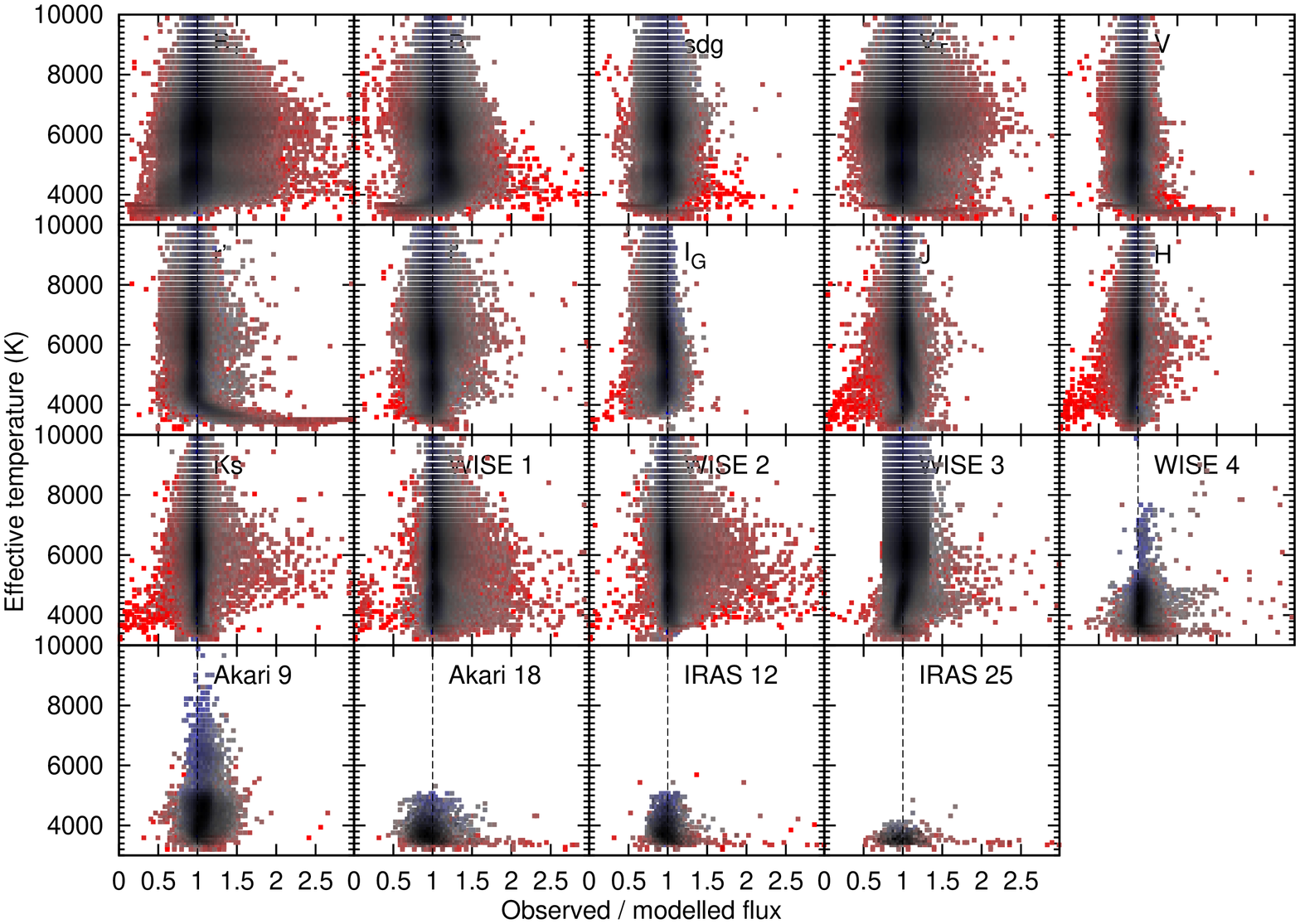}}
\caption{Final run. As Figure \ref{OEFluxFig}, but showing deviation from the stellar model versus effective temperature.}
\label{OETempFig}
\end{figure*}

\begin{figure*}
\centerline{\includegraphics[height=0.97\textwidth,angle=-90]{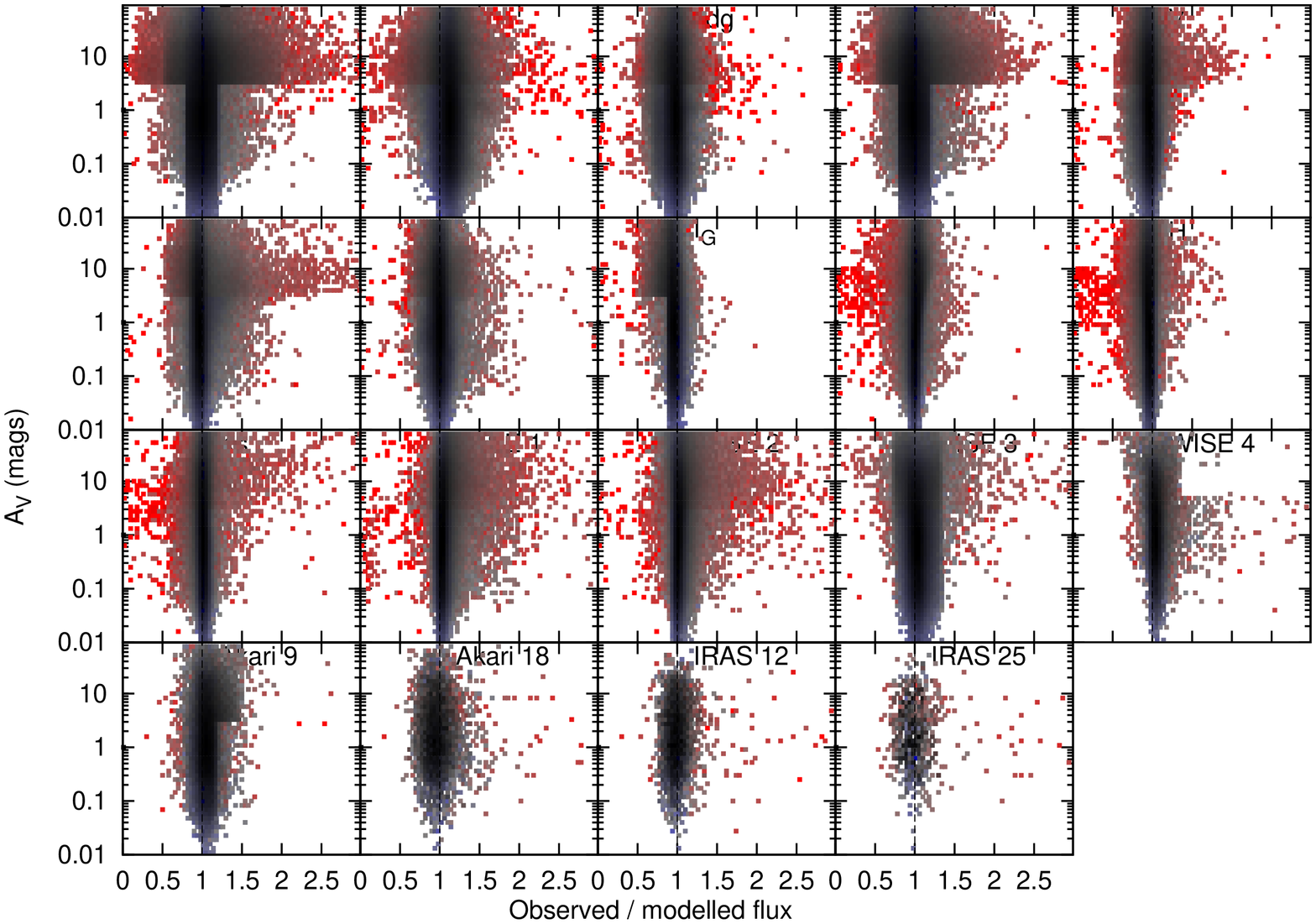}}
\caption{Final run. As Figure \ref{OEFluxFig}, but showing deviation from the stellar model versus interstellar reddening.}
\label{OEEBVFig}
\end{figure*}

The majority of badly fitting data has now been systematically removed from the catalogue. Figure \ref{OESkyFig} shows the goodness-of-fit statistic, $Q$, as a function of position on the sky. The final catalogue show average deviation from the model fit of between $Q = 0.04$ and 0.07. Sources within $\sim$5 degrees of the Galactic Plane typically have uncertainties which are a factor of $\sim$2 higher than this, as do sources in the Galactic Bulge, Cygnus, and the Orion complex. Small patches of badly fit data in the very south correlate with regions of extended dust emission.

Figures \ref{OEFluxFig} through \ref{OEEBVFig} detail the remaining deviations in each band, as a function of (respectively) catalogue flux, modelled luminosity, modelled effective temperature, and line-of-sight interstellar reddening. A variety of effects related to both saturation and sensitivity clearly remain, but at a much reduced level compared to the original dataset.



\section{Data flagging in \emph{Hipparcos} data}
\label{AppendixHip}

\begin{figure*}
\centerline{\includegraphics[height=0.97\textwidth,angle=-90]{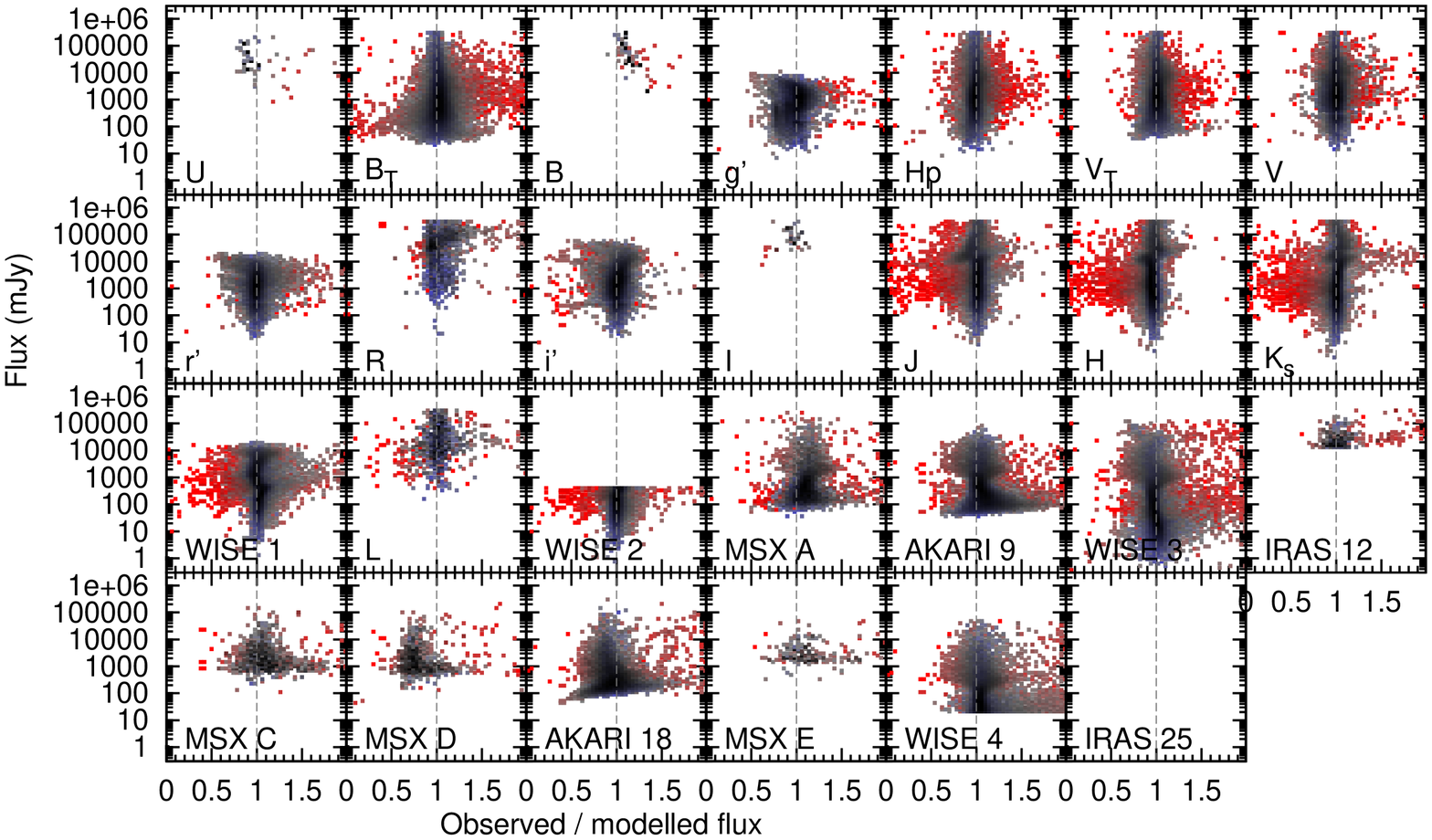}}
\caption{As Figure \ref{OEFluxFig} for the final run of the \emph{Hipparcos stars}.}
\label{OEFluxHipFig}
\end{figure*}

\begin{figure*}
\centerline{\includegraphics[height=0.97\textwidth,angle=-90]{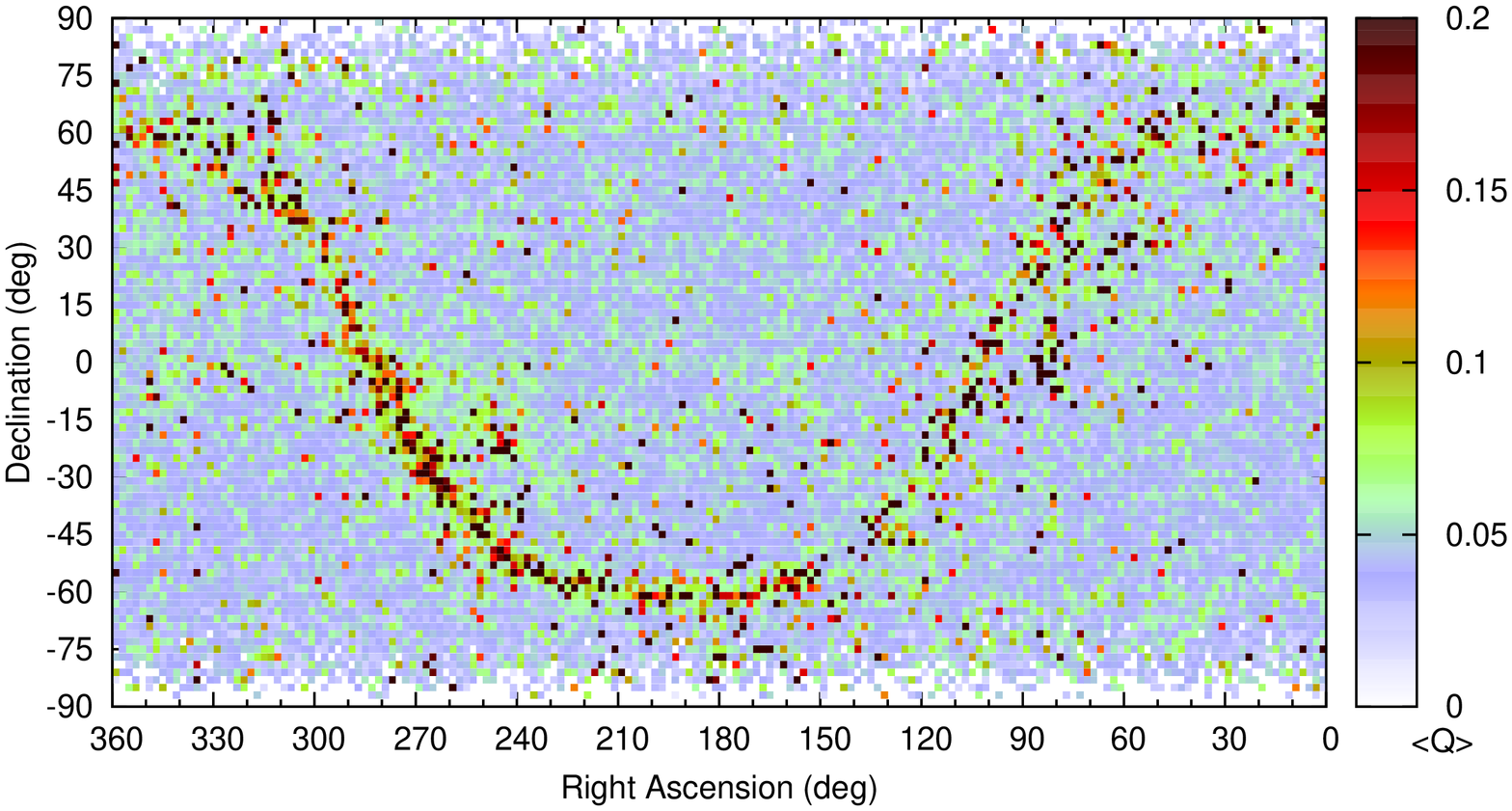}}
\caption{As Figure \ref{OESkyRun3Fig} for the \emph{Hipparcos stars}.}
\label{OESkyHipFig}
\end{figure*}

The comparative brightness of the \emph{Hipparcos} set of stars, and the larger number of catalogues available for them, provides greater reliability and redundancy in their SEDs. Consequently, bad data could be more easily recognised and removed. The smaller dataset is also quicker to compile and run, making iterative cuts easier. However, the \emph{Hipparcos} dataset generally contains more nearby stars, with larger proper motions. Since the cross-matching exercise was done without taking these into account, this has resulted in a greater fraction of missing data or false matches than could otherwise have been achieved. A more experimental basis was adopted, which let us decide on the following cuts, where magnitudes in systems without well-defined reference points (\emph{DIRBE, MSX, IRAS}) are quoted in $AB$ magnitudes\footnote{Vega magnitudes are used otherwise, except for the Sloan $ugriz$ filters, where $AB$ magnitudes are used by convention. Vega-magnitude zero points and filter transmission profiles were adopted from the Spanish Virtual Observatory's Filter Profile Service (http://svo2.cab.inta-csic.es/svo/theory/fps/index.php?mode=voservice) for the Sloan and Johnson--Cousins optical and near-infrared filters.}:
\begin{enumerate}
\item SDSS $u$-band and $z$-band data were removed. The original issue with this data was eventually traced to an ambiguity surrounding airmass correction in the filter transmissions, but the data were removed anyway because of strong saturation issues.
\item SDSS $g$ data were rejected if $H_{\rm p}-g < -1.0$ mag for similar reasons.
\item SDSS $r$ data were rejected if $H_{\rm p}-r < 0$ mag.
\item SDSS $i$ data were rejected if $H_{\rm p}-i < 0$ mag.
\item APASS $B$ magnitudes were rejected if $B_{\rm T}-B < -0.2$ mag, $H_{\rm p}-B < -0.6$ mag or $9.95 < B < 10.80$ mag. This removes saturated stars and bad matches, over ranges which take into account the likely photometric scatter due to uncertainties, circumstellar or interstellar reddening and companion objects. The final criterion specifically removes stars around 10$^{\rm th}$ magnitude saturation limit.
\item APASS $g$ data were similarly rejected if $H_{\rm p}-g < -0.2$ mag, or $9.95 < g < 10.80$ mag.
\item APASS $V$ data were similarly rejected if $H_{\rm p}-V < -0.2$ mag, or $9.95 < g < 10.30$ mag.
\item APASS $r$ data were similarly rejected if $H_{\rm p}-r < -0.2$ mag, or $9.95 < g < 10.15$ mag.
\item APASS $i$ data were similarly rejected if $H_{\rm p}-r < -0.5$ mag, or $9.95 < g < 10.15$ mag.
\item Tycho-2 $B_{\rm T}$ and $V_{\rm T}$ data were removed if $H_{\rm p}-V_{\rm T} < -0.3$ mag and $H_{\rm p}-V_{\rm T} < ((B_{\rm T} - V_{\rm T}) - 1) / -1.8$ mag. This removes unphysical magnitudes caused by false matches.
\item Mermilliod $U$, $B$ and $V$-band magnitudes were all removed if $B-V_{\rm T} > ((B_{\rm T} - V_{\rm T}) + 1) / 1.3$ mag for similar reasons.
\item Mermilliod $U$-band data were also specifically removed if either $U-B_{\rm T} > (B_{\rm T} - V_{\rm T}) + 0.5$ mag and $B_{\rm T} - V_{\rm T} < 2$ mag, or $U-B_{\rm T} < (B_{\rm T} - V_{\rm T}) - 1.5$ mag and $U - B_{\rm T} < 1$ mag.
\item DENIS $I$-band was found to be too heavily saturated for use in the catalogue. It was entirely removed.
\item UKIDSS and IPHAS data were similarly saturated and removed in their entirety.
\item \emph{DIRBE} [1.25] and [2.2] data were removed, respectively, if 2MASS $J$- and $K_{\rm s}$-band existed. \emph{DIRBE} data exhibit more scatter than 2MASS data, due to the lower signal-to-noise.
\item \emph{DIRBE} [3.5] and [4.9] data were both removed if both 2MASS $K_{\rm s}$-band and \emph{AKARI} [9] data existed.
\item \emph{DIRBE} [12] and [25] are respectively removed if they are $>$0.65 mag ($\lesssim$2000 Jy). This removes significant scatter in low signal-to-noise results.
\item \emph{IRAS} [12] and [25] are similarly removed if [12] $>$ 5.65 mag ($\lesssim$25 Jy) or [25] $>$ 3.15 ($\lesssim$200 Jy), to reduce scatter.
\item \emph{WISE} [3.4] data were removed if $W_1 < 3.0$ mag, to remove a systematic offset in saturated data.
\item \emph{WISE} [4.6] data were removed if $W_2 < 6.5$ mag, to remove an increasing offset in near-saturated data. The cut was chosen at the point where the systematic offset exceeds 5 per cent.
\item \emph{WISE} [11.3] data were removed if $W_3 < -1.5$ mag, to remove saturated data.
\item \emph{WISE} [22] data were removed if $W_4 < -2.3$ mag, to remove saturated data.
\item \emph{WISE} [22] data were also removed if $W_4 > 6.5$ mag, to remove highly scattered, low signal-to-noise detections.
\item \emph{WISE} [22] data were additionally removed if $K_{\rm s}-W_3 < 0.7$ mag, and $W_1 - W_3 < 0.7$ mag, and $W_3 - W_4 > 1.2$ mag, and (if it exists) \emph{AKARI} [9]--[18] $<$ 0.7 mag. This complex system of cuts ensures that stars with genuine infrared excess stay in the catalogue, but that stars where only \emph{WISE} [22] is in excess are removed.
\item \emph{MSX} $B_1$ and $B_2$ data were removed in their entirety, due to the large scatter in their goodness of fit.
\item \emph{MSX} $C$, $D$ and $E$ were respectively removed if $C > 8.15$ mag ($<$2 Jy), $D > 8.15$ mag ($<$2 Jy) or $E > 6.4$ mag ($<$10 Jy).
\item \emph{MSX} $C$ band was also removed if $W_3 - C > 2 (W_3 - W_4) - 4.8$ mag.
\item \emph{MSX} $D$ band was also removed if $W_3 - D > 2 (W_3 - W_4) - 5.2$ mag.
\item \emph{MSX} $E$ band was also removed if $W_3 - E > 2 (W_3 - W_4) - 6.0$ mag.
\item Johnson--Cousins optical data from APASS was used in preference to Mermilliod \citep{Warren91}, which was used in preference to \citet{MM78}.
\item Optical data from SDSS was used in preference to APASS in the Sloan filter sets.
\item Near-infrared data from 2MASS was used in preference to DENIS, which was used in preference to \citet{MM78}. Exceptions were made for sources above or close to the saturation limit ($J,H,K_{\rm s} < 5.6,\ 5.0,\ 4.7$ mag), where data from 2MASS and \citet{MM78} are averaged if both exist.
\item \emph{IRAS} data was used in preference to \emph{DIRBE} data at 12 and 25 $\mu$m.
\item Any datapoint with an error of $\delta M > 0.2$ mag was rejected, except for bright stars ($<$6th magnitude) where uncertainties up to $\delta M = 0.4$ mag were allowed. This restriction removes many uncertain detections, while retaining detections for saturated stars: this is particularly important when retaining 2MASS magnitudes for bright giants.
\end{enumerate}
Each of these cuts was tested indivdiually on the dataset, and manual inspection of a selection of both cut and retained objects was used to fine tune them. Since these cuts do not require the iterative processing done on the Tycho-2 data, they were performed in a single run of the data reduction pipeline.

Following this analysis, significant remaining colour terms were identified in the $U$-band and $u$-band data, in the APASS $B$ data, and in the $I$-band data. These were especially prominent in the cooler stars, with a marked temeprature dependence, suggesting a departure in the filter transmission function. The data were recomputed with these bands removed. The data were then recombined: if the source is below 5400 K (where these bands aren't important in constraining the SED), or otherwise if the goodness-of-fit parameter ($Q$) was more than halved, the $u$-, $U$-, $B$- and $I$-band data were removed.

The result of these cuts is a largely clean dataset. The majority of scatter from unity in Figure \ref{OEFluxHipFig} appears to be intrinsic to the sources in question. Photometric blending with very close background objects cannot be excluded, and the poor quality fits are highly concentrated in the Galactic Plane (Figure \ref{OESkyHipFig}). Typically, blending manifests itself as a discrepency between surveys with large beams (e.g.\ \emph{IRAS}) and those with small beams (e.g.\ \emph{WISE}). Such data are therefore typically excluded by the above cuts, so most of the scatter should not only be intrinsic to each detected source, but to each star in question.


\section{Exploring the Lutz--Kelker bias and related effects}
\label{AppendixLK}

\subsection{Theory and manifestations of the effects}
\label{AppendixTheoryLK}

The Lutz--Kelker bias \citep{LK73}, and the wider range of effects it produces, is a complex and often confusing problem (see, e.g., \citealt{Smith03} for a review of the subject). It is often not clear whether or not a bias correction needs applied to a given data set, and even less clear as to what that correction should be.

We can generalise the problem to a variable $x \pm \delta x$, mapped to another variable, $y^{+\delta_1 y}_{-\delta_2 y}$ as $y = x^{-1}$. For an arbitrary probability distribution function (PDF), inverting any given quantile on the PDF of $x$ gives the corresponding quantile on the PDF of $y$. Therefore, if one quotes the 16th, 50th and 84th centiles (to give the median of the PDF $\pm$ the 1$\sigma$ range), $x + \delta x = (y - \delta_2 y)^{-1}$ and $x - \delta x = (y + \delta_1 y)^{-1}$. In terms of uncertainties, one can then reduce this to $\delta x / x = \delta y / y$. However, if one prefers to obtain the maximum-likelihood estimator of $y$ (the peak of the PDF, appropriate for a single measurement of a single star, but not a single measurement within a catalogue of stars), the translation of $x \rightarrow y$ depends on the precise shape of the PDF.

Poisson noise in the detected stellar light, randomly moves the image centroid for the star in a Gaussian manner, hence the parallax PDF for most stars is normally taken to be close to Gaussian. The above formalism for translating $x \rightarrow y$ holds, provided $x - \delta x > 0$. If a non-negligible part of the parallax PDF extends below zero, a valid parallax measurement is translated into an invalid (negative) distance. For our nominal cut of $\delta \varpi / \varpi < 0.414$, $<$1 per cent of the PDF falls below zero. For an isolated star, our na\"ive translation of $\varpi = d^{-1}$ and $\delta \varpi / \varpi = \delta d / d = \sqrt{\delta L / L}$ is therefore a reasonable approximation, provided one treats the fractional errors appropriately.

Modification of the PDF is often performed to account for two factors: the distribution of stars and the distribution of apparent luminosities, namely the manifestations of the Lutz--Kelker and Malmquist biases \citep{Malmquist20}. Namely, the probability of finding a star at a given distance is not only a function of that star's parallax, but also of the distribution of stars with distance, and the probability of detecting that star at a given distance. Historically, parallax studies have been done in the solar neighbourhood, where the stellar density is roughly constant, so the distribution of stars with distance is $\propto d^2$. This distribution means that most stars are found towards the faint end of the signal-to-noise distribution and, as flux is $\propto d^{-2}$, so is detectability, and the two effects largely cancel.

In more advanced analyses, these two proportionalities no longer hold. Stellar distributions exhibit spatial variation, particularly regarding concentration in the Galactic disc. The detectability depends both on how close one is to the observational limit of detection (or saturation) and astrophysical parameters such as extinction along the line of sight. Many of these parameters can be accounted for using a 3-D stellar and extinction model of the Milky Way, as in the approach of \citet{ABJ16}, which allows recovery of distances for objects where the parallax PDF contains a non-negligible negative component.

The PDF can be arbitrarily multiplied by other PDFs, based on what is known about (e.g.) the star's kinematic properties, metallicity, abundances, inferred age, pulsation properties, or other information. A comparison between the temperatures derived from the na\"ive and ``ABJ'' methods shows little difference in most cases, except where the SED fitter is forced to make a choice between two similar $\chi^2$ minima. The cautious user is therefore advised to construct their own PDF for their object of choice, using all the information available to them, and perform the inversion themselves. Both the na\"ive and ``ABJ'' methods are only appropriate for single stars, and any use of these data on population studies should strictly require correction for that population's characteristics.

\subsection{Comparison of the na\"ive method and that of \citet{ABJ16}}
\label{AppendixCompareLK}

\begin{center}
\begin{table*}
\caption{Comparison of goodness of fit between a na\"ive inversion of parallax to obtain distance, and a full Lutz-Kelker correction as applied by \citet{ABJ16}.}
\label{LKTable}
\begin{tabular}{cccccccccc}
    \hline \hline
\multicolumn{1}{c}{Feature} & \multicolumn{1}{c}{Primary}    & \multicolumn{1}{c}{Luminosity}  & \multicolumn{1}{c}{Temperature} & \multicolumn{2}{c}{Objects in} & \multicolumn{4}{c}{One-dimensional standard deviation}\\
\multicolumn{1}{c}{\	}   & \multicolumn{1}{c}{constraint} & \multicolumn{1}{c}{range}       & \multicolumn{1}{c}{range}       & \multicolumn{2}{c}{region}     & \multicolumn{1}{c}{Na\"ive} & \multicolumn{1}{c}{ABJ} & \multicolumn{1}{c}{Na\"ive} & \multicolumn{1}{c}{ABJ} \\
\multicolumn{1}{c}{\	}   & \multicolumn{1}{c}{\ }         & \multicolumn{1}{c}{(L$_\odot$)} & \multicolumn{1}{c}{(K)}         & \multicolumn{1}{c}{Na\"ive} & \multicolumn{1}{c}{ABJ} & \multicolumn{1}{c}{$\log(L)$} & \multicolumn{1}{c}{$\log(L)$} & \multicolumn{1}{c}{$T_{\rm eff}$ (K)} & \multicolumn{1}{c}{$T_{\rm eff}$ (K)} \\
    \hline
Lower MS               & Luminosity                     & 0.3--0.7                        & 4000--6500                      & 16\,331 & 15\,787 &\nodata&\nodata&   281 &    317\\
Lower MS               & Luminosity                     & 1--2                            & 4500--7000                      & 52\,847 & 52\,202 &\nodata&\nodata&   308 &    343\\
Lower MS               & Temperature                    & 0.1--1                          & 4800--5200                      &  6\,120 &  6\,178 & 0.173 & 0.173 &\nodata&\nodata\\
Upper MS               & Temperature                    & 1--300                          & 7000--7500                      &  6\,191 &  6\,795 & 0.321 & 0.310 &\nodata&\nodata\\
Upper MS               & Temperature                    & 1--1000                         & 9000--9500                      &     209 &     225 & 0.360 & 0.373 &\nodata&\nodata\\
MSTO                   & Both                           & 2--7                            & 5500--6500                      & 80\,077 & 77\,064 & 0.149 & 0.149 &   251 &    259\\
Red clump (wide)       & Both                           & 15--150                         & 3650--5250                      & 32\,122 & 32\,489 & 0.220 & 0.219 &   256 &    278\\
Red clump (medium)     & Both                           & 20--100                         & 4050--5050                      & 26\,158 & 25\,535 & 0.194 & 0.190 &   217 &    223\\
Red clump (narrow)     & Both                           & 25--70                          & 4300--4900                      & 15\,924 & 14\,721 & 0.117 & 0.117 &   151 &    150\\
Upper RGB              & Luminosity                     & 200--300                        & 3650--4650                      &     139 &     137 &\nodata&\nodata&   233 &    240\\
\hline
\end{tabular}
\end{table*}
\end{center}

Although we emphasise the cautionary warnings above for the exact treatment of data, the magnitude of Lutz--Kelker effects in our data are relatively small. From the entire data set, 65 per cent of stars show no change in temperature and $<$0.1 per cent change in luminosity between the two datasets. The difference in fitting exceeds our quoted uncertainties in temperature and/or luminosity by 50 per cent in 19 per cent of cases, and by 100 per cent in 8.4 per cent of cases. Lutz--Kelker effects can therefore be considered an important contributor to the uncertainty budget in our derived parameters in $\sim$10--20 per cent of cases.

The correction applied to account for Lutz--Kelker effects depends implicitly on the assumptions made for the underlying population. Both the na\"ive method and the distance estimators for \citet{ABJ16} should properly only be used for single stars, and any extension to a population of stars should properly require a new correction to be applied based on that population's properties. However, for the purposes of this paper, we must firstly choose whether to apply that correction and, secondly, what that correction should be.

To test whether the \citet{ABJ16} results represent an improved derivation of stellar properties above our na\"ive parallax inversion, we take those 19 per cent of stars where the uncertainties exceed 50 per cent of our quoted uncertainties. From this, we select particular features of the H--R diagram where we except stars to fall on a particular, narrow sequence. If the distances of \citet{ABJ16} are a closer representation of the true distances, we should see the features in the H--R diagram become narrower, as stars become closer to their true luminosities.

Table \ref{LKTable} shows a number of features in the H--R diagram. A tight cut has been placed in luminosity and temperature, with the other parameter loosely constrained so as to remove significant off-sequence outliers. For the red clump, both parameters were constrained either loosely, moderately, or severely. The expectation is that the better-fitting dataset will provide a lower standard deviation in the loosely constrained parameter, plus have a larger number of stars falling in that region. Results can be affected on continuous distributions like the main sequence and giant branches by stars entering the selected region, which should reside in higher-source-density regions that bound it.

In general, there is very little to separate the results of the two different approaches. In general, the approach of \citet{ABJ16} most often produces a very similar luminosity constraint. However, it almost universally provides a worse fit in temperature. In most cases, it also provides a lower number of sources. The exception is on the upper main sequence where stars from cooler temperatures appear to scatter in from cooler temperatures, improving the source counts and reducing the standard deviation.

We therefore conclude that the approach of \citet{ABJ16} provides a worse fit to the main features of the H--R diagram, containing the majority of stars. This persists for different selections of $0 < \delta \varpi / \varpi \leq 0.414$. Nevertheless, we expect \citet{ABJ16} to provide a better fit in specific cases, particularly outside our fitted range ($\delta \varpi / \varpi > 0.414$), and for certain off-sequence regions (e.g.\ unexpectedly luminous stars, such as those in the Hertzsprung gap) where their method produces fewer scattered stars.

Simply providing a better fit to the H--R diagram does not mean that the na\"ive method is more valid for any given source. Nor does it mean that the Lutz--Kelker correction of \citet{ABJ16} (or any other study) should not be taken into account. However, for the purposes of simplicity, we have opted to explore the properties of the H--R diagram in our paper using the na\"ive method for determining distances. Any persons using this dataset are strongly advised to think carefully about how the Lutz--Kelker bias and related effects will affect their results.


\section{Discussion on interstellar extinction}
\label{AppendixEBV}

\begin{figure}
\centerline{\includegraphics[height=0.47\textwidth,angle=-90]{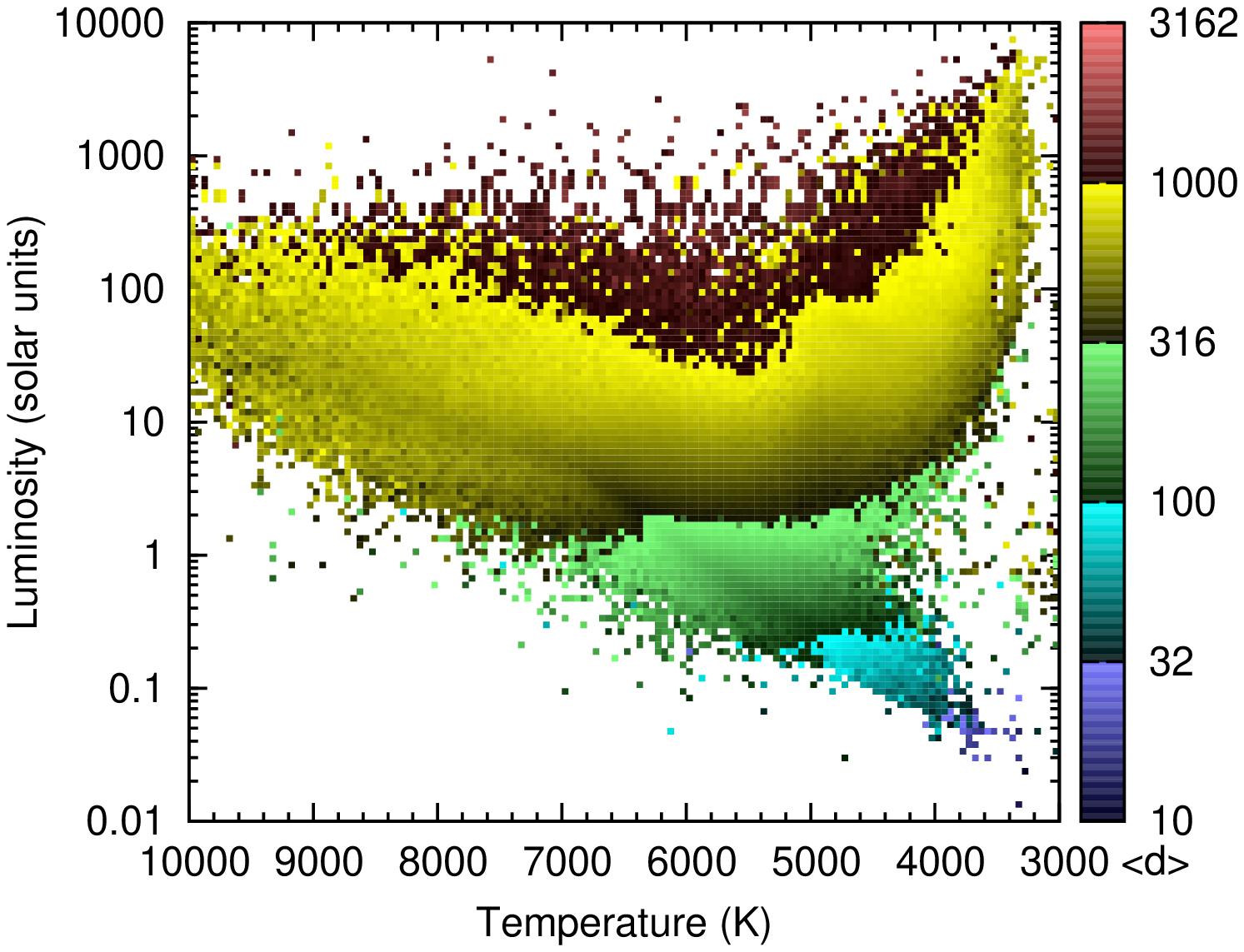}}
\centerline{\includegraphics[height=0.47\textwidth,angle=-90]{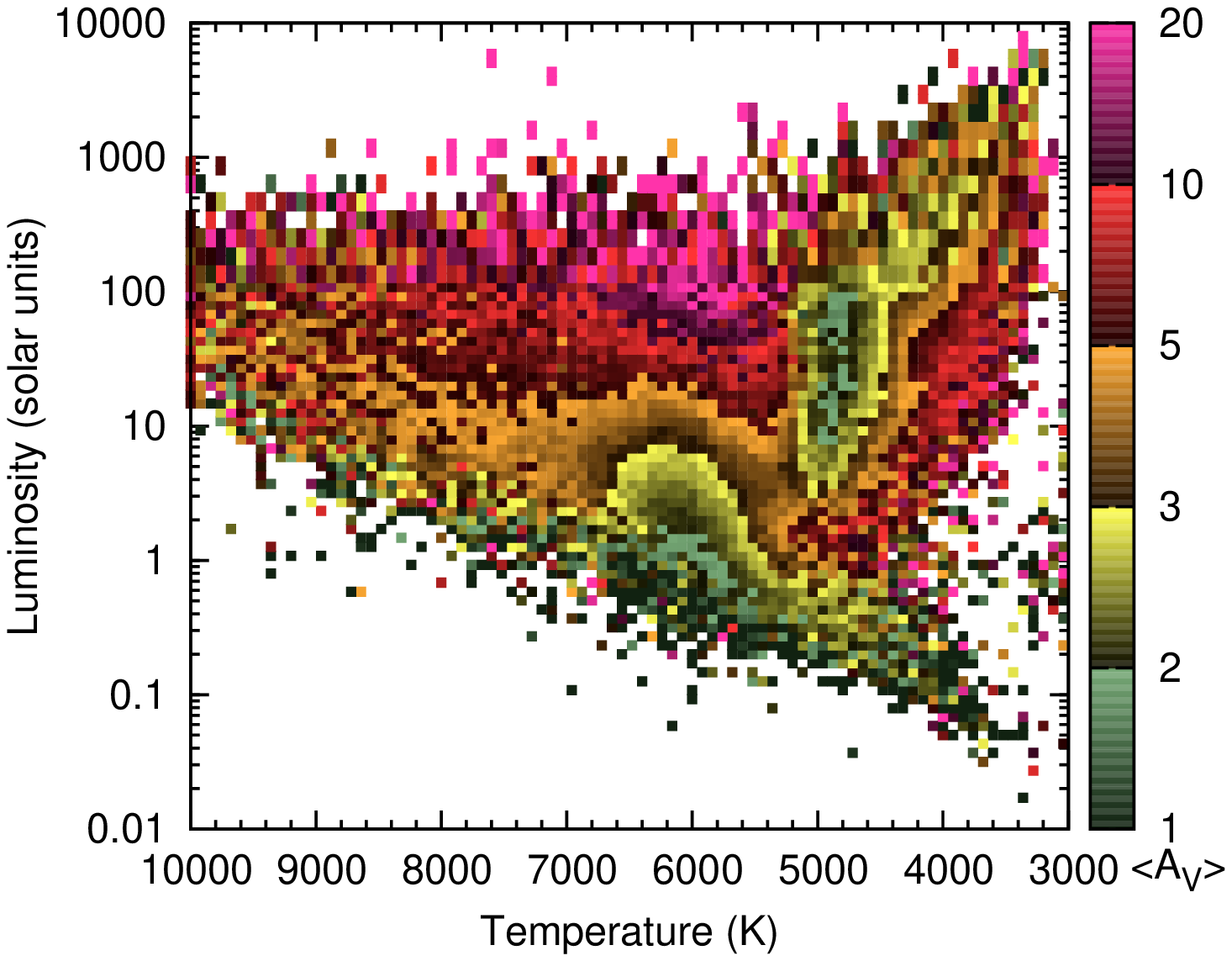}}
\caption{\emph{Top panel:} The average distance to stars of different regions in the H--R diagram. \emph{Bottom panel:} As top panel, for average line-of-sight interstellar extinction.}
\label{HRDAvFig}
\end{figure}

\subsection{General observations on extinction}
\label{DiscEBVGenSect}

Since the wavelength dependence of interstellar extinction is markedly different from that of a blackbody's Wien tail, the goodness-of-fit of a heavily extincted star should be significantly improved if the extinction is properly taken into account. We have established that the vast majority of stars which exhibit interstellar extinction will be made too optically faint to appear in our sample, and that extinction has a much more significant effect on the temperatures of warm stars than cool stars (Section \ref{HRErrorEBVSect}). For a careful selection of stellar types, it may therefore be possible to estimate which stars suffer how much extinction, and roughly where along the line of sight these extincting clouds exist.

Figure \ref{HRDAvFig} shows the average distances to stars of different fundamental parameters. Note that the median values may differ from the average, and that there is normally a substantial range within each bin. The plot against distance reveals several factors:
\begin{itemize}
\item Stars assigned to be more luminous tend to be at greater distances. This is expected, given the sensitivity limit of the observations, whereby luminous stars can be detected out to greater distances.
\item Stars above and below the main features of the H--R diagram (both the main sequence and the giant branches) tend to be at larger distances, resulting in vertical features in Figure \ref{HRDAvFig}. The fractional error in the parallax increases at larger distances, causing increased scatter in the luminosity. A manifestation of the Lutz--Kelker bias exists, whereby the scatter is preferentially towards higher luminosities, due to asymmetric errors in the distance \citep{LK73}.
\end{itemize}

Comparing this against the accompanying extinction plot, we can surmise the following:
\begin{itemize}
\item Cooler main-sequence stars lie along less-extincted lines of sight. This is expected as: (a) they are typically closer, hence the Galactic Plane extends to higher Galactic latitudes; and (b) they are typically older, hence come from populations with larger scale heights in the Galactic Plane.
\item Warmer giant stars lie along less-extincted lines of sight. This is expected, as warmer stars tend to be older and more metal-poor. Metal-poor stars have lower atmospheric opacity, hence they tend to be smaller and hence hotter at a given luminosity (e.g.\ \citealt{MGB+08}). However, stars may also scatter towards the cooler side of the giant branch if they exhibit interstellar (or circumstellar) extinction.
\item Stars within the Hertzsprung gap ($\sim$6000 K, 100 L$_\odot$) typically lie along very-high-extinction lines-of-sight. This may reflect the fact that the greater distances to these stars mean they almost invariably lie at low Galactic latitude, or it may be that these stars are scattered there by interstellar extinction.
\item Stars scattered away from the main sequence and giant branch are typically (though not universally) along highly extincted lines of sight.
\end{itemize}

\subsection{Extinction in the solar neighbourhood}
\label{DiscEBVMapSect}

\begin{figure}
\centerline{\includegraphics[height=0.47\textwidth,angle=-90]{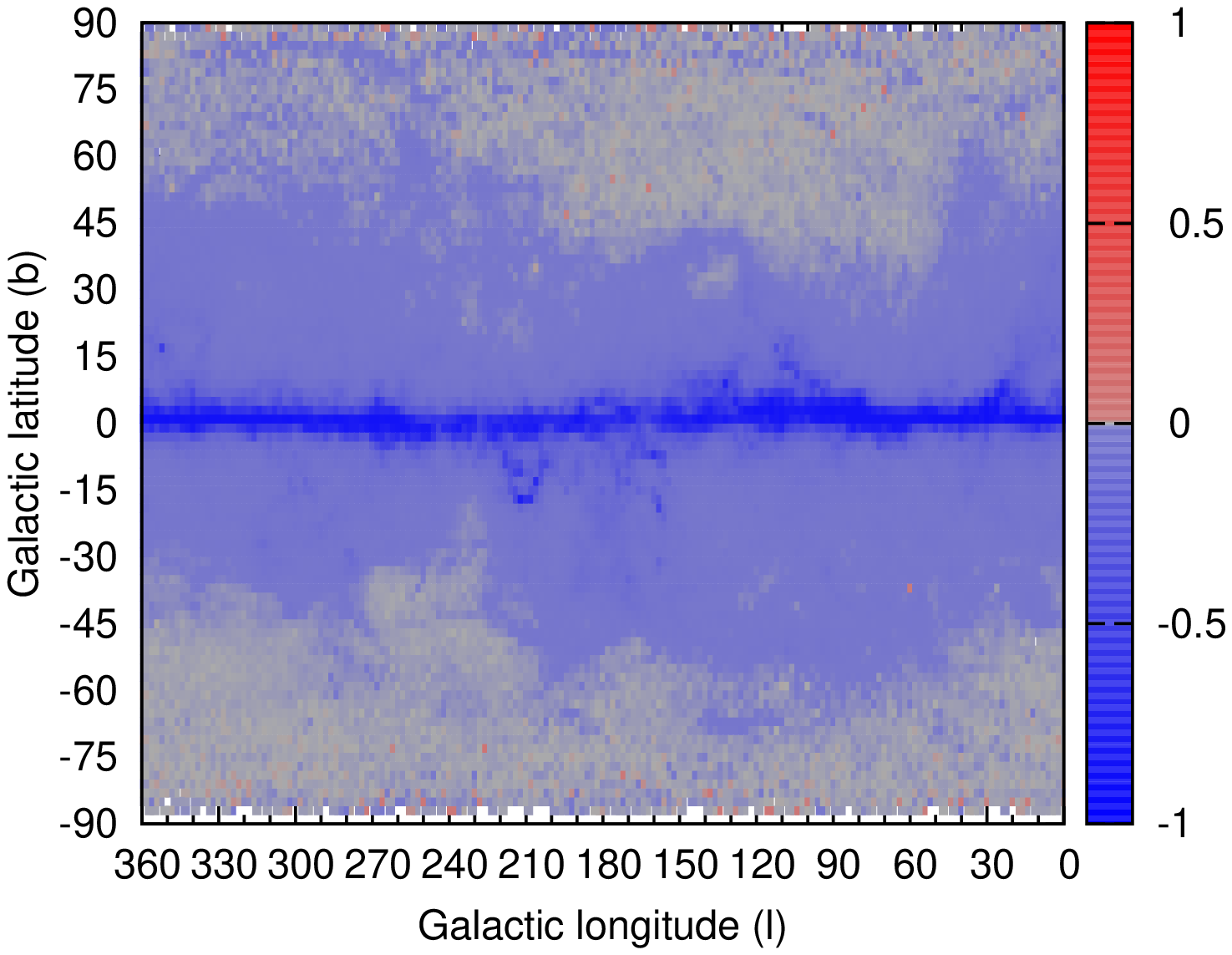}}
\centerline{\includegraphics[height=0.47\textwidth,angle=-90]{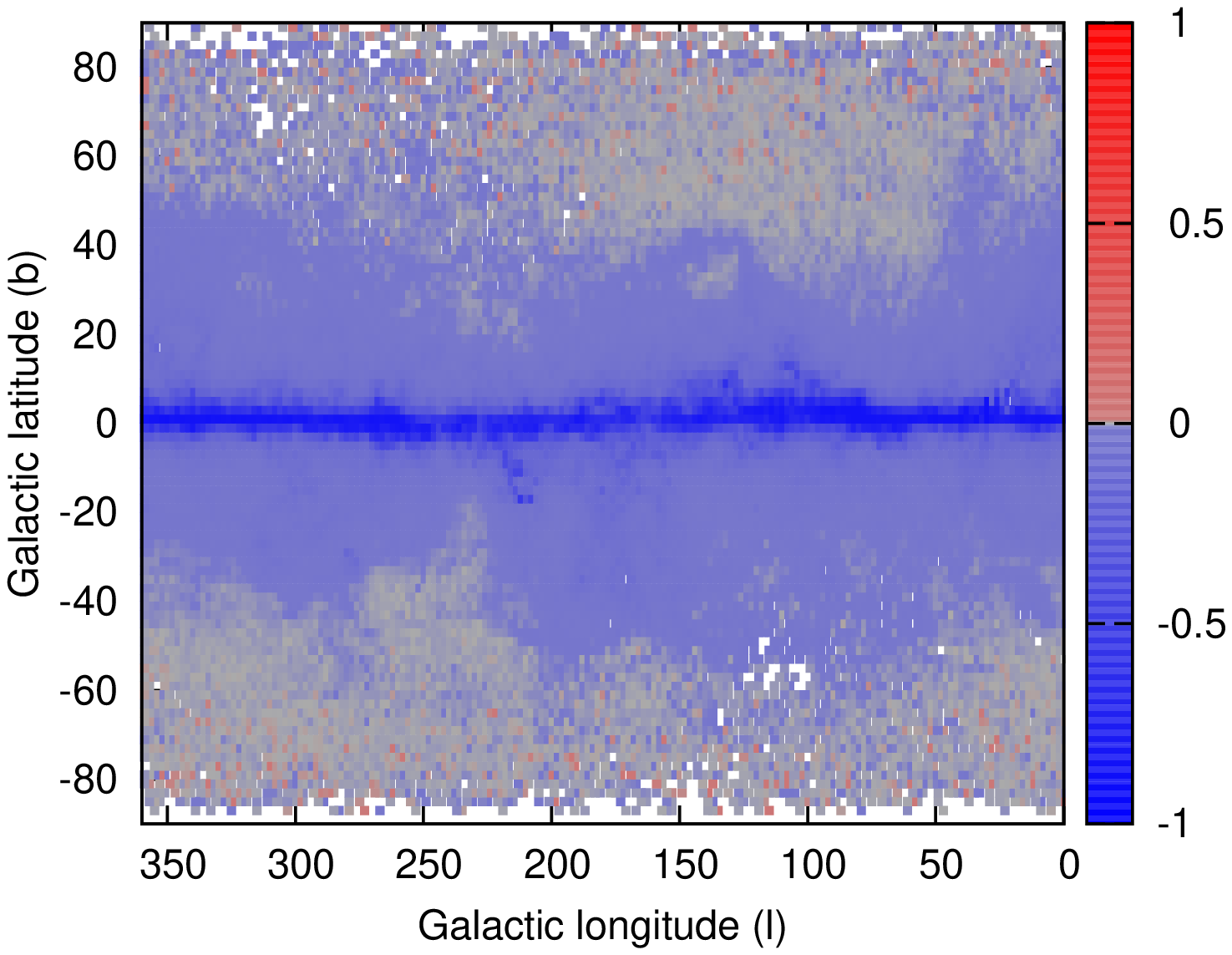}}
\caption{Map of extinction statistic $E$ (see Section \ref{DiscEBVMapSect}), averaged over square-degree bins of the sky. The colour scale runs from unextincted (dark blue, --1), probably unextincted (light blue, --0.01), no differentiation possible (grey, 0), probably extincted (light red, +0.01) and fully extincted (red, +1). The bottom panel shows only stars which are more than 500 pc from the Sun.}
\label{SkyAvFig}
\end{figure}

The \emph{Planck} line-of-sight extinction map can help determine which stars may suffer from extinction. These stars will be better fit by a model which has been reddened by the \emph{Planck} extinction measure than by the default assumption of zero extinction. To compute how well the star is fit, we can take the ratio:
\begin{equation}
E = \frac{Q_{\rm Av=0}-Q_{{\rm Av=}Planck}}{Q_{\rm Av=0}+Q_{{\rm Av=}Planck}}
\end{equation}
We expect $E = -1$ for a star with zero extinction, and $E = 1$ for a star suffering the full \emph{Planck} extinction.

Figure \ref{SkyAvFig} shows $E$ mapped out in square-degree bins on the sky. For regions with high extinction ($A_{\rm V} \gtrsim 1$), the stars always appear to be better fit by a model with less extinction, hence the stars must be predominantly in front of the extincting layer(s). At very high galactic latitudes, stars are frequently marginally better fit with an extincted model, suggesting that the dust causing the extinction is very local.

The bottom panel of Figure \ref{SkyAvFig} shows only stars more than 500 pc away from the Sun\footnote{Note that this sample is subject to a substantial Lutz--Kelker bias, and is presented for indication only.}. The same trends are seen here, although there is slight systematic shift to extincted models fitting better at all galactic latitudes, driven partly by the lower average fit quality of distant objects.

While these results do not substantially improve our understanding of extinction on their own, they do confirm our expectation that most stars in the \emph{Tycho} catalogue should not suffer substantial amounts of extinction. Better mapping of extinction could arise from comparing spectroscopically derived temperatures to the photometric temperatures computed here.


\section{Discussion on infrared excess}
\label{AppendixDust}

\subsection{Sources and spectral characteristics of infrared excess}
\label{DiscXSSourceSect}

Infrared excess is usually attributable to cool circumstellar material, e.g.\ companion stars, circumbinary discs, accretion or excretion discs, natal clouds of embedded sources in young clusters (e.g.\ the Pleiades), proto-planetary discs, ejecta of massive stars and cataclysmic variables, debris disks around main-sequence stars, and terminal winds of mass-losing stars, such as AGB (and potentially RGB) stars. Any remaining artefacts will also contribute, including spurious data, blended background galaxies, and poorly subtracted diffuse infrared backgrounds.

The variety of astrophysical categories makes it difficult to identify a single measure of infrared excess which maximises detection of astrophysically real sources, and minimises contaminants. Typically, though not exclusively, circumstellar material contains warm dust at temperatures up to the sublimation temperature ($\sim$1000 K; e.g.\ \citealt{GS99}). As our observational dataset cuts off near 25 $\mu$m, Wien's displacement law limits our sensitivity to dust at $\gtrsim$116 K. If we require two wavebands to show excess, this increases to $\gtrsim$132 K for \emph{WISE} [22], $\gtrsim$161 K for \emph{AKARI} [18] and $\gtrsim$256 K for \emph{WISE} [11.3].

Circumstellar dust is typically oxygen rich. Warm ($\gtrsim$few $\times$ 100 K) oxygen-rich dust exhibits strong Si--O stretching and O--Si--O bending modes near 9.7 and 19 $\mu$m, respectively: the precise wavelengths depend on the exact mineralogy of the dust. Carbon stars have their own features around 11.3 and 21 $\mu$m (e.g.\ \citealt{WOK+11,RKJ+15}), although an underlying continuum dominates their dust spectra. Thus, most circumstellar material will present a strong infrared excess longwards of 9 $\mu$m. The excess (as a fraction of the underlying continuum) typically increases with wavelength until at least 25 $\mu$m.

Many stellar types exhibit infrared excess at $\lambda < 9$ $\mu$m. If the star is still optically visible (hence detectable by \emph{Hipparcos}/Tycho-2), the excess is normally negligible at $\lambda \lesssim 4$ $\mu$m. In most cases, we therefore expect a slow rise in infrared excess between $\sim$4 and 25 $\mu$m, often including a sudden jump near 10 $\mu$m \citep[e.g.][]{BSvL+11,WOK+11,ASB+13,ITK+16}. There are exceptions. In some evolved stars, the 10 $\mu$m bump may be weak or absent \citep[e.g.][]{MSZ+10,MZS+16,SML+12,SKM+16}. In (proto-)planetary or circumbinary discs, and some other objects, emission may not become significant until 20 $\mu$m or longer \citep[e.g.][]{BFMK+13,DRSB+16}. Excreta or accreta of hot stars will typically not be dust rich: hot gas can exhibit substantial excess at $\lambda < 9$ $\mu$m \citep[e.g.][]{MBG+05,LCL16}. Hence, defining infrared excess to begin between $\sim$4 $\mu$m and $\sim$25 $\mu$m should identify most sources of infrared excess.

\subsection{Reddening of the central star and the role of geometry}
\label{DiscXSGeometrySect}

Infrared excess is usually attributable to stellar UV/optical light being absorbed by circumstellar material, especially dust, and reradiated in the infrared\footnote{Scattering of light by dust grains can also play a role, but this does not normally change the received stellar spectrum appreciably.}. Absorption by circumstellar dust mirrors interstellar reddening: absorption is stronger at shorter wavelengths. This reddens of the SED, lowering the photometric effective temperature\footnote{As stars become progressively obscured, the photosphere changes appreciably with wavelength, and the concept of a surface becomes ill-defined. This is particularly true of pulsating and aspherical stars. At some point there arises a distinction between the photosphere and temperature as traced by the SED, and those traced by optical or near-infrared spectroscopy.}. At this point, the spectroscopially and photometrically derived temperatures can deviate significantly from each other.

For completely obscured stars, the photosphere shifts into the dust envelope, the effective temperature declines below $\sim$3000 K, and the star disappears from the input \emph{Hipparcos}--Tycho-2 and \emph{Gaia} DR1 catalogues. The observational distinction between star and circumstellar material blurs. The photospheric flux normally becomes over-estimated in the mid-infrared, meaning the amount of infrared excess is underestimated. This effect is negligible for stars with mild infrared excess, but becomes significant for more obscured stars where the dust SED becomes comparable in flux density to the stellar SED and begins to affect the fitting procedure. Examples of such extreme sources are Herbig Ae/Be stars, like HIP 94260, and highly evolved AGB stars, like CW Leo.

The strength of this effect depends on the departure from spherical symmetry and geometric inclination, which dictate the obscuration in our line of sight. For example, face-on discs like HL Tau and TW Hya exhibit little extinction \citep{ALMABP+15,AWZ+16}, while edge-on discs like IRAS 04302+2247 and HK Tau exhibit very high extinction \citep{GWGR10,MDP+11}. Strong asymmetries also exist in some evolved stars, either as clumps or discs \citep[e.g.][]{RIH+14,LBC15,LKP+15,KHR+16}.


\section{Looking forward to future \emph{Gaia} releases}
\label{AppendixGaia}

This current paper serves in part to examine the challenges that must be solved to scale this work up to the full \emph{Gaia} sample of stars. \emph{Gaia} DR1 contains some 1.142 billion stars. An expected 200 million stars will have accuracies better than 10 per cent in the final \emph{Gaia} data release. The current work contains only 1.5 million stars. The challenges of this extra computation are not to be overlooked. The analysis for this paper took around 4.5 days per run on a modest eight-core workstation. While it is expected that efficiency savings can be made, it implies 4800 CPU-days will be needed for the entire \emph{Gaia} sample. Thankfully, the problem is largely parallelisable, but it is clear that a computing cluster or distributed computing will be necessary.

The photometric accuracy for well-behaved, single, unblended stars ($\pm\sim$120 K) is considerably better than can be achieved from integrated \emph{Gaia} BP/RP photometry alone ($\pm\sim$500 K\footnote{Gaia report GAIA-C8-TN-MPIA-DWK-001; http://www.rssd.esa.int/doc\_fetch.php?id=3168868}), but not as good as expected from detailed BP/RP spectroscopy ($\pm\sim$0.23 per cent at $G$ = 15 and $\pm\sim$1.3 per cent at $G$ = 18.5 mag, or roughly $\pm$12 K and $\pm$65 K on a typical 5000 K star\footnote{Gaia report GAIA-C8-TN-MPIA-CBJ-042; http://www.mpia.de/\~calj/ilium/itup\_tn.pdf}). If good enough photometric accuracy can be achieved on faint stars ($G \gtrsim 19$ mag), SED fitting should be more accurate than those currently employed by the \emph{Gaia} team. Due to the requirement for good photometry, this is likely to be limited to cooler stars ($\lesssim$4500 K). For stars with $G < 19$ mag, the difference between the photometric and spectroscopic errors can be used as an absolute calibration of interstellar reddening.

The key to obtaining good accuracy in temperature is good photometric input data. As future \emph{Gaia} data releases measure fainter stars, obtaining sufficiently high-quality photometry will become increasingly difficult. Obtaining a large quantity of good photometry is also necessary, so that one can identify and remove bad data, while keeping unusual but astrophysical sources. In this paper, this could be achieved for the \emph{Hipparcos} stars but not for the Tycho-2 sample. Particular challenges come from the southern hemisphere, which lacks SDSS data, and the Galactic Plane, where source confusion and high backgrounds hamper the accuracy of mid-infrared photometry. Several additional major surveys were not used in this work, but which could be used to improve the quality of the photometric fits. These include:
\begin{itemize}
\item The \emph{Spitzer Space Telescope} Legacy Programmes, especially the Galactic Legacy Infrared Midplane Survey Extraordinaire \citep[GLIMPSE][]{BCB+03,CBM+09} and the 24 and 70 $\mu$m Survey of the Inner Galactic Disk with the Multiband Imaging Photometer for \emph{Spitzer} \citep[MIPSGAL][]{CNCM+09} surveys, which contain higher-resolution infrared imagery of the Galactic Plane, which can reduce problems with high infrared background and stellar blending.
\item The European Southern Observatory (ESO) / Very Large Telescope Survey Telescope (VST) and ESO / Visible and Infrared Survey Telescope for Astronomy (VISTA) public photometric programmes. The VISTA Hemispheric Survey (VHS; \citealt{MBG+13}) will supercede the photometric depth and precision of 2MASS, supplemented in regions by the VISTA Kilo-Degree Infrared Galaxy Survey (VIKING; \citealt{ESK+13}) and VST optical surveys, notably the VST Atlas \citep{SBC+13} and VST Kilo-Degree Survey (KIDS; \citealt{dJKA+13}). The Galactic Plane will also receive substantial coverage from the VST Photometric H$\alpha$ Survey of the Southern Galactic Plane (VPHAS+; \citealt{DBF+13,DGSG+14}) in the optical and VISTA Variables in the Via Laceta survey (VVV; \citealt{MLE+10}) and its forthcoming extension\footnote{https://vvvsurvey.org/} in the near-IR. Early data releases are already available for some of the above surveys.
\item The Dark Energy Survey (DES; \citealt{ABC+06}) and Panoramic Survey Telescope \& Rapid Response System 1 (Pan-STARRS) `3$\varpi$' survey; \citealt{CMM+16}) should allow considerable improvement on the accuracy of optical photometry at high galactic latitude.
\item For bright stars, narrow-band surveys like the Javalambre Physics of the Accelerating Universe Astrophysical Survey (J-PAS; \citealt{BDM+14}) will provide great constraint on the optical SED of stars, allowing significant reduction of scatter.
\end{itemize}

In addition to data volume, the choice of quality cuts made in this work have often been semi-arbitrary, and not necessarily optimised. Many of these are related to our use of the VizieR cross-correlation tool, which was utilised for its speed in this work. This tool does not allow source matches based on anything other than simple proximity. Potential improvements for future data releases include:
\begin{itemize}
\item Better use of catalogue flags for identifying the correctly matching source. The ability to differentiate between (e.g.) parent and child objects within SDSS would improve the quality of the matching to those surveys.
\item Better use of catalogue flags for removing bad data. In this work, cuts were made to bad data based on goodness-of-fit. While catalogue flags for (e.g.) artefacts and saturated sources are not perfect, in many cases they may improve on these cuts.
\item Use of catalogue magnitudes to identify the best match. Many of the bad data flagged by our analysis was in catalogues where the correct source was identified as saturated and removed from the catalogue, while (fainter) nearby or child sources were identified to be the `correct' match by the VizieR algorithm. A check for a magnitude consistent with that of more complete samples would aid the photometric matching.
\item Accounting for proper motions of stars. Stars are assumed to be fixed for this work at the positions listed in the original \emph{Hipparcos} and Tycho-2 catalogues. The majority of sources with small proper motions ($\lesssim$80 mas yr$^{-1}$) should be matched in the majority of catalogues, as the majority of catalogues we use were published within a few years of those results. However, 12\,525 \emph{Hipparcos} sources and 66\,820 Tycho-2 sources have proper motions greater than this, and data on these sources may be missing from the merged catalogue. Propagation of source co-ordinates to the catalogue epoch should improve in substantial increases in photometric accuracy for this few per cent of nearby sources.
\end{itemize}

The combination of improved data volume and quality should allow a revision of this work to broadly match the accuracies achieved in this work, but on the much fainter stars which will be present in future \emph{Gaia} releases.


\label{lastpage}

\end{document}